\pdfoutput=1
\documentclass[acmsmall,table]{acmart}

\setcopyright{acmlicensed}
\copyrightyear{2018}
\acmYear{2018}
\acmDOI{XXXXXXX.XXXXXXX}

\acmJournal{JACM}
\acmVolume{37}
\acmNumber{4}
\acmArticle{111}
\acmMonth{8}
\usepackage{graphicx}
\usepackage{aas_macros}

\usepackage{amsmath, amsthm, amssymb, url, color, enumitem}
\usepackage{algorithm}

\usepackage{listings}
\usepackage{makecell}
\usepackage{multirow}

\theoremstyle{plain}
\newtheorem{design}      {Design Decision}
\newtheorem{rationale}   {Design Rationale}
\newtheorem{extension}   {C++ Annotation}
\newtheorem{hypothesis}  {Hypothesis}

\usepackage{hyperref}

\usepackage{graphicx}
\usepackage{xcolor}
\newcommand{\redcol}{\cellcolor{red!25}}
\newcommand{\greencol}{\cellcolor{green!25}}

\usepackage{xspace}
\newcommand{\swift}{{\sc swift}\xspace}
\renewcommand{\vec}[1]{{\mathbf{#1}}}
\renewcommand{\d}{{\rm d}\xspace}

\usepackage[final,commandnameprefix=ifneeded]{changes}

\definechangesauthor[name=Review1.1, color=gray]{R1}  
\definechangesauthor[name=Review1.2, color=gray]{R2}  
\definechangesauthor[name=Review1.3, color=gray]{R3}  
\definechangesauthor[name=Review1.4, color=gray]{R4}  
\definechangesauthor[name=Ours, color=gray]{Ours}     
\definechangesauthor[name=Review2.1, color=blue]{R21} 
\definechangesauthor[name=Review2.2, color=red]{R22}
\definechangesauthor[name=Ours2, color=green]{Ours2}

\newcounter{syntax}
\makeatletter
\newenvironment{syntax}[1][htb]{%
  \let\c@algorithm\c@syntax
  \renewcommand{\ALG@name}{Extended Syntax}
  \begin{algorithm}[#1]%
  }{\end{algorithm}
}
\makeatother

\newcounter{code}
\makeatletter
\newenvironment{code}[1][htb]{%
  \let\c@algorithm\c@code
  \renewcommand{\ALG@name}{Source code}
  \begin{algorithm}[#1]%
  }{\end{algorithm}
}
\makeatother

\begin{document}

\title[C++ with HPC extensions]{
 An extension of C++ with memory-centric specifications for HPC to reduce memory footprints and streamline MPI development
} 

\author{
 Pawel K.~Radtke
}
\email{
  pawel.k.radtke@durham.ac.uk
}
\affiliation{%
  \institution{Computer Science, Durham University}
  \city{Durham}
  \country{Great Britain}
}

\author{
 Cristian G.~Barrera-Hinojosa
}
\email{
  cristian.barrera@uv.cl
}
\orcid{1234-5678-9012}
\affiliation{%
  \institution{Instituto de Fisica y Astronomia, Universidad de Valparaiso}
  \city{Valparaiso}
  \country{Chile}
}

\author{
 Mladen Ivkovic
}
\email{
  mladen.ivkovic@durham.ac.uk
}
\affiliation{%
  \institution{Computer Science, Durham University}
  \city{Durham}
  \country{Great Britain}
}

\author{
 Tobias Weinzierl 
}
\email{
  tobias.weinzierl@durham.ac.uk
}
\authornotemark[1]
\affiliation{%
  \institution{Computer Science, Durham University}
  \city{Durham}
  \country{Great Britain}
}

\renewcommand{\shortauthors}{Radtke et al.}

\begin{abstract}
  \replaced[id=Ours2]{C++}{The C++ programming language} \replaced[id=Ours2]{leans}{and its cousins lean} towards a memory-inefficient storage of structs:
The compiler inserts \replaced[id=Ours2]{padding}{helper} bits\deleted[id=Ours2]{ such that individual attributes fit into bytes, and it adds bytes aligning attributes with cache lines}, while it is not able to exploit knowledge about the range of integers, enums or bitsets.
Furthermore, the language provides \replaced[id=R21]{no support for}{neither support for data exchange via MPI nor for} arbitrary floating-point precisions.
We propose a \deleted[id=Ours2]{C++} language extension based upon \deleted[id=Ours2]{C++} attributes through which
developers can guide the compiler what memory arrangements would be beneficial:
Can multiple booleans \added[id=Ours]{or integers with limited range} be squeezed into one bit field, 
do \replaced[id=Ours]{floating-point numbers}{floats} hold fewer significant bits than in the IEEE standard, \replaced[id=R21]{and is a programmer willing to trade attribute ordering guarantees for a more compact object representation}{or does the code \replaced[id=Ours]{benefit from a}{require a user-defined} MPI datatype for \deleted[id=Ours]{certain} subsets of attributes}?
The extension offers the opportunity to
fall back to normal alignment \deleted[id=Ours]{and padding rules} \added[id=R21]{and native C++ floating point representations} via plain C++ assignments,
no dependencies upon external libraries are introduced, and the resulting code
remains \added[id=R22]{(syntactically)} standard C++.
\added[id=R21]{
 As MPI remains the de-facto standard for distributed memory calculations in C++,
 we furthermore propose additional attributes which streamline the MPI datatype modelling in combination with our memory optimisation extensions.
}
Our work implements the language annotations within LLVM and demonstrates their potential impact\deleted[id=Ours2]{, both upon the runtime and the memory footprint,} through smoothed particle hydrodynamics \deleted[id=Ours2]{(SPH)} benchmarks.
They uncover the potential gains in terms of performance and development productivity.

\end{abstract}




\maketitle



\section{Introduction}

%
%
Switching from a low-level (machine) programming language to a generic high-level
language such as \deleted[id=R22]{C, } C++\deleted[id=R22]{,} or Fortran makes programming more
efficient, increases the code performance, and it introduces
machine-portability:
Source code is not tied to one architecture's instruction set anymore as long as
fitting compilers are available, while the translator can take over the lion's share of work to make the low-level code
fast.
This tuning includes \replaced[id=R4]{hardware-optimised}{proper} memory alignment and padding\added[id=R4]{ within the limits set by \replaced[id=R22]{the}{C++’s} Application Binary Interface (ABI) \added[id=R22]{for the respective translation toolchain}}.
Some developers have the skills and knowledge to tweak the memory layout\added[id=Ours]{ manually}
and, through this, to produce faster code than a compiler, but it is generally
difficult to compete with a good compiler which has access to heuristics reflecting the internals of a machine.

%
%
One dominant high-level language family in scientific computing is \deleted[id=R22]{C/}C++ with its cousins \added[id=R22]{C,} CUDA and SYCL~\cite{Reinders_Ashbaugh_Brodman_Kinsner_Pennycook_Tian_2021}. 
Fortran remains the other prominent language to realise core software in high-performance computing (HPC). Our work focuses on C++ and starts from the identification of some shortcomings within C++ which adversely affect HPC developers.

First, the C++ language yields classes with a large memory footprint.
Since we are interested in data arrangements, we use struct and
class as synonym from hereon, assuming that a class is a struct with different
default visibility constraints\deleted[id=R4]{ plus, in some cases, a pointer to a virtual
function table \cite{Hyde:2017:WriteGreatCodeVol2}}.
A struct's members are aligned in memory by introducing padding bytes. 
Further to that, the smallest memory unit that can store a variable is a byte,
which provides a poor information density for a boolean.
As it can only hold true or false, one bit would be enough to encode its
information.
Enumerations suffer from this over-provisioning of memory, too.

Second, the C++ language lacks support for a ``continuous'' range of data
precisions.
It offers datatypes which are natively supported by hardware, yet does not allow
programmers to express further knowledge about the value ranges of integers or
the actual accuracy of a numerical datatypes (number of significant bits). This
again affects the memory footprint of applications and makes programming for
different datatypes (mixed precision programming) \cite{Higham:2022:MixedPrecision} 
laborious.

Finally, the C++ language does not offer built-in support for distributed 
memory parallelisation through the Message-Passing Interface (MPI)
\cite{Gropp:2014:AdvancedMPI}.
MPI remains the de-facto standard to program supercomputers.
If developers want to map C++ structs onto MPI, they have to translate the
struct's \replaced[id=R1]{instance variables}{attributes} manually into memory addresses and trigger some address
arithmetics.
This quickly becomes error-prone
and time consuming, notably once we want to support different MPI types per
struct which exchange different subsets of \replaced[id=R1]{a struct's instance variables}{attributes}.

Our work is driven by the hypothesis that these shortcomings of C++ often have 
a negative impact on the quality of scientific software design and its
performance.
In an era where the CPU--memory gap is
widening~\cite{Dongarra:2011:ExascaleSoftwareRoadmap}, memory modesty gains importance.
Codes with small memory footprint have reduced memory bandwidth
requirements and are able to retain more data within the caches close to the
core.
They perform better.
In an era where the memory per core is stagnating, weak scaling per node is
constrained.
Codes with \added[id=Ours]{a} small memory footprint can squeeze larger problems onto a single
node and hence run into strong scaling saturation later.
In an era where the energy consumption of computers---a metric determined by memory movements---gains importance, the science per moved byte,
i.e.~the information density, deserves particular attention.
Codes should use every single bit to hold meaningful information.
In an era where the interconnect bandwidth struggles to keep pace with the
per-node performance, it is important to minimise the memory footprint per information exchanged
between nodes.

%
%
We propose novel C++ annotations to address the language's shortcomings.
We also prototype a LLVM modification supporting
the new annotations.
Our annotations, firstly, allow developers to mark booleans, enumerations or
integers with constrained ranges to indicate that they should be packed into one
large bitfield within the struct.
Our compiler automatically supplements accesses to struct \replaced[id=R1]{members, i.e.~its instance variables and class variables,}{attributes} with the 
required bit shift operations.
Secondly, 
we propose that floating-point data are annotated with \deleted[id=Ours]{information}
density information:
The struct's floating-point data are held in a compressed bit representation
(smaller than built-in hardware datatypes) and mapped to and from native
datatypes throughout computations.
Finally, 
our compiler extension accepts MPI datatype annotations for those
struct members that are to be exchanged via message passing.
Different MPI views, i.e.~subsets of \replaced[id=R1]{instance variables}{attributes} that are to be exchanged, can be specified straightforwardly.
\added[id=R21]{These MPI extensions are optional and work with or without the memory layout modifications.}

%
%
\replaced[id=R21]{
 C++ attributes allow users to write code that remains valid even when the attributes are unrecognised by the compiler, as unknown attributes are simply ignored by the translator.}{C++ annotations allow users to stick to code which complies with the C++ standard. If a compiler does not ``understand'' the annotations, they are
ignored.}
Annotations can be applied incrementally, i.e.~do not require a code
refactoring/rewrite to unfold their potential.
We also do not introduce any dependencies on external libraries\added[id=R21]{ besides the explicit connection to MPI for the bespoke MPI attributes}.
Realised as additional code transformation pass, our language extensions 
play in a team with other compiler optimisations, while 
hiding how data are internally encoded from the user.
This provides a ``native'' way for developers to toggle between various data
representations:
Simple assignments \replaced[id=R4]{to built-in variables switch}{ change} from \added[id=R4]{our} memory-optimised to \replaced[id=R4]{the ABI's}{default (performance-optimised)} data representations which are subject to proper alignment, padding and mapping onto hardware-supported data formats.
\deleted[id=R4]{Any compiler will eliminate these assignments if our extensions are not
supported.}

C++ provides means to eliminate padding and to control alignment.
They overwrite built-in compiler heuristics~\cite{Hyde:2017:WriteGreatCodeVol2}.
However, they \deleted[id=R4]{do not operate on the bit-level,} require manual intervention and \replaced[id=R4]{do not allow developers to work outside the ABI's memory arrangement guarantees even if intended}{quickly break code}.
MPI provides the means to wrap C++ classes into bespoke MPI datatypes.
However, no genuine C++ integration exists\added[id=R4]{ \cite{Hillson:2000:Cpp2MPI}}, i.e.~defining MPI datatypes requires \replaced[id=R21]{byte-level}{bit-level} address manipulation on the developer side and introduces significant syntactic overhead~\cite{Renault:2006:MPIPreProcessor}.
\added[id=R21]{
 Such manual address and offset arithmetics are also incompatible with attributes which alter and permute the internal memory representation of structs under the hood. In particular, any manual offset computation becomes invalid once the compiler is permitted to reorder, compress, or bit-pack fields.
}
The C++ language offers a small number of floating-point data types.
Symbolic, high-level programming environments such as Matlab or NumPy support
generic, flexible precisions such that
developers can focus on methodological challenges
\cite{Carson:2023:MixedPrecisionSPAI,Higham:2022:MixedPrecision}.
However, it remains unclear to which degree the developed algorithms 
translate one-to-one into production-ready C++ code.
Most multi-precision codes therefore stick to built-in precisions, i.e.~rely on ``specialisations'' of generic algorithmic building blocks for few
hardware formats
(cmp.~for
example
\cite{Abdelfattah:21:SurveyMixedPrecision,Abdulah:2022:GeostatisticalModeling,Cao:2022:GordonBell,Carson:2023:MixedPrecisionSPAI,Doucet:2019:MixedPrecision,Langou:2006:Fp32ToFp64Precision,Ltaief:2023:GordonBell}).
In C++, \replaced[id=R1]{templates provide}{template meta programming provides} a mechanism to write \added[id=R1]{such} precision-generic realisations.
However, \replaced[id=Ours]{template programming}{it} works only over types which offer all operators used within the
templated code\added[id=R22]{, while the produced code has to operate within the ABI constraints}. 
Template \deleted[id=R1]{meta} programming also \deleted[id=R1]{introduces syntactic overhead compared to
plain realisations with built-in data types, and it} increases compile times\replaced[id=R1]{---although this might not be a major stumbling block anymore on today's systems and with today's compiler generations---}{.
Introducing templates typically} requires \deleted[id=R1]{major} code rewrites and ripples through the implementation.

Finally, we can use bespoke libraries to provide support for multifaceted
or flexible precision \cite{Lindstrom:2014:ZFP, Flegar2019FloatX, Fousse2007MPFRAM}.
However, switching to user-defined data
types (bespoke classes with higher information density and non-native floating
point formats) runs risk to make a code incompatible with third-party libraries
if they are not prepared to utilise different data types\added[id=R3]{ and requires us to maintain these data types}, while any \replaced[id=R3]{embedding of bespoke types into user classes might}{ introduction
of such a type can} hinder the compiler to perform further optimising memory
rearrangements \cite{Hyde:2017:WriteGreatCodeVol2}.
Our approach \replaced[id=R22]{avoids many of these disadvantages, at the cost of requiring compiler support for the proposed annotations and violating ABI compatibility}{has none of these disadvantages} and indeed can be read as an embedded
domain-specific language (DSL) or language extension, which streamlines the
development of numerical HPC codes.


%
%
In the long term, we expect many supercomputing projects to benefit from our
ideas.
For the present paper, 
we illustrate the potential impact by means of a simple smoothed particle
hydrodynamics (SPH) code
inspired by~\cite{Schaller:2016:Swift,Schaller:2023:Swift}.
SPH is one well-established method for simulating fluids of complex
structure~\cite{Lind:2020}
or a vast dynamic range~\cite{Price:2012:SPH} with moving particles.
They are typically administered within a dynamically adaptive mesh.
Maintaining the dynamically adaptive mesh plus the particle-mesh relations
induces an (integer) data overhead.
As the particles move, SPH requires frequent spatial resorting of particles between MPI ranks~\cite{Oger:2016}, while
particles interact between MPI ranks each and every time step. 
The resorting typically requires the migration of the whole particle, while the
exchange of few particle properties suffices to realise the particle-particle
interactions in most SPH steps.
We need different MPI data views, i.e.~exchange different \replaced[id=R1]{variable}{attribute} subsets depending on the algorithmic context.
SPH often suffers from strong scaling limitations.
We have to keep the particles' memory footprint low to allow for bigger
simulations.
In addition, application scientists face pressure in their domains for ever-increasing
simulation sizes and resolutions, which directly translates to the total number
of particles they are able to fit onto a machine. That number is currently in the
order of hundreds of billions and growing \cite{schayeFLAMINGOProjectCosmological2023}.
As such, keeping the particles' memory footprint as low as possible is of vital concern.
In this context, empirical evidence suggests that SPH particles hold some
floating-point quantities which do not require full single or double precision
\cite{hosonoEfficientImplementationLoworderprecision2024}.
On the whole, we consider SPH as a prime example of
an application that benefits from our proposed C++ annotations.
Many other application domains face similar challenges.

%
%
Our work is organised as follows:
We first present our SPH use case in
Section~\ref{section:use_case_SPH}.
The rough algorithmic sketch of SPH principles highlights some fundamental
challenges arising from such codes.
In Section~\ref{section:language}, we discuss properties of a direct translation
of the algorithmic steps into plain C++, its properties, and what a
better-suited implementation would look like.
This allows us to introduce our new C++ annotations as well as the underlying
code transformations triggered by them.
The manuscript continues with a discussion of how these code transformations are
realised within LLVM (Section~\ref{section:translation}).
We return to the SPH demonstrator in
Section~\ref{section:demonstrator} for a review of the potential impact of the
extensions, before we assess the observed impact in
Section~\ref{section:results}.
A brief outlook and summary in
Section~\ref{section:conclusion} close the discussion.

\section{Use case: Smoothed Particle Hydrodynamics}
\label{section:use_case_SPH}

\begin{figure}[htb]
 \begin{center}
  \includegraphics[width=0.32\textwidth]{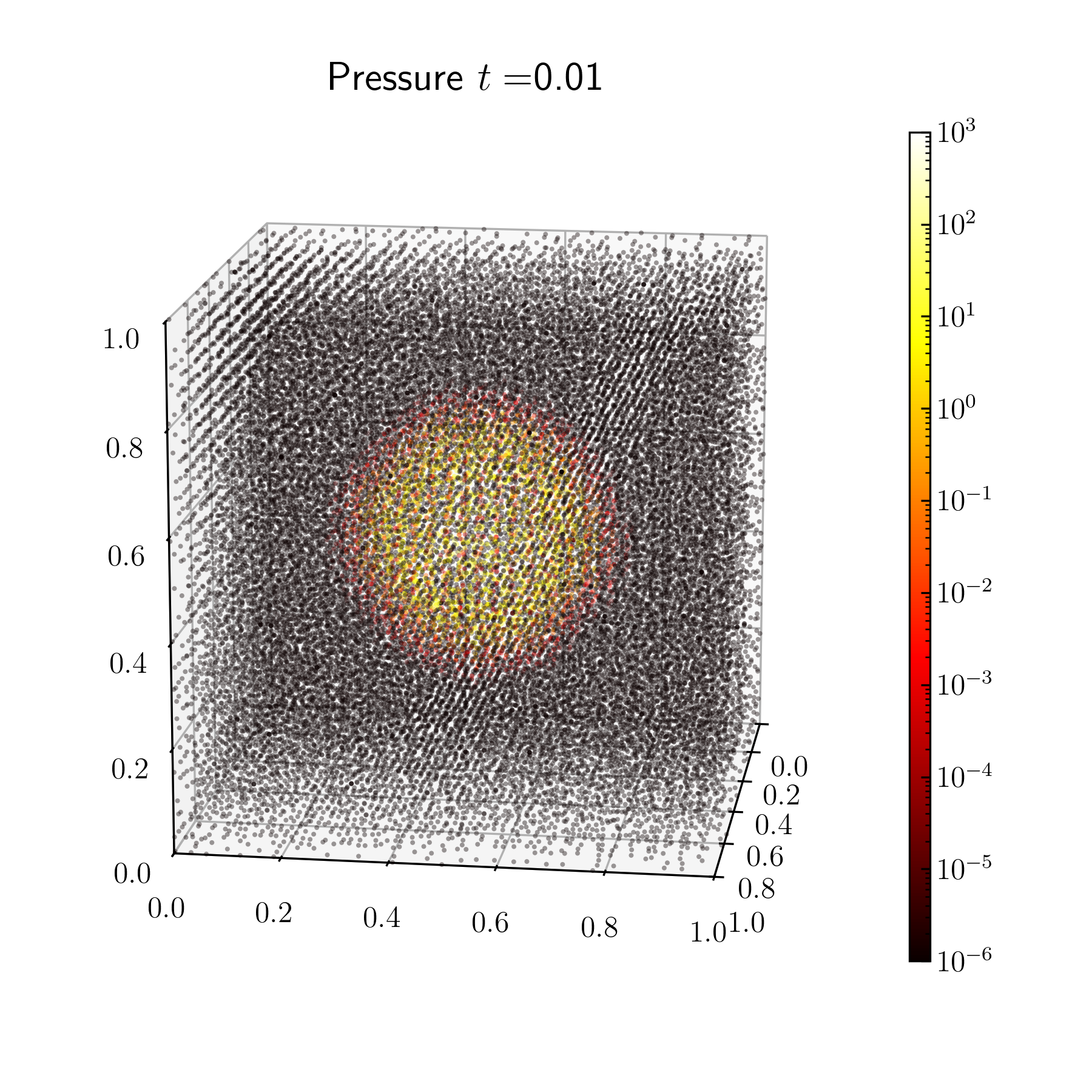}
  \includegraphics[width=0.32\textwidth]{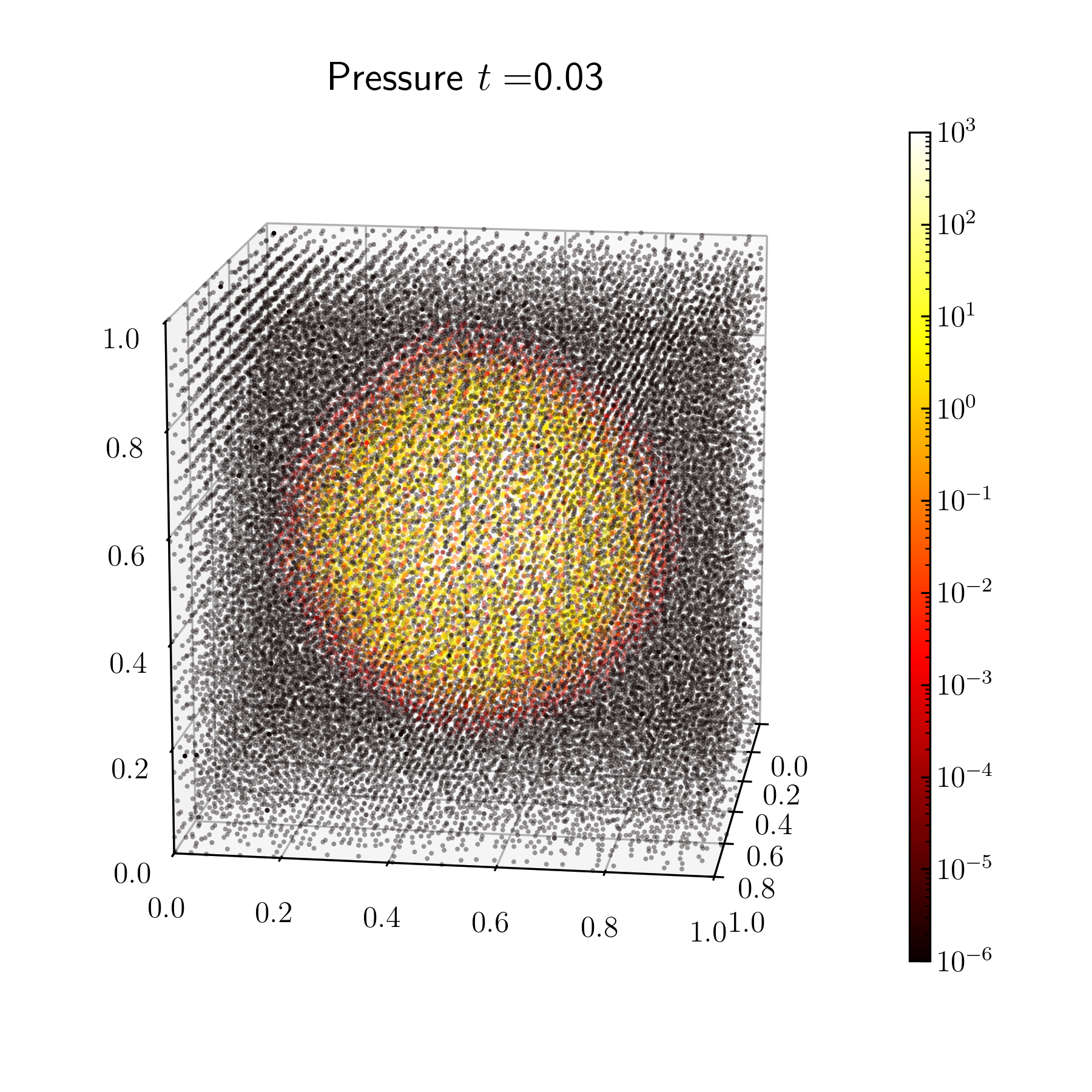}
  \includegraphics[width=0.32\textwidth]{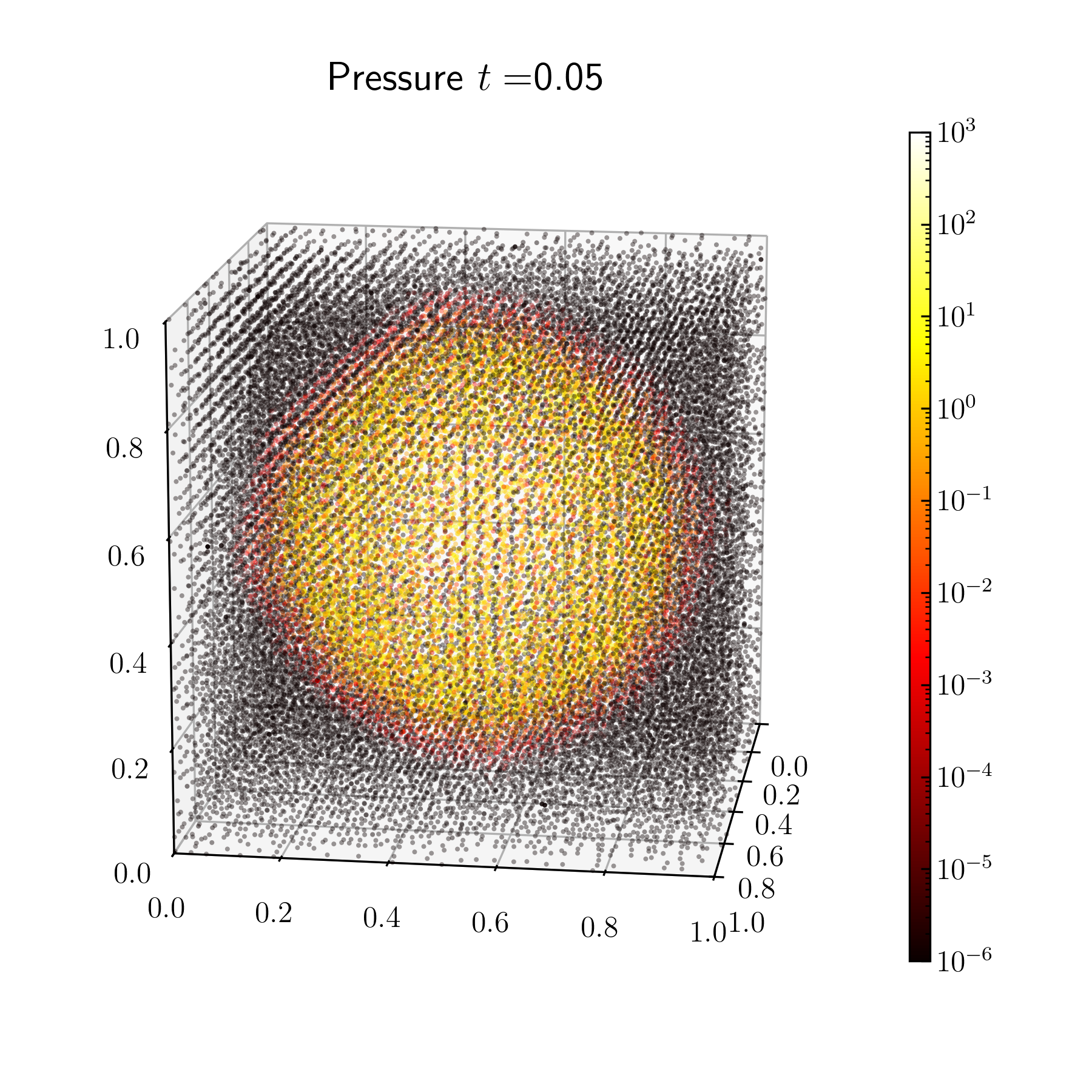}
 \end{center}
 \caption{ 
   \replaced[id=Ours]{
    The Sedov blast is a classic SPH benchmark: 
    Particles are initially scattered homogeneously over the domain with only one particle in the domain centre assigned high energy.
    This leads to a radial pressure shock expanding over $t \in \{0.0101,0.0301,0.05\}$ (from left to right).
   }{   
   The Noh problem is a classic SPH benchmark:
   All fluid particles are initially positioned in a Cartesian layout
   (left) and are given a velocity towards the box center.
   The flow therefore develops a high-density region in
   the centre (middle), which eventually develops into a shock pushing fluid
   outwards again (right).
   Two-dimensional toy setup with $64^2$ particles over a regular,
   time-invariant $3 \times 3$ grid.
   }
   \label{fig:noh-problem-setup}
  }
\end{figure}

%
%
Smoothed Particle Hydrodynamics (SPH) is used to model complex physical systems
in a wide domain of computational sciences ranging from engineering to
astrophysics~\cite{Gingold:77:SPH,Lind:2020,Monaghan:92:SPH,Price:2012:SPH}. \deleted[id=Ours]{See~\cite{Lind:2020}
and~\cite{Price:2012:SPH}, respectively, for state-of-the-art reviews on these
two application domains.} 
In SPH, a fluid of interest is discretised in terms of particles suspended in a
computational domain\added[id=Ours]{ (Figure~\ref{fig:noh-problem-setup})}, and the dynamics of the system are described by
a set of equations for the interaction and evolution of these particles.
SPH codes' physics are encapsulated within the implementations of the 
particle-particle interaction and the particles' evolution.
\deleted[id=Ours]{As we have a finite number of particles, SPH discretises the continuous fluid flow into a finite number of equations.}
As each \added[id=Ours]{SPH} particle is equipped with a finite search radius and only particles within each \replaced[id=Ours]{others's}{other} search radius interact, the arising discretised system of equations is sparse.


%
%
Conceptionally, SPH boils down to a temporal combination and arrangement of
relatively simplistic steps per time step per particle:

\begin{enumerate}
  \item 
    The density field of the fluid at the particle's position $x_i$ is calculated based on the local
    distribution of particles. Around a given particle, only a compact set of
    particles, i.e. a neighbourhood, contributes to the value of the density
    $\rho(x_i)\equiv\rho_i$.
    Optionally, the size of the neighbourhood can be adjusted by solving a nonlinear implicit
    equation per particle that depends on the density.
  \item 
    The particle's acceleration due to pressure gradients as well as the
    change in its internal energy are calculated. These
    calculations require information from the neighbours around each particle.
    Hence, their algorithmic intensity depends on the size of the neighbourhood.
  \item 
    Finally, the particle's position, velocity, and internal energy are updated
    by integrating the equations of motion forward in time. Unlike the
    previous two steps, these updates do not require any exchange of information with any neighbour.
\end{enumerate}

\noindent
These three basic steps are typically complemented with some global reduction and
broadcast phases, e.g.~to identify the global admissible time step size.

%
%
Although only particle data structures are needed for SPH, most simulations
use a grid---among other meta data such as Verlet lists or
Cell Linked Lists \cite{Dominguez:2011:NeighbourLists}---as
a helper structure to find neighbours efficiently.
Binning the particles into a mesh allows us to search only through a small set
of particles per time step for potential interaction partners:
Two particles interact if and only if they are held within the same or two
adjacent, i.e.~vertex-connected, mesh cells.
We use the grid as lookup mechanism.
To make this work, the mesh cells have to have a size that is at least twice the
maximal interaction radius of all particles held within the cell.
Once the interaction radius, i.e.~neighbourhood, changes, the grid should
change, too, leading to the following additional algorithmic steps:

\begin{enumerate}
  \setcounter{enumi}{3}
  \item As the particles move, we have to update the particle-mesh association.
  We have to resort.
  \item As the particles move and their density and interaction radius change,
  the mesh has to be adapted, i.e.~refined and coarsened locally.
\end{enumerate}

\noindent
Our prime area of interest is cosmological simulations, where some particles move
quickly, i.e. with a relative difference of several orders of magnitude compared to other particles contained in the simulation domain.
These particles have to be resorted frequently. 
Therefore, we commit to an array of structs (SoA) data layout where the
particles are administered within an adaptive Cartesian mesh\added[id=R1]{, and the physical attributes of a particle are mapped onto the particle struct's instance variables, while the mesh entities, i.e.~vertices and cells, are modelled as structs, too.
A commitment to SoA does not imply that we stick to this data format globally:
Some compute kernels dynamically resort to structures of arrays (AoS), i.e.~might rearrange data temporarily and locally \cite{Radtke:2025:AoStoSoA}}.


The algorithmic sketch outlined above highlights that

\begin{itemize}
  \item the adaptive mesh is a pure helper data structure lacking physical
  quantities. Storing, manipulating and maintaining these meta
  data are dominated by integers, enumerations and booleans.
  \item the attributes of a particle require various precisions. The cut-off radius determining the neighbourhood for example does not have to
  be very accurate, as it bounds the maximum particle-particle
  distance. We can equip it with a hard-coded safety factor.
  At the same time, domain experience 
  suggests that the particles' position requires double precision, as the positions feed into complex non-linear evolution equations, while the densities vary by a factor of up to $10^{11}$ throughout the domain. 
  \item we need at least three types of MPI data exchange: The density update requires us to exchange the density and neighbourhood search radii
  between ranks, the force calculation determines the acceleration and physical properties, and the actual particle update and resorting send
  whole particles comprising all fields around.
\end{itemize}

\section {An extended C++ language}
\label{section:language}

C++ introduces an upper bound on the information density of structs that is
significantly lower than the theoretical maximum.
We work with minimal memory chunks of bytes.
If a byte, an 8-bit entity, stores a boolean value which could be represented by
one bit (on/off), the information density is only 12.5\% (1/8).
If that boolean is followed by another struct \added[id=R4]{member} aligned at eight-byte addresses,
it is attached an additional seven bytes, lowering its information density \replaced[id=R4]{further}{to
1.56\% (1/64)}.
We can make an analogous case for floating-point data \replaced[id=Ours]{where}{were} the actual number of
meaningful, i.e.~significant, bits \added[id=Ours]{is} known to the developer.
\replaced[id=Ours]{Whenever we do not exploit the full mantissa, the}{The} information density will \deleted[id=Ours]{likely} stay under 100\%, i.e.~not all bits hold meaningful information.

Memory alignment and the padding are key to make code fast:
They ensure that loads and stores hit memory entries exclusively, 
they ensure that the machine can work efficiently with cache lines, 
and they ensure
that vector operations can load small vectors en bloc into the respective vector
registers.
However, many HPC codes are notoriously memory-bound, or they suffer from cache
and memory latency due to scattered data access.
A lower information density makes this situation worse.

The C++ language and compiler vendors offer ways to increase the
information density of user-defined data structures. 
On the one hand, the language supports
the notion of bit-fields with a user-defined number of bits. 
The \replaced[id=R1]{\texttt{std::bitset} container}{\texttt{bitfield} class} allows developers to pack multiple booleans together
into one primitive datatype\replaced[id=R1]{. It is up to the implementation to decide which integral type (e.g.~unsigned integer) to use for storage, but it is always a multiple of whole bytes.}{(integer)}.
It tackles an extreme case of memory ``waste''.
On the other hand, many compilers (gcc and Clang/LLVM, among
others) support the \replaced[id=R4]{\texttt{\_\_attribute\_\_((packed))}}{\texttt{\_\_attribute\_\_((packed(N)))}} and
\texttt{\_\_attribute\_\_((aligned(N)))} syntax to allow users to manually
control memory alignment\replaced[id=R1]{. Since C++11, the language supports}{, and newer C++ generations even support} explicit
alignment\replaced[id=R1]{ }{, e.g.~}through \texttt{alignas}  \deleted[id=R22]{and \texttt{std::aligned\_alloc} } \added[id=R1]{, i.e.~alignment is now integral part of the language and not a ``mere'' annotation via attributes. \replaced[id=R22]{Both}{All three} variants---the keyword, \added[id=R22]{and the} attribute \deleted[id=R22]{and wrapper around the C call \texttt{aligned\_alloc}} ---are semantically equivalent}.
\replaced[id=Ours]{Yet, these solutions only}{The solutions} tackle special cases (sets of booleans) or provide byte-level
control over some memory arrangements to the user.

We propose to extend C++ such that developers can squeeze out fill-in bits and
bytes, exploit knowledge of potential value ranges of integers and inform the
compiler of the required floating-point accuracy.
This works per \replaced[id=R1]{member variable}{attribute}.
\deleted[id=R21]{Further to that, we propose to add support for the explicit modelling of multiple MPI data views over each struct.}

\begin{design}
  \label{design:language:augmentations}
  Our extensions are language augmentations. They neither introduce
  dependencies on external libraries nor require rewrites of the underlying
  code.
\end{design}

\noindent
\added[id=R21]{Further to that, we propose to add support for the explicit modelling of multiple MPI data views over each struct.}

\begin{design}
  \label{design:language:mpi-data-types}
  \added[id=R21]{
   If the MPI datatype annotations are employed, we extend the code to use standard MPI routines only.
   No further dependencies are added, and the MPI datatype generation under the hood is fully aware of any other memory reorganisation.
  }
\end{design}

\noindent
Design \replaced[id=R21]{decision~\ref{design:language:mpi-data-types}}{decision~\ref{design:language:augmentations}} assumes MPI to be
omnipresent on \replaced[id=R21]{HPC systems}{an HPC system and does not count it in as an external library}.

\begin{design}
  \label{design:language:semantics}
  \deleted[id=R22]{Our extensions are semantics-preserving:} If the assumptions expressed
  through an annotation remain valid, the extensions do not alter the code's
  behaviour\added[id=R4]{ as long as the program logic does not depend on \replaced[id=R22]{the bit-wise data layout as determined by the ABI}{struct-level data layout}}.
\end{design}

\noindent
For integer data, Design \replaced[id=Ours]{Decision}{decision}~\ref{design:language:semantics} is strict.
For floating-point data, we have to discuss the precise meaning of
semantics-preserving\added[id=R22]{, since the annotations change the bit-wise outcome of floating point operations within the boundaries laid out by the developer.
The data layout remark highlights that the extensions are not semantics-preserving from an ABI point of view.
If a code relies on certain memory layouts, the extensions break the code.
If code is memory layout-agnostic, the extensions are semantics-preserving}.
\added[id=R4]{
 Memory layout-agnostic code comprises all implementations that do not explicitly use the memory arrangement of the underlying data structures.
 It includes code with pointer arithmetics over arrays where the increment is implicitly determined by the compiler, but excludes, for example, codes relying on certain values returned by \texttt{sizeof} over a struct augmented with our annotations or the exact ordering of member fields.
}
\added[id=R22]{
 It includes code that uses the extensions only internally and compiles all translation units with the extensions enabled,
 but it excludes code that calls external libraries that have not been translated with the extensions, i.e.~stick to the native ABI.
 It however comprises codes which manually reconstruct (copy) all data into an ABI-compatible form prior to external method calls, 
 i.e.~code that uses the extensions exclusively internally.
}

\begin{design}
  \label{design:language:ignore}
  Our extensions are optional. If they are not supported by the C++ toolchain,
  they are ignored.
\end{design}

\noindent
We realise our language extensions through C++ attributes. 
If a compiler is unaware of particular attributes, they are simply ignored.
We define additional compiler passes which map the attributes onto plain C++ instructions internally.
No external libraries are required.
As the extensions are prototypically implemented within \replaced[id=R1]{LLVM's Clang frontend, we embed them into the \texttt{clang::} namespace}{LLVM, we make the attributes start with the \texttt{clang::} prefix}\footnote{\added[id=Ours]{It would be reasonable to embed the extensions into an experimental subnamespace \texttt{clang::exp} or similar. However, nested attribute namespaces are not supported by the language at this point.}}.

\begin{rationale}
 \label{language:rationale:acceptance}
 The acceptance of C++ language extensions hinges upon the fact whether users
 can introduce and benefit from the extensions without (a) code rewrites and (b)
 tying their code to external libraries.
 Since the annotations approximately preserve the semantics of the plain underlying code
 and as they are optional, developers can evolutionary augment their code in a
 trial-and-error fashion.
\end{rationale}

\begin{rationale}
 \label{language:rationale:compiler-integration}
 Interpreting all extensions through additional compiler passes
 ensures that their realisation (a) benefits from all optimisation know-how
 within the compiler and (b) remains agnostic of the target architecture,
 i.e.~back-end.
\end{rationale}

\noindent
Our objectives, i.e.~an increased information density and reduced memory footprint, could also be achieved through a library hosting tailored C++ classes.
However, this approach is not minimally invasive, i.e.~entails \deleted[id=R1]{major} code
rewrites compared to vanilla C++ \added[id=R1]{over built-in datatypes} \added[id=R3]{and introduces third-party dependencies} (cmp.~Rationale
\ref{language:rationale:acceptance}).
With \replaced[id=R1]{templates}{template meta programming}, developers can write type-generic algorithm
realisations\added[id=R1]{ themselves without relying on external libraries}.
However, \deleted[id=R1]{this introduces additional syntactic (template) overhead, it prolongs
compile times, and requires the library's high-density classes to offer each and
every operand that might be used over built-in types.
The} templating \deleted[id=R1]{also} quickly ripples through the code base, as all functions used
have either to be specialised or generalised to offer support for \replaced[id=R1]{the}{user-defined}
data types\added[id=R1]{ injected into the templates}.
\replaced[id=R1]{Templating is not minimally invasive.}{As an example, a simple use of \texttt{std::min} quickly fails for bespoke user
types, and becomes nasty once a user-defined class is used in combination with a
built-in type.
Finally, a realisation through a library implies that certain compiler
optimisations become less straightforward.
Compilers still struggle to optimise very large code blocks introduced by
templated functions as optimisation spaces grow exponentially, and simple
techniques such as attribute reordering \cite{Hyde:2017:WriteGreatCodeVol2}
might become obstructed.}
\added[id=R1]{
 The biggest challenge behind a template- or library-based realisation of flexible precision arises from setups where we arrange several variables with different relevant bits and types within one struct.
 A compiler-based approach can minimise the memory footprint over the whole struct.
 A library or templated solution can tailor the memory footprint per variable, but a holistic optimisation for all possible combinations of variables, variable types and variable precisions is non-trivial.
}
\added[id=R3]{
 Besides its development streamlining and avoidance of mandatory external dependencies, 
 our compiler approach hence widens the potential optimisation opportunities and hence goes beyond existing mixed and reduced precision approaches.
}

\subsection{Memory compactification}
\label{subsection:memory-compactification}

%
%
C++ programmers pick a well-suited built-in datatype for their
integers, i.e.~a qualified variant of \texttt{int}.
Once developers know the range of values encoded within an integer,
we can argue whether a choice of a variant of  \texttt{int} is valid, i.e.~large
enough to host all potential values' bit codes.
Enums \replaced[id=R4]{introduce a finite number of named enumerators over an integral type.
As long as we assume that exclusively the named numerators are used to assign a value to an enumeration variable,}{are symbolic identifiers for integer values. Consequently,}
 we know the exact number of bits required to encode any element of
an enum declaration.
Booleans are integers with a value range from $\{0,1\}$.
They carry one meaningful bit.

%
%
\paragraph{Motivation}
Whenever we work with integers of limited range, \replaced[id=R4]{symbolic enumerators}{enumerations} or small bit sets,
the information density of plain C++ code is low.
Scientific simulation software is often not integer-heavy.
Yet, \deleted[id=Ours]{we note that} integer arithmetics are used within meta data structures,
e.g.,~trees, containers, search algorithms, lookup tables, which are performance
critical.
Data access latency penalties caused by memory footprints of \added[id=Ours]{the arising} structs \deleted[id=Ours]{that are
larger than a cache line therefore} are problematic\added[id=Ours]{ and we hence have to attempt to fit as many of them as possible into one cache line}.

Further to that, \deleted[id=Ours]{we note that} integer data scattered among structs dominated
by floating-point data have the potential to inflate struct
encodings~\cite{Hyde:2017:WriteGreatCodeVol2}, as they \added[id=R4]{might} insert
padding bytes just before the floating-point numbers
\replaced[id=R4]{
 to align each number fitting to the machinery used and its ABI,
 and it might insert fill-in bytes at the end of the struct to ensure that the next struct instance within an array is well-aligned, too. 
 Besides the increase of the memory footprint, the arising
}{
In such a case,} arrays of structs become \added[id=Ours]{excessively} hard to vectorise \replaced[id=R4]{once}{as}
gather and scatter operations start to span many different cache lines.
\added[id=Ours]{
 Few integers can harm the efficiency of a lot of floating-point arithmetics. 
}

Reducing the memory footprint of integer-heavy structs hence is important.
Since integers often feed into control logic, any reduction has to preserve all 
data \replaced[id=R4]{bit-wise}{bit-wisely}.

%
%
\paragraph{Lossless compression}
Let \emph{packing} be a lossless compression, where multiple integer-valued
 variables are stored within one large bitfield.
 This bitfield has no bits without semantics.

\begin{extension}
\label{extension:pack-integer}
We introduce C++ attributes that label 
integer, boolean and enumeration members of structs as candidates for packing.
For integers, a range of values, i.e.~upper and lower limits can be specified.
\added[id=R4]{For enumerations, the labelling indicates that the programmer uses exclusively enumerators which are explicitly introduced in the \texttt{enum}'s declaration.}
\added[id=Ours]{C++'s ABI guarantees on the members' ordering can be violated once a struct hosts at least one packed member.}
\end{extension}

%
%
\noindent
Given a set of labelled integer \replaced[id=R1]{variables of}{attributes with} known range\added[id=R1]{ within a struct}---this includes
enumerations and booleans---a compiler can construct one large bitset which
encodes this sequence of values with high information density. 
The \added[id=R1]{new C++} attributes apply to \added[id=R1]{both} scalars and \replaced[id=R1]{(multidimensional)}{multidimensional} arrays with known array
ranges (\replaced[id=R4]{Syntax}{Algorithm}~\ref{algorithm:language:packed-integer}).

\begin{syntax}[htb]
  \caption{
    Syntax of the C++ extensions for integers, enums and booleans in pseudo
    Backus–Naur form. 
    C++ keywords are set bold, while non-bold
    names are examples of identifiers following the ISO C++
    conventions. Entries \texttt{( . | . | . )} enlist alternatives,
    while double square brackets \texttt{[[.]]} embrace C++ annotations. 
    The new attributes are called \texttt{pack} and
    \texttt{pack\_range}.
    Uppercase identifiers \texttt{L}, \texttt{M}, \texttt{N}, \texttt{MIN},
    \texttt{MAX} are to be replaced with compile-time constants in real code.
    \label{algorithm:language:packed-integer}
    }
 {\footnotesize
\lstset{language=C++}
\begin{lstlisting}
(struct | class) Data {
  [[clang::pack]]
  (bool | enum) field1;

  [[clang::pack]]
  (bool | enum) field2[M][N];
	
  [[clang::pack_range(MIN, MAX)]]
  ( (signed | unsigned) (char | short | int | long | long long) )  field3;
	
  [[clang::pack_range(MIN, MAX)]]
  ( (signed | unsigned) (char | short | int | long | long long) )  field4[L];
}
\end{lstlisting}
 }
\end{syntax}

\begin{itemize}
  \item The \texttt{[[clang::pack]]} attribute applies to datatypes that can be
  packed automatically without any additional user-supplied information.
  These are booleans \deleted[id=R4]{and enumerations} as well as fixed-sized arrays of these.
  We implicitly know their underlying integer ranges.
  \item \added[id=R4]{
    The \texttt{[[clang::pack]]} attribute applies to scoped and unscoped enumerations and it flags that exclusively the values enlisted as enumerators within the \texttt{enum} declaration are used. Therefore, we can count the number of enumerators and deduce the number of bits required to encode this finite set.
  }
  \item The \texttt{[[clang::pack\_range(MIN,MAX)]]} attribute controls packing
  of integer \replaced[id=R1]{variables}{attributes} and constant-sized arrays thereof. Since the compiler
  does not know a priori how many bits such a field uses, we ask users to
  manually provide a range of values that a field must be able to support.
\end{itemize}

\paragraph{Mapping onto plain C++}
%
%
Our compiler determines the number of bits that are required
to store packed values without losing any data.
That is one bit for each
boolean and $\lceil \log _2(n) \rceil$ bits for enumerations with $n$
\replaced[id=R4]{enumerators}{alternatives}.
Integer datatypes which are annotated with
\texttt{[[clang::pack\_range(MIN, MAX)]]} can be packed
losslessly with a bit footprint of $\lceil \log _2(\texttt{MAX}-\texttt{MIN}) \rceil$.

\begin{rationale}
 The technical details how integers of limited range are packed into one big
 bitset have to be hidden (cmp.~Rationale~\ref{language:rationale:acceptance}).
\end{rationale}

\noindent
Throughout the compilation, our compiler ``removes'' packed \replaced[id=R1]{members}{attributes} from
their struct, and it inserts one large bitfield that can accommodate all of
their required bits.
Code access to these values are wrapped
into appropriate bit masking, i.e.~we pick the right bits from the large bitset.
Arrays of packed integers can be mapped onto sequences of entries within the
bitset as long as the array size is known at compile time.
If enumerator values are manually assigned yet do not span a continuous range,
the compiler generates a lookup table to map them onto a compact range prior to
packing.
\added[id=R4]{An \texttt{enum}'s base type, if specified, is ignored.}

%
%

\paragraph{C++ context}
Packed values can coexist with unpacked values within one data structure. 
\deleted[id=R4]{None
of the characteristics of unpacked values are changed by the presence of packed
values.
Notably, only packed values lose their native memory alignment. }
However, the packing can permute the ordering of the struct \replaced[id=R1]{members, i.e.~instance and class variables,}{attributes} in the
memory:
\replaced[id=R1]{In C++, the}{The}
order of \replaced[id=R1]{members}{attributes} of a struct in memory \deleted[id=R1]{usually} follows the order of their
declaration\added[id=R1]{ in source code}.
Consequently, performance guidebooks recommend to order \replaced[id=R1]{instance variables}{the attributes} large-to-small\added[id=Ours]{ to avoid excessive padding}~\cite{Hyde:2017:WriteGreatCodeVol2}.
Our annotations make the compiler extract \replaced[id=R1]{members}{attributes} from the struct.
They are, eventually, inserted via a large bitset at the end of the struct's memory.
\replaced[id=R4]{The (partial) order over all unpacked members remains intact, and the packed bitset also preserves the declaration order of all packed attributes. However, we}
{We} change the \deleted[id=R1]{attribute}\added[id=R4]{total member} ordering.

\replaced[id=Ours]{Further to that weakening of the C++ ABI, taking}{Taking} pointers or references of packed \added[id=Ours]{member} fields is not possible and fails with a
compilation error\replaced[id=Ours]{ while an}{. 
An} attempt to store a value that falls outside of the specified range of a
packed field is undefined\replaced[id=Ours]{. In this case,}{, as} the annotation's underlying assumptions are
violated\replaced[id=R1]{ and C++ in general does not check for over- and underflows.}{ (cmp.~Rationale~\ref{design:language:semantics})}.
\deleted[id=Ours]{Packing
is only supported for built-in types.
}

Structs within structs cannot be annotated with packing, although their built-in \replaced[id=R1]{members}{attributes} in turn can be subject to packing.
\added[id=Ours]{ 
 Our reference implementation of the packing supports such a well-constrained composition of classes hosting packed data:
 If a class with packed instance members holds a member object which in turn owns compressed data, both types will end up with a bit field for ``their'' members independently.
 If a class with packed instance members is a subtype of another type with packed instance members, both types will end up with a bit field for ``their'' members independently.
 In theory, it would however be possible to fuse all packed data into one unified packed field accross multiple superclasses and aggregates.
 This would reduce the memory footprint of complex composite data structures with deep inheritance.
}

\added[id=R1]{
 Reordering of members could potentially break code semantics even though our language extension converts data into native formats prior to calculations, and pointer arithmetics over whole structs remain supported. 
 The extension's limitations have no impact on our case studies, where all code is recompiled from scratch, but they prevent users from using pre-compiled code that employs structs and pointer arithmetic.
 A production-level compiler hence might prohibit users from using external libraries or at least issue a warning. 
 Alternatively, it might be reasonable to require developers to explicitly enable the reordering throughout compilation via a dedicated compiler flag. 
 Similar to some fast mathematics extensions that aggressively reorder arithmetic operations and might violate numerical stability constraints, developers would then need to acknowledge that they take ownership of all implications.
 If we want to combine our memory compactification extensions with third-party libraries relying on structs, developers must replicate the struct without the language attributes in the baseline version and provide copy constructors and assignment operators to map one struct onto the other. 
 The structs without annotations then remain compatible with libraries; we meanwhile work with the annotated version throughout ``our code'' but convert them into a vanilla version before using any external library.
While such a conversion could be streamlined in the compiler via additional loop transformations (compare work in \cite{Radtke:2025:AoStoSoA}), its cost must be carefully evaluated on a case-by-case basis.
}

\added[id=Ours]{
Packing is natively supported through bitfields in C++.
They assign a class data member an explicit size in bits, and therefore also allow the compiler to pack variables into a smaller memory footprint.
Bitfield annotations work exclusively over integral types, booleans and enums, and it is up to the compiler how to exploit the augmented memory footprint information.
Our approach differs from the language feature in three ways:
First, our annotations are imperative if the compiler supports them, i.e.~they always lead to code transformations and are not declarative heuristics or suggestions.
Second, our annotations provide a higher level of abstraction as they denote ranges rather than bit counts. 
The compiler has the ownership to map these ranges onto proper bit representations, while the ranges themselves can be parameterised over compile-time constants. 
Finally, our annotations work over arrays of packed variables. 
}

\added[id=R1]{
 Syntax~\ref{algorithm:language:packed-integer} refers exclusively to built-in data types and native C-style arrays.
 A generalisation of our techniques to comprise \texttt{std::bitset} and \texttt{std::array} as well is possible and logically straightforward yet not featured by our implementation yet.
 At the moment, the attributes to not apply to those STL containers.
}

\paragraph{Impact}
The information density of a struct hosting a packed \replaced[id=R1]{variable}{attributes} is
equal or higher than the struct realisation in plain C++. 
As the memory footprint is reduced, we expect memory-bound compute
kernels to benefit from better cache utilisation.
However, any access to packed data is subject to additional conversion effort.
We assume that the underlying bit shifts and masking operations are fast on
modern hardware.
Yet, it is not clear a priori what performance impact the padding has.

Our approach facilitiates explicit unpacking and packing.
A simple \texttt{int a = packedA} over a packed integer
\texttt{packedA} will convert the bitset information into a native \texttt{int}.
Subsequent accesses to such a variable \texttt{a} will not suffer from any conversion penalty.
\replaced[id=R4]{If \texttt{a} is a manual, temporary copy of \texttt{packedA}, the}
{The} synchronisation back into \texttt{packedA} \deleted[id=R4]{however} remains with the
user, i.e.~the user has to manually copy the updates value back.
The corresponding \texttt{packedA = a} \replaced[id=Ours]{lets}{however will let} the compiler automatically \replaced[id=Ours]{inject}{introduce} all required packing operations. 
\added[id=Ours]{
 The pattern generalises to structs as discussed above and allows us to remain compatible with third-party code.
 However, manual conversion also can be useful for compute-heavy kernels which access the same struct members multiple times. 
}

\added[id=Ours]{
 Beyond (manual) hotspot conversion, the impact of the compiler's reordering of struct members on the performance is an aspect separate to conversion overheads and better bandwidth utilisation:
 Some codes deliberately classify struct variables into hot and cold depending on their frequency of use in particular application phases, and place hot variables next to each other to ensure high cache usage.
 Our compiler extension could break such manual optimisation.
 However, it is not clear if this is a major concern for scientific code bases, where fast codes tend to favour structs-of-arrays (SoA) for compute-intense kernels anyway, or explicitly convert data into SoA prior to usage~\cite{Radtke:2025:AoStoSoA}.
}

\subsection{Floating point storage precision}
\label{subsection:floating-points}

%
%
C++ programmers pick a well-suited built-in datatype for their floating-point
numbers when they write numerical algorithms.
Traditionally, this is either \texttt{float}, \texttt{double} or
\texttt{long double}, though alternatives such as fixed width
floating-point types or library-induced further types become
increasingly popular.
The choice is guided by forward/backward stability arguments and the
precision required in the output.

%
%
\paragraph{Motivation}
Supercomputers broke through the exascale wall twice:
First in half precision and later in double.
The higher throughput in half precision results
from improved vector computing capabilities, but also 
from a reduced pressure on the memory subsystem due to a smaller 
memory footprint.
As machines yield significantly higher performance for reduced precision, 
new (competing) floating-point formats become supported by hardware
\cite{Lindstrom:2018:CompetingFPStandards}, and scientists recast algorithms
into mixed precision formulations, where as many computationally expensive steps as possible are rewritten with lower precision
data types
(cmp.~\cite{Bornemann:2004:SIAM100DigitChallenge,Carson:2018:MultiPrecision,Carson:2023:MixedPrecisionSPAI,Higham:2022:MixedPrecision,Ichimura:2018,Langou:2006:Fp32ToFp64Precision,Murray:2020:LazyIntegration}
and many others).
Nevertheless, computationally intense compute kernels remain notoriously
memory-bound, while we continue to work with
overspecified data formats in many cases.
The number of native floating-point formats within the language is too small to tailor the memory footprint of each variable precisely to its significant bits.
We ``over-invest'' in bits.

At the same time, projects start to identify cases where data
logically does not exhibit the information density provided by native
floating-point formats:
Some data arising in intermediate compute
steps~\cite{Bungartz:2008:DaStGen,Bungartz:2010:DaStGenAndHierarchicalStorage,Carson:2023:MixedPrecisionSPAI}
or streamed into post-processing~\cite{Diffenderfer:2019:ZFPErrorsForFP,Lindstrom:2014:ZFP} do not
``need'' all the significant bits, i.e.~many bits carry no physical meaning
\cite{Abdelfattah:21:SurveyMixedPrecision,Tsai:2019:CompressObjects}.

Reducing the memory footprint of floating-point data beyond the few
available formats made available by the hardware hence is timely and important
to decrease the bandwidth requirements of codes further and to release stress from
the last-level cache.

%
%
\paragraph{Lossy compression}
Since we work with numerical approximations of real numbers,
we leverage any user intelligence on the mantissa's information density.

\begin{extension}
\label{extension:pack-floating-points}
We introduce a C++ attribute that enables developers to specify the number
of significant, i.e.~relevant bits in the mantissa of a floating-point value within a struct.
This qualifies the floating-point variable for packing.
\end{extension}

%
%
\noindent
For floating-point variables with known significant bits, 
a compiler can extract these significant bits from the floating-point
representation and store the bits within a bitset rather than the full
\texttt{float} or \texttt{double}.
In our C++ augmentation, the attribute
\replaced[id=R4]{\texttt{[[clang::mantissa(BITS)]]}}{\texttt{[[clang::truncate\_mantissa(BITS)]]}} specifies, for any native C++
floating-point type, that the actual mantissa can be stored with only \texttt{BITS} bits.
The exponent and the bit for the sign are preserved with their original bit
counts.
Our attribute applies to scalars and multidimensional arrays with known array
ranges (\replaced[id=R4]{Syntax}{Algorithm}~\ref{algorithm:floating-point-truncation}).
\added[id=R1]{
 Again, the extension to support \texttt{std::array} would be straightforward yet is not implemented yet.
}

\begin{syntax}[htb]
  \caption{
    \replaced[id=Ours]{Attribute syntax}{Syntax} of the C++ extension for the mantissa (exponent) truncation of
    floating-point numbers.
    \texttt{BITS}, \texttt{M} and \texttt{N} are integer constants known at
    compile time.
    \label{algorithm:floating-point-truncation}
    }
 {\footnotesize
\lstset{language=C++}
\begin{lstlisting}
(struct | class) Data {
	[[clang::mantissa(BITS)]]
	(float | double | long double) field1 [, fieldArr[M][N]..., ...];
}
\end{lstlisting}
 }
\end{syntax}

\paragraph{Mapping onto plain C++}
The extracted mantissa bits plus the sign bits and the exponents 
are packed into a large bitset together with all the enums, booleans
and integers which carry a \texttt{[[clang::pack]]} attribute.
Our floating-point packing integrates seamlessly with the integer packing.

\begin{rationale}
 For performance reasons, calculations have to stick to built-in data formats
 (cmp.~Rationale~\ref{language:rationale:compiler-integration}).
 However, developers often have expert insight how many significant bits their data really encode in-between
 calculations.
\end{rationale}

\noindent
We continue to run all calculations in native precision:
Our extension specifies how data are stored, but these formats are converted
back into the native C++ datatype prior to
calculations.
Therefore, our approach is lossy and realises a compress-decompress pattern.

\begin{rationale}
 Whenever external functions\added[id=R4]{ over built-in types} are invoked, the compressed data are
 automatically converted into native floating-point numbers.
\end{rationale}

\noindent
The compression therefore does not propagate through the \deleted[id=Ours]{code.
Major rewrites do not ripple through the} callstack.
 \added[id=R4]{
  Passing pointers to compressed floating-point numbers remains unsupported unless a function operates over instances of a struct, has been translated with the same compression (and packing) features, and we work with struct instances as atomic entities, i.e.~pass around references, copies or pointers to whole structs.
 }

\paragraph{C++ context}
As we store all compressed floating-point values internally within bit fields,
we inherit all properties of the packed integers, including the fact that
referencing via pointers is
not possible and fails with a compilation error. 
As we preserve the range of the exponent, it is impossible to create an
additional overflow compared to the baseline code version ignoring the C++
attribute.
However, the attribute can introduce additional underflows for very small
quantities, and it can amplify the truncation error.

Further to the reduced precision and potential losses of significant bits, the
packing has the potential to change the semantics of codes which employ logic
over floating-point data:
C++ guarantees $ a \not < b \Rightarrow a \geq b $.
Let $\hat a$ and $\hat b$ be compressed variants of $a$ or $b$, respectively,
and let 

\[
 p(a,b) = \left\{ 
  \begin{array}{rcl}
   p_{<}(a,b) && a < b \\
   p_{\geq}(a,b) & \text{if} & a\geq b
  \end{array}
 \right.
\]

\noindent
be written down as C++ if-else statement.
For $a<b$ reasonably close, we might preserve $\hat a < \hat b$ or end up in a situation where $\hat a = \hat b$ due to the truncation.
The truncation shifts and reduces the representable data points of
\texttt{double} and \texttt{float} within $\mathbb{R}$.

\paragraph{Impact}
The language extension realises a lossy compression.
Among such techniques, there are approaches which preserve all the bits of the exponent \cite{Tagliavini:2018:bfloat16},
and approaches which also reduce the bits per mantissa (compare
IEEE's half precision vs.~single).
Our approach preserves the exponent to be able to cover the same
range as the original data format.
We hence spread out the discrete
data points within $\mathbb{R}$ that can be represented compared to the baseline
type.


For selected problems, mainly from the linear algebra world, one can show that 
sophisticated rewrites with reduced sample accuracy over $\mathbb{R}$ do not
compromise the solution
\cite{Bornemann:2004:SIAM100DigitChallenge,Carson:2018:MultiPrecision,Higham:2022:MixedPrecision}.
For other problems, empirical data suggest that reduced precision is sufficient
\cite{Eckhardt:15:SPH,Fortin:2019:NbodyGPUDualTreeTraversals,Weinzierl:18:AlgebraicGeometric}.
In general, stability and error propagation have to be studied carefully.

Similar to integer data packing, floating-point packing induces operations overhead.
We have to unpack it from the input bitfield and befill the floating
point registers prior to the actual computation.
In particular, floating point sequences cannot be loaded ``en bloc'' from the
memory into vector registers due to this conversation and the fact that we miss
out on alignment or padding.
We assume that savings in memory transfers and bandwidth have the
potential to compensate for this penalty; notably if developers read from and write to packed structs carefully.

Vendors add support for reduced precision calculations
to their chips.
This is primarily driven by artificial intelligence
\cite{Agrawal:2019:Fp16}.
Our extension does not advocate for reduced precision calculations, since
it continues to work with standard C++ types for all calculations.
It however works hand in hand with modifications of the core calculations or
\replaced[id=R1]{templates facilitating}{template meta programming to facilitate} precision-generic codes.
\added[id=R1]{
 Our approach does not introduce new ideas along the lines of mixed- or reduced-precision algorithmics, but it simplifies the programming of such algorithms and widens the range of available precision formats as compared to industry standards.  
}

\subsection{\added[id=R21]{Optional extension:} MPI datatypes over structs}

C++ developers pick a well-suited distributed programming model for their code
manually, as C++ has no built-in support for this.
MPI remains the de-facto standard for distributed memory codes in
high-performance computing.
It ``natively'' facilitates the exchange of scalars of built-in types, as
well as arrays of these.
For more complex data structures, manual work is required.

\paragraph{Motivation}
%
%
Modern MPI supports user-defined datatypes
\cite{Gropp:2014:AdvancedMPI}.
They cover structs hosting scalars and arrays of
different types which are not contiguous
in memory.
We can also define an MPI datatype over a struct which covers only some
of its \replaced[id=R1]{instance variables}{attributes}.
This allows developers to exchange structs and arrays thereof partially, instead
of serialising and exchanging all information independent of whether data is
needed or not.
It allows developers to maximise the information density on a communication
stream.

%
%
Introducing user-defined datatypes requires developers to use low-level
operations:
We create an instance of the struct of interest, extract the
relative addresses of the struct's \replaced[id=R1]{instance variables}{attributes} into a table, and commit the table 
to MPI as a new datatype\added[id=R4]{ \cite{Renault:2006:MPIPreProcessor}}.
The datatypes encodes how a struct is serialised, i.e.~how its
bitstream is broken down into arrays of primitive types. 
Changes in a struct's number of \replaced[id=R1]{variables}{attributes}, types, 
their order or the underlying type inheritance hierarchy require the maintenance 
of all ``derived'' MPI datatypes.
It is laborious.
\added[id=R21]{
 Once a compiler is allowed to reorder memory layout, to pack or to use non-IEEE data formats, it is not possible anymore to construct MPI datatypes
 manually without equipping code with information about how data is arranged internally due to the C++ annotations.
 This contradicts Design Decision~\ref{design:language:augmentations}.
}

Creating and using bespoke MPI datatypes that only exchange required information\added[id=R21]{ and support our annotations is not only mandatory but also} is timely.
Networks on supercomputers notoriously suffer from congestion and bandwidth restrictions, and hence throttle scientific codes.
\added[id=R21]{Facilitating a smaller memory footprint addresses these bottlenecks.}

\paragraph{Embedded MPI datatypes}
We introduce a C++ attribute that enables developers to automatically create a
factory method~\cite{Gamma:94:DesignPatterns} returning an MPI datatype.
This datatype may encode an arbitrary subset of \replaced[id=R1]{instance variables}{attributes} of the struct.

\begin{extension}
\label{extension:mpi}
We introduce a C++ attribute that enables developers to annotate a
member function of a struct to highlight that this function returns an MPI
datatype.
The function's existing implementation---if available---is \replaced[id=R1]{swapped out for}{replaced by} a
generated routine.
\end{extension}

%
%
\noindent
Our C++ extension (\replaced[id=R4]{Syntax}{Algorithm}~\ref{algorithm:language:mpi}) streamlines the
construction of MPI datatypes:

\begin{itemize}
  \item A function annotated with \texttt{[[clang::map\_mpi\_datatype]]}
  has to return an \texttt{MPI\_Datatype} and may not accept any
  arguments.
  It has to be \texttt{static}\added[id=R1]{ as the concept of a datatype is tied to the class and not to instances of it}.
  \item 
  If our compiler encounters a method annotated with this attribute, it
  generates an implementation of a factory method \cite{Gamma:94:DesignPatterns}. Upon its first
  invocation, the routine constructs an MPI datatype. After that, it returns
  this datatype. 
  \item An existing function implementation is replaced by the compiler.
  \item Via \replaced[id=R1]{\texttt{[[clang::map\_mpi\_datatype(a,b,\ldots)]]}}{\texttt{[[clang::map\_mpi\_datatype("a","b",\ldots)]]}}, developers can
  instruct the compiler that the generated \texttt{MPI\_Datatype} should cover
  only some \replaced[id=R1]{instance variables}{object attributes} \texttt{a}, \texttt{b}, \ldots. 
  The subtypes have to be primitive, i.e.~have to be supported by MPI natively.
  Without these selectors, the factory method's return datatype comprises all
  \replaced[id=R1]{variables}{attributes} of a struct, i.e.~it serialises the whole struct.
  \item If structs are contained within structs (nested), the default
  MPI datatype covers the whole conglomerate. Users however can pick
  \replaced[id=R1]{variables within}{subattributes of} arbitrarily nested structs through
  \texttt{struct1.struct2.attribute}\added[id=Ours]{ since attribute arguments accept expressions}.
  \item If the function attribute enlists one packed integer or
  floating-point attribute, all packed \replaced[id=R1]{instance variables}{attributes} of the struct are subject to
  the generated MPI datatype.
\end{itemize}

\begin{syntax}[htb]
  \caption{
    Struct with MPI augmentation. 
    The compiler generates the implementations of an augmented routines,
    replacing their user implementations.
    Both generated routines return an \texttt{MPI\_Datatype} which can directly
    be used with \texttt{MPI\_Send} or \texttt{MPI\_Recv} or any MPI routine.
    The first datatype exchanges all \replaced[id=R1]{variables hold by}{attributes of} \texttt{Data}, while the
    second datatype exchanges only two \replaced[id=R1]{instance variables}{attributes} of the struct.
    \label{algorithm:language:mpi}
    }
 {\footnotesize
\lstset{language=C}
\begin{lstlisting}
struct Data {
  [[clang::map_mpi_datatype]]
  static MPI_Datatype getMyFullMPIDatatype();

  [[clang::map_mpi_datatype(field1, field2.subfield1)]]
  static MPI_Datatype getDatatypeForSubset();
}
\end{lstlisting}
 }
\end{syntax}

\paragraph{Mapping onto plain C++/MPI}
The generated code takes care of all address arithmetics and the construction of
helper data structures to describe the MPI datatype.
Hiding the technical complexity behind MPI datatypes is not a new endeavour or
idea, and there are different ways to achieve this:
Boost for example supports data exchange of structs via byte code 
serialisation.
Here, the serialisation is realised through routines of a pre-defined name
which are injected into the struct.
This is an aspect-oriented approach.
Our approach does not serialise the objects directly, but instead maps the
\replaced[id=R1]{struct's variables}{struct attributes} onto a MPI dataype, i.e.~the actual serialisation is delegated
to the MPI library.

We define the construction of the MPI datatypes to be lazy, i.e.~they are
generated upon the first invocation of the routine.
This ensures that the MPI datatype construction does not precede any
MPI initialisation.
Even if the datatype is hosted within a library, its construction happens upon
the first invocation of the factory method, i.e.~after the code using the
library has established the MPI environment.
It remains the responsibility of the developer to clean-up (free)
user-defined MPI datatypes created via our factory methods.
\added[id=R4]{
 An automatic clean-up in a destructor would leave the decision on the destruction order to the linker.
 For some MPI implementations, this can lead to complications if MPI is shut down prior to freeing user-defined datatypes. 
}

\paragraph{C++/MPI context}
Since we extract the MPI datatypes from the source code at
compile time, data format changes are automatically reflected within the MPI
datatype generation.
The extension implicitly flattens any inheritance hierarchy, although
it does not support any polymorphism within MPI.
The MPI datatype construction masks out the \texttt{vtable}\replaced[id=Ours]{, and}{, but} it does not
distinguish any particular subtypes.
As a static routine, the resulting MPI datatype is tied to one particular class.

Our extension assumes that developers continue to work with MPI
directly.
Users have to know which datatype they send and receive in turn.
The augmentation provides data types only and no other MPI features.
\added[id=R4]{
 This mindset is similar to other approaches introducing a separate pre-compiler to assemble the MPI datatypes \cite{Hillson:2000:Cpp2MPI,Renault:2006:MPIPreProcessor}.
 However, our approach integrates directly into the translation process, and we do ignore C++'s visibility annotations and instead introduce an orthogonal concept:
}

By default, all fields are included in the generated \texttt{MPI\_Datatype}
instance. 
However, developers can  explicitly specify which fields should be
included by listing them as attribute arguments.
This enables developers to create multiple tailored \texttt{MPI\_Datatype}s per
struct, since we tie the datatype construction to a static member function
rather than the struct itself.
Developer can create multiple \emph{views} over their structs:

\begin{rationale}
 \label{rationale:mpi}
 To keep data consistent between ranks, many codes have to exchange some
 \replaced[id=R1]{instance variables}{attributes of structs} only. 
 Which \replaced[id=R1]{variables}{attributes} to pick can depend on the context (\added[id=Ours]{e.g., }algorithmic phase\deleted[id=Ours]{, e.g.}).
\end{rationale}

\noindent
As the annotation triggers the compiler to replace any existing implementation
of the annotated function, users can guarantee that
their code continues to be correct even if the annotations are not supported
(cmp.~Rationale~\ref{design:language:semantics} and
Rationale~\ref{design:language:ignore})\replaced[id=R1]{, as long as they}{.
For this, they have to} provide a dummy realisation of the static function\replaced[id=R1]{:}{, e.g.
serialising the whole struct independent of the view chosen.}

\paragraph{Impact}
%
%
The MPI annotation \deleted[id=R3]{in first place} is convenient for developers and \added[id=R3]{it} can be used independently of the packing. 
\added[id=R3]{
  While MPI offers user-defined datatypes exposing the data structure layout explicitly, 
  our annotations hide a struct's interna and therefore streamline their use:
  Address arithmetics become obsolete, and adding additional class members does not require updating and maintaining the underlying MPI datatype, as it is constructed by the compiler behind the scenes.
  Complementary, the opportunity to provide a default implementation which is swapped out for the generated code gives developers the opportunity to write a (generic) baseline implementation in case that the annotations are not supported.
  In many codes, we find such baseline implementations exchanging structs bit-wise serialising and transferring the whole struct inclusive all padding bytes. 
}.

While the ease argument is important in itself, \replaced[id=R3]{the annotation-based approach}{it} becomes particularly important once \replaced[id=R3]{developers use}{we support the previously introduced} 
packed integer and floating-point data types, for which no MPI data type equivalent exists.
It frees developers from the duty to care about the existence and implications
of packing. 
\added[id=R3]{
 An automatic construction of the underlying MPI data type is particularly important here, since we allow packing to reorder a struct's variables, which adds further complexity to a manual mapping of struct variables onto native MPI datatypes.
}

\replaced[id=R3]{Still, programmers could lower the packed struct onto a plain bit representation and transfer raw bytes. Compared to this, our} 
{The} concept of views \replaced[id=R3]{reduces}{aims to reduce} the bandwidth pressure on the node
interconnects, as \replaced[id=R3]{we can pick}{
 we eliminate waste bytes or increase the information density
of the exchanged data stream, respectively.
Picking} individual \replaced[id=R1]{instance variables}{attributes} that are to be exchanged from a struct or an array
of structs\added[id=R3]{. We eliminate fill-in bytes and increase the information density
of the exchanged data stream---a feature that complements packing itself. Again, the maintenance of multiple views is streamlined for programmers and does not require them to manually manicure MPI datatypes as structs evolve. However, using views} means that MPI has to gather and scatter data from
the memory.
It is not clear how expensive these steps are\added[id=R3]{ with today's MPI implementations}.

\added[id=R21]{
 The optional MPI extension differs from the other proposed attributes from Section~\ref{subsection:memory-compactification} and \ref{subsection:floating-points} in that it is not strictly affecting C++ semantics but provides an alternative API for MPI datatype generation which is more expressive, and reconciles our memory-layout and packing extensions with the way MPI reasons about data layout.
}

\replaced[id=R3]{\replaced[id=R21]{Yet, our}{Our} attributes' semantics and realisation rely upon}{However, our annotations wrap around} \replaced[id=R21]{pure MPI only}{MPI and do not interfere in any way with the standard}.
Therefore, any MPI optimisation carries over to our annotated code directly.
Notably, concepts such as message compression
\cite{Filgueira:2012:MPICompression,Ke:2004:MPICompession} or sophisticated
message buffering are not compromised by our techniques.

\section{ \added[id=Ours2]{Related solutions} }

\subsection{\added[id=R22]{C++26 reflections}}

\added[id=R22]{
  Static reflection, as specified in P2996R13 for C++26, offers compile-time inspection of types, including their data member names, types, sizes, offsets, and qualifiers.
  The same mechanism may also be used to inspect member function signatures and enumerations.
  This functionality is likely sufficient (though there is limited compiler support so far for these features) to implement the MPI datatype generation, i.e.~the factory mechanism expressed through Annotations~\ref{extension:mpi}.
  We however hypothesise that the construction of various views requires significant developer effort and code (see Rationale~\ref{rationale:mpi}).
}

\added[id=R22]{
 The interplay between reflection and our data storage and layout manipulation remains unclear.
 While C++'s static reflection mechanism is fundamentally observational in nature, Proposal P3394 introduces field-level annotations to augment data structures with user-defined metadata.
 Such metadata can carry information on value ranges, mantissa sizes, or MPI datatype generation, such as the augmented information injected through attributes in our proposal.
 It is not clear how and how easy programmers can use the metadata to construct stand-in types approximating the functionality proposed in this paper, i.e.~variants of existing structs with a packed data representation:
 It is likely that compile-time sorting and packing of attributes make it possible to synthesise such a packed stand-in type and let the compiler replaces sets of attributes with their packed counterparts.
 Obviously, compressed floating-point attributes only can also be modelled through dedicated types of their own which become aggregates of the generated stand-in types.
}

\added[id=R22]{
 The reflection mechanism's inability to synthesise member functions means that any getters, setters, constructors, or operators over the original aggregate type would not apply to the stand-in type.
 The routines would have to be represented using metaprogramming techniques or a decorator pattern~\cite{Gamma:94:DesignPatterns} to make them applicable to the stand-in type, too.
 Templates historically suffer from syntactic overhead.
 Although significantly reduced through concepts in C++20, they remain non-trivial to many developers.
 Metaprogramming introduces call indirection which can have negative implications for the performance.
 The approach proposed in this paper preserves the original struct signatures and facilitates aggressive compile-time optimisations, and it implies that the programmer is not required to do major code refactoring when the extensions are applied retroactively to large existing code bases.
 The disadvantage that it remains C++ yet breaks ABI semantics would also hold for reflection-based techniques.
}

\subsection{\added[id=R21]{Reduced and mixed precision}}

\added[id=R21]{
 Mixed- and reduced-precision computations beyond IEEE-754  
 have been studied extensively in numerical linear algebra with comprehensive analyses of stability and convergence properties~\cite{Carson:2018:MultiPrecision, Higham:2022:MixedPrecision, Abdelfattah:21:SurveyMixedPrecision, Higham:2021:StochasticRoundingErrorAnalysis}. 
 They demonstrate how lower precision arithmetics can achieve higher performance whilst preserving accuracy, for example by combining 32-bit arithmetic with refinement to obtain 64-bit results~\cite{Langou:2006:Fp32ToFp64Precision}, or by dynamically varying precision during solver execution~\cite{Ichimura:2018, Murray:2020:Delayed}.
 Alternatively, exploiting knowledge about the information density within a (linear) system can enable algorithms to yield high accuracy at high performance as we parts of some equation systems do not require full 64-bit precision~\cite{Ltaief:2023:GordonBell}.
 Orthogonal work investigates stochastic rounding and alternative rounding modes as a means of controlling numerical error when operating at reduced precision~\cite{Fasi:2022:StochasticRoundingSurvey,Fasi:2021:StochasticRoundingArithmetic,Higham:2021:StochasticRoundingErrorAnalysis}.
 The agenda overall is performance-driven, and the precision is a parameter of computation, not a property of storage layout or data structures.
}




\added[id=R21]{
 On the storage side, floating-point compressors such as ZFP~\cite{Lindstrom:2014:ZFP} operate over small tensors or structured grids, optimising memory footprint and bandwidth over arrays. 
 Related work introduces universal coding of real numbers~\cite{Lindstrom:2018:CompetingFPStandards} or object-based compressed memory hierarchies~\cite{Tsai:2019:CompressObjects}.
 Originally, these techniques are tailored towards I/O due to better usage of memory hierarchies and interconnect bandwidth, 
 but are intentionally decoupled from core compute algorithms.
}

\added[id=R21]{
 Library-based approaches such as MPFR~\cite{Fousse2007MPFRAM}, Boost.Multiprecision or FloatX~\cite{Flegar2019FloatX} offer non-standard precision implementations to the user via bespoke scalar types. 
 Domain-specific formats such as DLFloat~\cite{Agrawal:2019:Fp16} target bespoke (low-power) systems~\cite{Tagliavini:2018:bfloat16}.
 While these libraries and formats can support both extended and reduced precision, their integration model relies on explicit type replacement, i.e., users have to refactor their code. 
 Analogous work has been proposed for the aforementioned I/O libraries.
}

\added[id=R21]{
  Our work does not target the instruction-level parallelism, since we let the compiler work with the specified C++ standard types for the core calculations.
  Vector-optimisation techniques such as writing bespoke kernels over different C++ precisions hence are independent of the proposed language extensions.
  However, the extensions help to tune bandwidth-bound codes and performance on heterogeneous nodes~\cite{Radtke:2025:AoStoSoA}, as they unlock a whole spectrum of encoding precisions for the user code.
  The language extensions facilitate the implementation of higher-level compression concepts such as ZFP and rapid precision refactoring.
  Their optimisation over multiple attributes per struct is another key advantage not found in other approaches which either work on scalars or arrays of homogeneous base type.
  Previously offered as source-to-source precompiler~\cite{Bungartz:2008:DaStGen}, the present approach migrates all datatype construction and conversions into the compiler, 
  hiding technical complexity and facilitating aggressive optimisation passes.
}

\section{Realisation within LLVM}
\label{section:translation}

LLVM is the baseline of
many vendor-specific mainstream compilers (Intel, NVIDIA and AMD).
Due to its clear separation-of-concerns, its explicit intermediate program
representations, and a clear data flow through translation passes, Clang/LLVM
is a natural candidate to realise our extensions.

We fork \replaced[id=R1]{LLVM~21.0.0}{LLVM~13.0.0}.
Within this fork, all language extension are supported by default, though we can
instruct the compiler to ignore them through \texttt{-fno-hpc-language-extensions}.
Depending on the invocation, the compiler defines or undefines the symbols
\linebreak \texttt{\_\_PACKED\_ATTRIBUTES\_LANGUAGE\_EXTENSION\_\_}\replaced[id=R1]{  and }{,}
\texttt{\_\_MPI\_ATTRIBUTES\_LANGUAGE\_EXTENSION\_\_} \deleted[id=R1]{and
\texttt{\_\_AOSSOA\_ATTRIBUTES\_LANGUAGE\_EXTENSION\_\_}}
such that users can mask out code fragments through \texttt{ifdef} guards.

\subsection{Extension architecture}

%
%
LLVM is a modern compilation framework breaking down the translation into 
stages or phases.
For our work, Clang serves as compiler frontend.
It translates the (annotated)
C++ source code into LLVM's intermediate language/representation (LLVM IR).
This LLVM IR then is subject to optimisation passes and eventually streams into
the (multi-target) machine code generation.

%
%
Many embedded DSLs add an additional level of abstraction on top of the generic
programming language C++ and hence require front-end, i.e.~lexer and parser,
modifications.
Our language extensions use C++'s annotations.
We can therefore stick to an unmodified font-end to build up the abstract syntax
tree (AST), and manipulate this AST before we lower it into plain LLVM IR.


\begin{rationale}
 Since we realise our extensions through an additional compile pass following
 the parsing, they become independent of both the 
 IR optimisations and target-specific machine code production, as well as any
 C++ front-end modifications.
\end{rationale}

\noindent
Clang’s high-level architecture follows a textbook compiler structure \cite{Kruse:2021:ClangLoopTransformations}. 
A SourceManager and FileManager handle file-related operations.
The Preprocessor and Lexer run through the files' byte streams
and produce tokens which are used to identify syntactic elements.
They are handed
over to the semantic analysis (Sema) which yields an abstract syntax tree (AST).
The Sema's TreeTransform helper mechanism adds additional AST nodes besides
those corresponding directly to parsed tokens:
Each implicit template instantiation for example creates its own copy of the
AST subtree into which it substitutes template parameters.
We use an analogous mechanism to realise the transformations triggered by the
annotations.
If \replaced[id=R1]{a struct's variable}{an attribute} is marked as packed, we replace all follow-up accesses with the
corresponding packing or unpacking code.

Yet, Clang favours forward propagation of information in line with LR(k)
grammars and the C++ language which is static and strongly typed, i.e.~requires
all types and variables to be well-declared prior to their first usage.
It is tied to single-pass translation.
Consequently, Clang's tree transformations support localised alterations,
such as changing AST nodes as they are created or unfolding
of subtrees.
Cross-references are eliminated in the tree generation phase by
replicating information (such as datatype, type size or memory alignment) where required.
Altering declarations in hindsight is not possible.

For our language modifications, we have to add or remove 
struct fields, or change types of declared variables.
The exact bitfield layout, for example, is only known after we have parsed the
whole underlying struct.
At this point, we might already have processed (in-line) source
code snippets. 
Our extensions potentially require non-local changes rippling through
many data copies within the AST.

Our realisation therefore abandons the single-pass paradigm and instead uses
in-memory pretty printing:
In a preparatory phase, we traverse the tree and search for our domain-specific
attributes.
The set of attributes yields a source transformation plan, i.e.~recipies which
fragments of the underlying source code have to be changed.
With these rules, our compiler extension reparses the code and builds the AST
again.
This time, it alters AST nodes that need to be changed immediately, and therefore propagates the changes to all replica of AST parts
or code using the altered AST segments.
This happens entirely in memory.

\subsection{Packing realisation}

We store packed integer data as sequence of values with $a_i$ bits, where $a_i
= \lceil (\log _2(n_i)) \rceil$.
$n_i$ is the number of potential values of each variable.
The resulting memory footprint is $\sum _i \lceil (\log_2(n_i)) \rceil
\leq \lceil \log_2 \prod _i (n_i) \rceil $.
We refrain from ``merging''
the ranges of multiple variables, as this would introduce additional arithmetic overhead when we
generate the data access operations.
Without further assumptions, the information density within the
packed bitset is therefore not optimal.
We choose simplicity over the theoretical maximum of the information density.

Our floating-point annotations support the C++ floating-point formats
\texttt{float}, \texttt{double} and \texttt{long double}.
Besides scalar \replaced[id=R1]{variables}{attributes}, the compiler can handle constant-sized arrays of
arbitrary dimensionality over these types. 
Bit-shifting and bit-masking over floating-point values are not natively
supported by the \deleted[id=R22]{C/}C++ language.
Our tool overcomes this obstacle by ``dereference casting'' of the
floating-point values to and from integer representations. 
Throughout this process, mantissa bits are cut or added.
The implementation of the conversion is realised as part of the same
compiler pass that handles the \texttt{[[clang::pack]]} and
\replaced[id=R4]{\texttt{[[clang::pack\_range(MIN,MAX)]]}}{\texttt{[[clang::pack\_range(BITS)]]}} attributes, i.e.~floating point
manipulations are directly forwarded into the logic handling integer packing.

Our conversion is a plain truncation, i.e.~we chop the digits after the
\texttt{BITS}th position off.
When the truncated representation is retranslated into a native format, the 
previously truncated bits are set to 0 in the reconstructed value.
Such a strongly biased conversion can lead to accumulation effects and make
numerical implementations unstable.
We recognise that techniques such as stochastic rounding
\cite{Fasi:2022:StochasticRoundingSurvey} could mitigate this phenomenon \cite{Fasi:2021:StochasticRoundingArithmetic,Higham:2021:StochasticRoundingErrorAnalysis} yet are out of scope here.

\begin{rationale}
  Numerical accuracy or stability considerations are out of scope for the
  present work, i.e.~we solely rely on the user to keep track of these
  phenomena.
\end{rationale}

\noindent
Within the translation pipeline, 
any access to a packed variable \texttt{a} results in some implicitly generated conversion code
from or to \texttt{a}'s packed data representation.
Any explicit copy \texttt{b=a} implies however that \texttt{b} is \emph{not}
packed anymore.
A statement \texttt{a++} over a packed variable hence unpacks and packs implicitly, i.e.~
synchronises the packed variable with its temporarily unpacked variant 
(cmp.~Sections~\ref{subsection:memory-compactification} and
\ref{subsection:floating-points}).

\begin{rationale}
  \label{rationale:packing-realisation:propagate}
  The packing attributes do not propagate through in the code, i.e. they do not apply to copied
  variables.
\end{rationale}

\noindent
\added[id=R22]{
 Whenever a compressed floating point entry is copied into a native C++ floating point variable, the target variable's encoding follows the local ABI and C++ conventions.
 It is free of compression.
 Whenever a bit is copied into a \texttt{bool} variable, the \texttt{bool} is a plain C++ variable and not one bit cut out of a larger bit sequence.
 A copy of a compressed and packed varaible does not ``inherit'' the source encoding.
}
\replaced[id=R22]{This}{The} ``do not propagate'' policy allows developers to eliminate any packing from
\replaced[id=R1]{variables}{attributes explicitly} by simply copying them into temporary variables.
Compilers remove such helper copies if the new \added[id=R1]{language} attributes are not supported\deleted[id=R1]{, unless they are used within very large compilation units where the additional helper assignments push the total code size beyond the heuristics from which on the compiler stops optimising}.

\added[id=Ours]{Further to facilitating manual unpacking, the propagation policy has major implications for the work with functions:}
Implicit conversion for read and write accesses \replaced[id=Ours]{means}{imply} that our approach works
seamlessly for third-party functions accepting built-in datatypes. 
A simple \texttt{data.a = foo(data.b, data.c)} over packed floating-point values \texttt{a}, \texttt{b} or \texttt{c} triggers
two unpack and one pack operation in the background, yet does not require any
bespoke realisation of \replaced[id=Ours]{\texttt{foo}}{\texttt{std::min}}.
It also continues to work if either \texttt{a}, \texttt{b} or \texttt{c} are
unpacked, native data.

\replaced[id=R1]{Likewise, the native type is used where function templates are instantiated implicitly, i.e. through usage. Standard library functions such as \texttt{std::min} are defined over a pair of references of the same type. This requires that both arguments are of the same type, and implicit type conversion based on the \texttt{operator type()} syntax does not prevent compilation errors. Our approach presents the native type to the template instantiation machinery inside the compiler, thereby avoiding compilation errors. This is in contrast to library- or template-based approaches which rely on custom scalar types. Naive mixing of native and custom scalar data types leads to compilation errors in standard library function template instantiations where all arguments are references of a single template parameter type. Naturally, such compilation errors may be fixed using explicit casting wherever an error occurs, but this changes the nature of the solution from a strictly localised source code change to a potential project-wide refactoring.}{This makes our approach differ from a template-based solution, where
\texttt{std::min} would have to be overloaded for packed
specialisations and combinations of packed and unpacked data.}

\subsection{MPI datatype mapping}

%
%
The MPI code generation triggered by \texttt{[[clang::map\_mpi\_datatype]]}
invokes \linebreak \texttt{MPI\_Type\_create\_struct}. 
Prior to this, it gathers the block lengths, i.e.~the continuous
occurrences (array lengths) of a given type, relative offsets of these arrays
over primitives within the memory, and the (MPI) types themselves into a map.
To populate the map, we recursively traverse the AST, starting from the
\texttt{CXXRecordDecl} node that describes the struct which declares the mapping
method.


%
%
The \texttt{MPI\_Datatype} created by the mapping methods is cached within the
generated routine in a \texttt{static} local variable, such that the actual call
to create the MPI datatype, \texttt{MPI\_Type\_create\_struct}, and the
corresponding \texttt{MPI\_Type\_commit} happen only once regardless of the
number of invocations of the mapping method~\cite{Hillson:2000:Cpp2MPI}.
\replaced[id=Ours]{
 The manual free employs \texttt{MPI\_Type\_free}.
}{
 Users have to ensure themselves that they invoke \texttt{MPI\_Type\_free}
appropriately.
}


\subsection{Overhead in machine code}

\begin{code}[htb]
  \caption{
    Top: We label \replaced[id=R1]{variables}{attributes} as \texttt{pack}, while all \deleted[id=R1]{remaining} accesses to
    the \replaced[id=R1]{variable}{attribute} remain unchanged (compare left to right side).
    The compiler maps all packed \replaced[id=R1]{variables}{attributes} into one big bitset and
    automatically replaces access to these fields with the corresponding bit arithmetics.
    Bottom: 
    Assembly instructions of baseline code (left) vs.~the packed variant (right) as emitted by our compiler with the \texttt{-O3} flag.
    \label{algorithm:language:packed-integer-access}
    }
 {\footnotesize
\lstset{language=C++}
\begin{minipage}{0.49\textwidth}
\begin{lstlisting}
struct Data {
   bool b;
};
bool getB(Data &data) {
   return data.b;
}
void setB(Data &data, bool val) {
   data.b = val;
}
void invertB(Data &data) {
   data.b = !data.b;
}
\end{lstlisting}
\hrule
\begin{lstlisting}
getB(Data&):
       mov     al, byte ptr [rdi]
       ret


setB(Data&, bool):                      
       mov     byte ptr [rdi], sil
       ret


invertB(Data&):                  
       xor     byte ptr [rdi], 1
       ret

\end{lstlisting}
\end{minipage}
\begin{minipage}{0.49\textwidth}
\begin{lstlisting}
struct Data {
   [[clang::pack]] bool b;
};
bool getB(Data &data) {
   return data.b;
}
void setB(Data &data, bool val) {
   data.b = val;
}
void invertB(Data &data) {
   data.b = !data.b;
}
\end{lstlisting}
\hrule
\begin{lstlisting}
getB(Data&):
       mov     al, byte ptr [rdi]
       and     al, 1
       ret
setB(Data&, bool):
       mov     al, byte ptr [rdi]
       and     al, -2
       or      al, sil
       mov     byte ptr [rdi], al
       ret
invertB(Data&)
       xor     byte ptr [rdi], 1
       ret
\end{lstlisting}
\end{minipage}
 }
\end{code}

\begin{code}[htb]
  \caption{
    Packing and unpacking (top, left vs.~right) introduce only few additional
    operations in the resulting machine code (bottom). 
    \label{algorithm:language:packed-float-access}
    }
 {\footnotesize
\lstset{language=C++}
\vspace{-10mm}
\begin{minipage}{0.49\textwidth}
\begin{lstlisting}
struct Data {

   float f;
};
float getF(Data &data) {
   return data.f;
}
void setF(Data &data, float val) {
   data.f = val;
}
void add(Data &data, float val) {
   data.f += val;
}
\end{lstlisting}
\hrule
\begin{lstlisting}
getF(Data&):                          
       movss   xmm0, dword ptr [rdi]           
       ret

       
setF(Data&, float):                         
       movss   dword ptr [rdi], xmm0
       ret

       
add(Data&, float):                          
       addss   xmm0, dword ptr [rdi]
       movss   dword ptr [rdi], xmm0
       ret


       
\end{lstlisting}
\end{minipage}
\begin{minipage}{0.49\textwidth}
\begin{lstlisting}



struct Data {
  [[clang::mantissa(7)]] 
  float f;
};
float getF(Data &data) {
   return data.f;
}
void setF(Data &data, float val) {
   data.f = val;
}
void add(Data &data, float val) {
   data.f += val;
}
\end{lstlisting}
\hrule
\begin{lstlisting}
getF(Data&):
       movzx   eax, word ptr [rdi]
       shl     eax, 16
       movd    xmm0, eax
       ret
setF(Data&, float):
       movd    eax, xmm0
       shr     eax, 16
       mov     word ptr [rdi], ax
       ret
add(Data&, float):
       movzx   eax, word ptr [rdi]
       cvtsi2ss        xmm1, eax
       addss   xmm1, xmm0
       cvttss2si       eax, xmm1
       mov     word ptr [rdi], ax
       ret
\end{lstlisting}
\end{minipage}
 }
\end{code}

Packing and unpacking translate into few machine
instructions that are inlined into the resulting code. 
In the case of a packed boolean, the
overhead is just one x86\_64 machine code instruction for reading (unpacking),
and three instructions for writing (packing), and no overhead for in-place
inversion (negation) (\replaced[id=R4]{Source Code}{Algorithm}~\ref{algorithm:language:packed-integer-access}).

For floating-point data, we obtain two extra x86\_64 machine code instructions for
either a read or a write operation, or three extra x86\_64 instructions for an
``in-place'' arithmetic operation, i.e.~a read immediately followed by a write
(\replaced[id=R4]{Source Code}{Algorithm}~\ref{algorithm:language:packed-float-access}).

\section{The SPH demonstrator}
\label{section:demonstrator}


\subsection{Governing equations}
%
%

The Lagrangian philosophy behind SPH---in which the fluid is mapped onto
particles---recasts the partial differential equations governing the dynamics
of the system into a set of coupled ordinary differential equations. They
describe the interaction and evolution of these particles. In this Section,
we present only the core equations of the numerical method we use, while the
full description of the governing equations is presented in Appendix~\ref{app:SPH}.

\noindent
At the heart of SPH is the smoothing operation which is used to estimate
scalar fluid quantities such as the density $\rho_i$ for each particle $i$.
Giving each particle some (constant) mass $m_i$, the smoothed density is
obtained via
\begin{equation}
  \rho_i = \sum_j m_j W_{ij}(H_i) \label{eq:SPH_density}
\end{equation}
where $W_{ij}(H_i) = W(\vec{x}_j - \vec{x}_i, H(h_i))$ is called the \emph{kernel}.
It's a smooth, differentiable, spherically symmetric, and monotonically decreasing
function with compact support of radius $H$. In practice, kernels are
computationally inexpensive polynomials.
Although the sum in~\eqref{eq:SPH_density} runs, in principle, over all particles
$j$ in the domain, the finite $H(h)$ reduces it to a loop over neighbours around
$\vec{x}_i$.

The smoothing length, $h$, plays a central role in SPH. It determines the compact
support $H$ and hence defines the concept of neighbours, i.e.~it defines the group
of particles which are close enough to contribute towards the value of the field,
Hence it determines the number of neighbouring particles included in smoothing
operations such as \eqref{eq:SPH_density}. Furthermore, it also specifies the
spatial resolution of the simulation \cite{dehnenImprovingConvergenceSmoothed2012}.

For the present SPH demonstrator, we consider an inviscid fluid in the absence
of gravity and external forces or energy sources. Hence, the individual particles tracking
the fluid evolve according to the Euler equation,
\begin{align}
\frac{\d \vec{v}_i}{\d t} &=
  -\sum_j m_j \left[
      f_i \frac{P_i}{\rho_i^2}\nabla W_{ij}(H_i)
    + f_j \frac{P_j}{\rho_j^2}\nabla W_{ij}(H_j)
  \right] + \vec{a}^{\rm AV}_i\,, \label{eq:Euler-equation-particles-maintext}
\end{align}

\noindent
while the thermodynamic internal energy per unit mass of the fluid, $u_i$, evolves
according to
\begin{align}
\frac{\d u_i}{\d t} &=
  f_i\frac{P_i}{\rho_i^2}
  \sum_j m_j( \vec{v}_i - \vec{v}_j) \cdot \nabla W_{ij}(H_i) + \dot{u}^{\rm AV}_i\,.
\label{eq:internal-energy-evolution-maintext}
\end{align}

\noindent
$\vec{v}$ is the velocity field, $P$ is the pressure and
${\bf\nabla}\equiv\partial/\partial{\vec{x}}$ is the spatial gradient.
The system is closed by specifying the equation
of state of the fluid, $P=(\gamma - 1)u\rho$, in which $\gamma$ is the adiabatic
index.

The used equations include physical quantities of the fluid plus
terms that are intrinsic to the SPH method. The scalar field

\begin{equation}
f_i = \left(1 + \frac{h_i}{3\rho_i}\frac{\partial\rho_i}{\partial h_i} \right)^{-1}
\quad
\text{with}
\quad
	\frac{\partial\rho_i}{\partial h_i}= 	\sum_j m_j\frac{\partial W_{ij}(H_i)}{\partial h_i}
	\label{eq:drho_dh-maintext}
\end{equation}

\noindent
represents the spatial fluctuations in the smoothing length $h(\vec{x})$
(typically known as `grad-$h$' terms). They have to be taken into account whenever
$h$ is allowed to change over space and time. Formulations with such
variable $h$ are crucial in astrophysical applications, where the fluid can be
strongly compressed (over a range of several orders of magnitudes).

Finally, following~\cite{Monaghan:92:SPH,Balsara:95:SPH}, and \cite{Price:2012:SPH},
we add an artificial viscosity (AV) to the (physically inviscid) fluid in order to
resolve potential discontinuities (e.g.~due to shocks) that
could develop in the fluid. In particular, we adopt the AV model used by
the {\sc gadget}-2 code~\cite{Springel-g2:2005}. The additional terms are reflected
as $\vec{a}_i^{\rm AV}$ and $\dot{u}_i^{\rm AV}$ in
eqns.~\eqref{eq:Euler-equation-particles-maintext} and
\eqref{eq:internal-energy-evolution-maintext}, respectively, and are fully described
in Appendix~\ref{app:SPH}.

\subsection{Particle organisation within a dynamically adaptive Cartesian mesh}

The elegance of SPH results from the localization of the interaction: Particles
closeby exchange information, particles that are far away from each other do
not. To exploit this in a
code, it is crucial to evaluate neighbourhood queries (which particle is
close) efficiently:

We employ a dynamically adaptive Cartesian mesh based upon
a spacetree~\cite{Weinzierl:2019:Peano} as meta data to speed up the
neighbourhood search.
Our computational domain is embedded into a cube. We cut this cube into 27
subcubes. Per subcube, we decide recursively and independently whether to refine further. This yields a tree
hierarchy of adaptive Cartesian meshes. Within this hierarchy, we make each
particle belong to the finest cube resolution hierarchy with a cube length of at
least $C H_i(t)$ with a hard-coded constant $C$, and assign it to the closest cube vertex
\cite{Weinzierl:2015:pidt}. This assignment scheme yields a natural refinement and
coarsening criterion: A cube is refined further if one of its vertices hosts at
least $K$ particles which would fit into the next finer resolution level, too.
Cubes are removed if a set of $3^d$
children of one large cube host fewer than $K$ particles. $K$ is a tuning
parameter. 

Since our code hosts particles
on the finest spacetree level which can accommodate their $H_i$, the smoothing
kernel domain never spans more than two mesh cubes in any direction on the
respective mesh level. 
To evaluate a sum over all neighbours of a particle it is hence sufficient to
loop over all particles which are contained in the same cube as the particle of
interest or in any vertex-connected neighbour cube.
However, we also have to extend this argument recursively over 
coarser and finer mesh resolutions.


\begin{table}[htb]
 \caption{
  Attributes per grid vertex.
  Some additional vertex properties required for the parallelisation and data
  exchange are omitted from the table.
  Magic range constants can be change by user (default shown).
  \label{table:demonstrator:vertex-data-model}
 }
 {
  \begin{tabular}{ |p{4cm}|l|p{3cm}|  }
 \hline
 Property & Data type & Range \\
 \hline
 refinement status  &  enumeration  & \{refined, unrefined, will be
 refined, will be coarsened\} \\
 is vertex local   & boolean  & \\
 particle pointers & pointers (via linked list, e.g.) & \\
 level             &   int array $\mathbb{N}$   & $\in(0,63)$\\
 local             &   boolean & \\
 hanging           &   boolean & \\
 neighbour ranks   &   int array $\mathbb{N}^{2d}$ & $\in(0,65536)$\\
 \ldots & \ldots & \ldots \\
  \hline
\end{tabular}

 }
\end{table}

Our implementation uses the spacetree as meta data structure to organise the
particles.
Once we linearise the tree along a space-filling curve, an enumeration per
vertex is sufficient to encode the
whole tree structure and to derive any spatial, geometric cell information. Said enumeration signals whether adjacent cubes within the spacetree are unrefined,
refined, will be refined, or will be coarsened.
Few additional bits and counters for the parallelisation plus 
pointers from and to particles supplement the vertex data type.
The mesh data has a small memory footprint and hosts primarily enums, booleans
and integers (Table \ref{table:demonstrator:vertex-data-model}).

\begin{hypothesis}
 \label{demonstrator:hypothesis:mesh}
 We assume that the mesh data resides in close cache and is used frequently as
 lookup mechanism.
 If the integer packing through Annotation~\ref{extension:pack-integer}
 introduces algorithmic latency, it will slow down the code.
\end{hypothesis}

\subsection{Parallelisation}

Our experiments employ a very simple domain decomposition method:
The mesh is split up into equidistant chunks along the Peano space-filling
curve.
Each chunk is deployed to one rank, i.e.~each rank gets a unique sequence of
cubes from the spacetree.
Each chunk furthermore is cut again into subchunks along the curve, such that
each thread is given a chunk of its own.
The particle distribution follows this non-overlapping domain decomposition of
the tree:
Each particle is owned, i.e.~stored and updated, by the thread
which owns the cube that overlaps with the particle centre.

To allow the individual chunks to update their particles independently, we
supplement each mesh with ghost cubes.
Due to the definition of $H$, one layer of ghost cubes on each spacetree level
is sufficient.
Particles falling into a ghost cube are replicated on neighbouring domain
subpartitions.
This requires synchronisation of data and yields a certain memory overhead, but
it allows the individual threads to process their particle data without any
sychronisation, as long as we ensure that all data (replica) are made consistent
after each algorithmic step.
More sophisticated, task-based
formalisms \cite{Schaller:2016:Swift} exploiting shared memory exist, yet are out
of scope here.

\subsection{Data model and data access pattern}
\label{sec:algorithmic-blueprint}

\begin{table}[htb]
 \caption{
  Core (physical) data per particle data model with known ranges and accuracies.
  \label{table:demonstrator:particle-core-data-model}
 }
 {
  \begin{tabular}{ |p{3cm}|l|l|p{2.0cm}|  }
 \hline
 Property & Symbol & Data type & Range \\
 \hline
 mass
  & $m$ & double & {\rm const.} \\
 smoothing length
  & $h$ & double & $\in(h_{\rm min},h_{\rm max}]$ \\
 position & ${\bf x}$ & double array $\mathbb{R}^d$& $\in(0,1]$ \\
 velocity & ${\bf v}$ & double array $\mathbb{R}^d$ &  \\
 acceleration & ${\bf a}$ & double array $\mathbb{R}^d$ & \\
 density & $\rho$ & double & $\in(0,\infty]$ \\
 pressure & $P$ & double & $\in(0,\infty]$ \\
 internal energy      & $u$ & double & $\in(0,\infty]$  \\
 time derivative of $u$ & $\dot{u}$ & double & \\
 \hline
\end{tabular}

 }
\end{table}

Our SPH implementation follows few well-trodden paths. 
As the particles may move in each and every simulation time step and hence have to be resorted into
the spacetree frequently, we hold them as an array-of-structs (AoS).
The particles'
core data model (Table \ref{table:demonstrator:particle-core-data-model}) stores
nine physical variables per particle and updates them along the following
scheme:

%
%

First, the algorithm calculates the density and smoothing length per particle.
The latter determines the shape of
$W_{ij}$, i.e.~the smoothing kernel $W$ associated with $i$ yet depending on the
distance to particle $j$. 
For the underlying iterative scheme, the algorithm reads
the particles' mass, density, position and smoothing length, and it updates
their $\rho$ and $h$ iteratively. Note however that during this particle-particle
interaction loop, for any given particle only the neighbouring
particles' masses and positions are required to be read, but not their $h$.

Second, the algorithm ``prepares'' each particle to evaluate its acceleration
and internal energy evolution using the updated values of $h$ and $\rho$, i.e.
it calculates and stores most of the terms going into the sum in the right-hand-side
of~\eqref{eq:Euler-equation-particles-maintext} and~\eqref{eq:internal-energy-evolution-maintext},
but the actual sum (loop) over $j$ is calculated later. In particular, this step
calculates $f_i$, $P_i$, as well as individual AV terms such as
\eqref{eq:balsara-minimal-sph}.

Third, a second loop is performed to calculate the acceleration that is exerted on
each particle by its neighbours via~\eqref{eq:Euler-equation-particles-maintext}
and the AV terms. The internal energy
transfer among them via~\eqref{eq:internal-energy-evolution-maintext} is also
performed. This step collects the terms evaluated and stored in the previous step.

The core algorithmic steps read and write different
subsets of the particle properties.
Prior to each algorithmic step, data read has to be sent from particles to their
halo copies on other ranks.

\begin{hypothesis}
 \label{demonstrator:hypothesis:subsets-attributes-compute-intense}
 The compute-intensive steps, i.e.~the non-linear density solve and force
 calculation, should benefit from packing, as they can hold more data in closeby
 caches.
 However, the same packing might constrain the vector efficiency.
\end{hypothesis}


\noindent
Finally, the code integrates the equations of
motion~\eqref{eq:Euler-equation-particles-maintext}
and~\eqref{eq:internal-energy-evolution-maintext} to update the particles'
position, velocity and internal energy. 
In this step, the spatial particle topology, i.e.~the association of
particles to mesh cells, can change, and particles can leave their subdomain,
i.e.~travel between cores and ranks.

\begin{hypothesis}
 \label{demonstrator:hypothesis:mpi-data-exchange}
 While the spatial particle topology (spatial arrangement) remains invariant for
 most compute steps, particles can eventually travel between ranks and hence
 require the exchange of all of their \replaced[id=R1]{instance variables}{attributes} via MPI.
 Here, we expect the code to benefit from an reduction of the memory footprint
 as we stress the interconnect's bandwidth.
 For all other algorithm steps, we expect to benefit from the fact that we can define
 views on data types and exchange only some particle \replaced[id=R1]{properties}{attributes}.
\end{hypothesis}

\noindent
Particles hold predominantly floating-point data.
Some \replaced[id=R1]{of their variables}{attributes} have temporal access character, i.e.~are only used for some
algorithm steps, while other properties such as the particle positions are needed
in each and every algorithm step.
We also store some secondary data such as gradients within each particle, i.e.~quantities that are derived from other data yet cannot be recomputed quickly on-the-fly when we need them later on.
This eliminates the need to reconstruct them expensively.
We end up with a significant memory footprint per particle
(Table \ref{table:demonstrator:particle-technical-data-model}).

As we commit to AoS as storage format and as we deal with huge numbers of
particles, we may assume that we have to read them from the main memory in each
and every compute step.

\begin{hypothesis}
 \label{demonstrator:hypothesis:bandwidth}
 We assume that the computationally cheap compute kernels suffer from bandwidth
 restrictions and hence benefit from the floating-point compression.
\end{hypothesis}

\subsection{Floating point accuracy}

\begin{table}[htb]
 \caption{
  Excerpt of additional \replaced[id=R1]{variables}{attributes} per particle which are required to keep the
  data consistent throughout the explicit time steps and the evaluation through
  multiple compute kernels.
  \label{table:demonstrator:particle-technical-data-model}
 }
 {
  \begin{tabular}{ |p{5cm}|l|l|p{3cm}|  }
 \hline
 Property & Symbol & Data type & Range \\
 \hline
 \hline
 \multicolumn{4}{|l|}{Variable smoothing length terms} \\
 \hline
 `grad-$h$' term & $f$ & double & \\
 density $h$-gradient & $\partial_h\rho$ & double & \\
 \hline
 \multicolumn{4}{|l|}{Artificial viscosity scheme} \\
 \hline
 Balsara switch & $B$ & double & $\in(0,1)$ \\
 signal velocity & $v_{\rm sig}$ & double & $\in(0,\infty]$ \\
 velocity curl & $\nabla\times{\bf v}$ & double array $\mathbb{R}^d$ &  \\
 velocity divergence & $\nabla\cdot{\bf v}$ & double array $\mathbb{R}^d$ & \\
 \hline
 \multicolumn{4}{|l|}{Newton-Raphson iterative solver} \\
 \hline
 old smoothing length & $h_{\rm old}$ & double & $\in(h_{\rm min},R_{\rm cutoff}]$ \\
 iteration count & $N_{\rm iter}$  & int & $\in(1,N^{\rm max}_{\rm iter}]$ \\
 has particle converged &  & bool & $\in\{0,1\}$ \\
 \hline
 \multicolumn{4}{|l|}{Time integration} \\
 \hline
 CFL time-step size & $\Delta t$  & double & $\in(0,\infty]$ \\
 has particle been kicked &  & bool & $\in\{0,1\}$ \\
 Move state &  & enum & $\in\{0,1,2\}$ \\
 Parallel state &  & enum & $\in\{0,1,2\}$ \\
 New Parallel state &  & enum & $\in\{0,1,2\}$ \\
 \hline
\end{tabular}

 }
\end{table}

It
is not clear in which precision different fields have to be stored: Even if we
assume that double precision is required for primary, physical quantities,
properties such as the smoothing length carry a lower information density: A
difference in $h$ in the order of floating-point accuracy most often does not
include more particles into the underlying truncated sum, while even additional
particles do not affect the algorithm outcome negatively. 
If in doubt, we can always make $h$ slightly larger.

There is, to the best of our knowledge, no formal proof which accuracy is
required for \replaced[id=R1]{particle variables}{attributes} which carry physical meaning.
Empirical evidence and comparisons to other codes from the field suggest that
we cannot make compromises on the particle positions and density which feed into non-linear follow-up calculations, but can compress other
quantities to single precision or beyond.

\begin{hypothesis}
 \label{demonstrator:hypothesis:floating-point-precision}
 Our baseline code is written over doubles, while other codes employ a mixture of double and single precision. Yet, not all variables might even require single precision.
\end{hypothesis}

\section{Results}
\label{section:results}

%
%
To assess the impact and potential of our language extensions, we rely on 
various benchmarks which highlight different extension properties.
We run all benchmarks on several test platforms.

Durham's Hamilton 8 supercomputer is a cluster hosting AMD EPYC 7702 64-Core processors, i.e.~the AMD K17
(Zen2) architecture, where the 2$\times $64 cores per node are spread over two sockets.
Each core has access to 32 kB exclusive L1 cache, and 512 kB L2 cache.
The L3 cache is (physically) split into chunks of 16 MB associated with four
cores.
Infiniband HDR 200GB/s serves as interconnect.
A second machine is an AMD EPYC 9654 (Genoa) testbed. 
It features $2 \times 96$ cores over $2 \times 4$ NUMA domains spread over
two sockets, hosts
an L2 cache of 1,024 KByte per core and a shared L3 cache with 384 MByte per
socket.
Our third system hosts an Intel Xeon Gold 6430 (Sapphire Rapid).
It features $2 \times 32$ cores over two sockets.
They form two NUMA domains with 
an L2 cache of 2,048 KByte per core and a shared L3 cache with 62 MByte per
socket.

We use Intel MPI (version 2021.4) for the distributed memory
parallelisation and realise all shared memory parallelism through OpenMP.
The experiments rely on the most aggressive generic compiler optimisation level
and code generation for the specific target instruction set.
All results are conducted with the Peano AMR framework
\cite{Weinzierl:2019:Peano} handling all the meshing, domain decomposition and data handling, while the SPH compute
kernels stem from the \swift software \cite{Schaller:2016:Swift,Schaller:2023:Swift}.
The
particle administration within the mesh follows the particle-in-dual-tree
concept \cite{Weinzierl:2015:pidt}.

\subsection{Lossless compression of integer data, enums and booleans}
\label{sec:orbit_experiment}

\begin{figure}[htb]
 \begin{center}
  \includegraphics[width=0.3\textwidth]{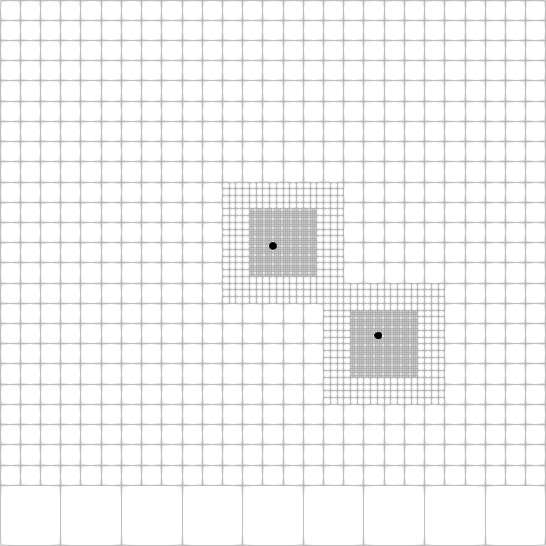}
  \hspace{0.1cm}
  \includegraphics[width=0.3\textwidth]{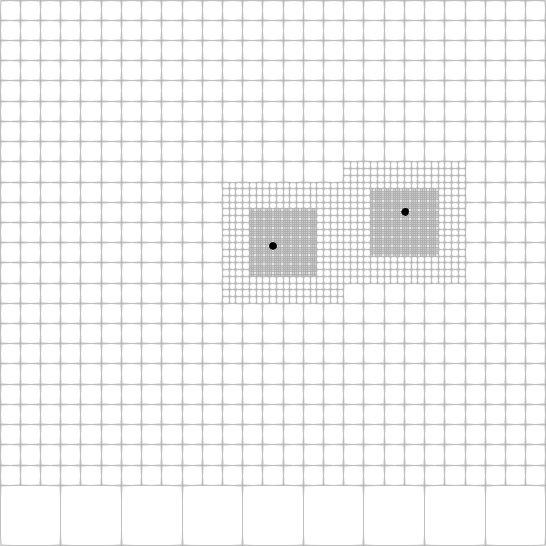}
  \hspace{0.1cm}
  \includegraphics[width=0.3\textwidth]{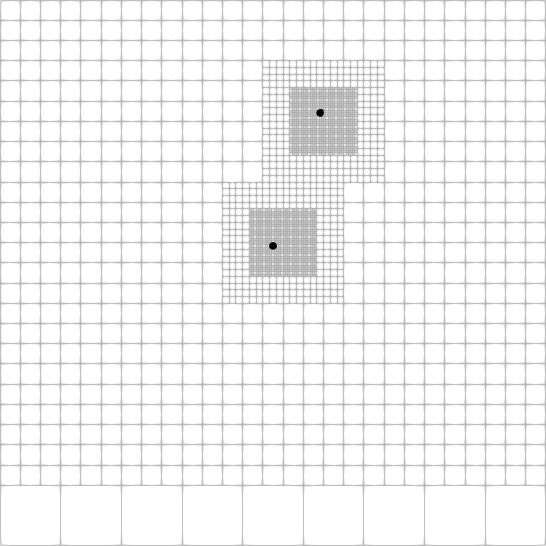}
 \end{center}
 \caption{
  Earth and Sun orbit around each other. Our adaptive mesh zooms into and
  follows the two objects.
  \label{results:experiments:earth-sun} 
 }
\end{figure}

%
%
In order to study the impact of our compression on integers, enums, and
booleans, we run a two-body problem where we disable all SPH-specific numerics, i.e.~density and force calculations.
Instead, we fix one object at the centre of the domain and send the second body
on a stable orbit by hard-coding its centripetal force\added[id=R1]{ (Figure~\ref{results:experiments:earth-sun})}.
We eliminate all compute load from the setup and instead focus on the meshing data structure which holds no floating point data.

Around the two objects, we resolve the computational domain 
with various maximum AMR depths:
The coarsest mesh is fixed before we refine recursively for a fixed number of
times around each particle.
Since one particle moves, the mesh moves, too.
The vertex's booleans and enums are annotated with
\texttt{[[clang::pack]]}, whereas integers are annotated with \linebreak
\texttt{[[clang::pack\_range(MIN,MAX)]]} and make use of known value ranges 
(Table~\ref{table:demonstrator:vertex-data-model}).

%
%
Our studies focus on single-core runs only, and we keep track of the
time-to-solution and number of instructions retired to assess the overhead introduced by the bit packing.
Further to that, we measure the L2 cache miss rate, i.e.~the number of cache misses vs.~the
number of instructions retired, as well as the L2 cache miss ratio, i.e.~the number of cache misses vs.~the number of cache
requests.
All data reported in Table~\ref{table:results:integer-compression:hw-counter} are
normalised against measurements from the unmodified code.
They are hardware counters obtained through Likwid~\cite{Hager:2010:Likwid}.

\begin{table}[htb]
 \caption{
  Impact of the integers compression on system
  characteristics for the simulation of the orbiting particle.
  The higher the depth, the more accurate (finer) the dynamically adaptive mesh. 
  \replaced[id=R4]{All quantities are}{We present baseline (uncompressed) data vs.~data} normalised against the
  baseline measurements \added[id=R4]{without any packing} \replaced[id=Ours]{on}{from} the EPYC 7702.
  \label{table:results:integer-compression:hw-counter}
 }
 {
   \begin{tabular}{r|rrr|r}
  AMR   & Instr.~retired & L2 cache miss rate & L2 rate
  cache miss ratio & Runtime
  \\
  \hline
  0 & 1.07 & 0.92 & 0.93 & 1.08\\
  1 & 1.12 & 0.91 & 1.00 & 1.12\\
  2 & 1.10 & 0.92 & 0.97 & 1.12\\
  3 & 1.12 & 0.92 & 0.93 & 1.11\\
\end{tabular}

 }
\end{table}

%
%
We work with a cache-oblivious AMR code~\cite{Weinzierl:2019:Peano} where the
mesh code of tree is linearised into one big stream, while 
the total memory footprint of the setup is small.
\replaced[id=R4]{The runtime increases by 8\% on a regular mesh up to 12\%~for setups with dynamic adaptivity}{Once we encounter a reasonable number of refinements around the particles, 
the	runtime increases by around 10\%} due to the
compression~(Table~\ref{table:results:integer-compression:hw-counter}).
This correlates directly to the number of instructions retired relative to the
baseline code.
The packing/unpacking introduces additional instructions.
\replaced[id=R4]{
 This happens even though the
}{
 All the data resides within the L2 cache most of the time. The
}
packing improves the cache access characteristics\replaced[id=R4]{, i.e.~we reduce the L2 and L3 misses by around 8\% or 0.07\% respectively. However}{, but} this effect is
marginalised\replaced[id=R4]{ as the baseline code only has an L2 miss rate of 0.02\% (not shown) and hence }{ and} cannot compensate for the additional instructions.

%
%
Our storage format modifications come not for free:
They require the compiler to introduce additional bit shifts and bit masking.
While these operations are cheap, they nevertheless increase the computational
load of the generated code compared to the baseline and make the code slightly
slower.
This confirms Hypothesis~\ref{demonstrator:hypothesis:mesh}
experimentally.

\subsection{The impact of mantissa compression on the accuracy of the solution}
\label{section:results:solution-accuracy}

%
%
In order to investigate how the accuracy of SPH is
affected by the mantissa truncation,
we run the Noh benchmark \cite{Noh:1987} using various precisions (valid bits)
for the particles' floating-point data.
Analytical solutions for the radial profiles of the density and velocity fields
are known for Noh.
At $t=0.1$, we expect a strong shock front at a circle of radius $r\approx0.032$ around the
centre and a rather smooth solution otherwise.
These profiles are calculated as circular averages over the solution.
SPH will yield oscillations around the shock and will deliver underestimated
densities inside the shock region.
Both are well-documented for traditional SPH schemes such as the one implemented
in our code.
Yet, it is not clear how the compiler's additional truncation amplifies or damps
these numerical artefacts.

%
%
A systematic study of admissible precisions per \replaced[id=R1]{variable}{attribute} is beyond scope, as
it would involve multiple long-term accuracy and stability studies.
We also note that there are many different combinations of accuracies,
as we can set the number of valid mantissa bits per particle \replaced[id=R1]{property}{attribute}.
This yields a large configuration space.
Therefore, we initially pick the same number of valid bits for each and every
floating point \replaced[id=R1]{value}{attribute}.
The only exception is the particle position $\vec{x}$, which we always store in 
double precision.
We present profiles for 52 mantissa bits (native double precision), 23 mantissa
bits (single precision) and 10 mantissa bits.
The latter is equivalent to half precision.
All calculations remain coded in double precision.

\begin{figure}[htb]
 \begin{center}
   \includegraphics[width=0.39\textwidth]{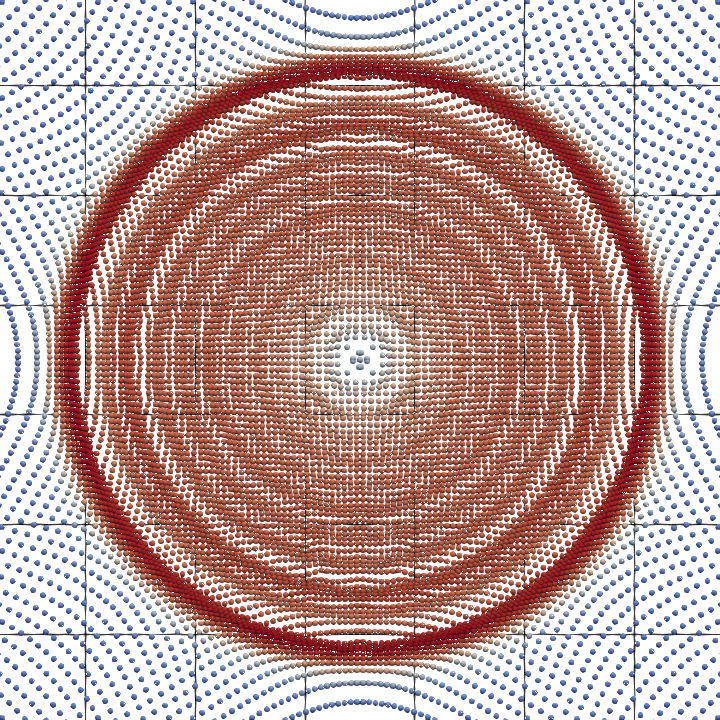}
   \includegraphics[width=0.39\textwidth]{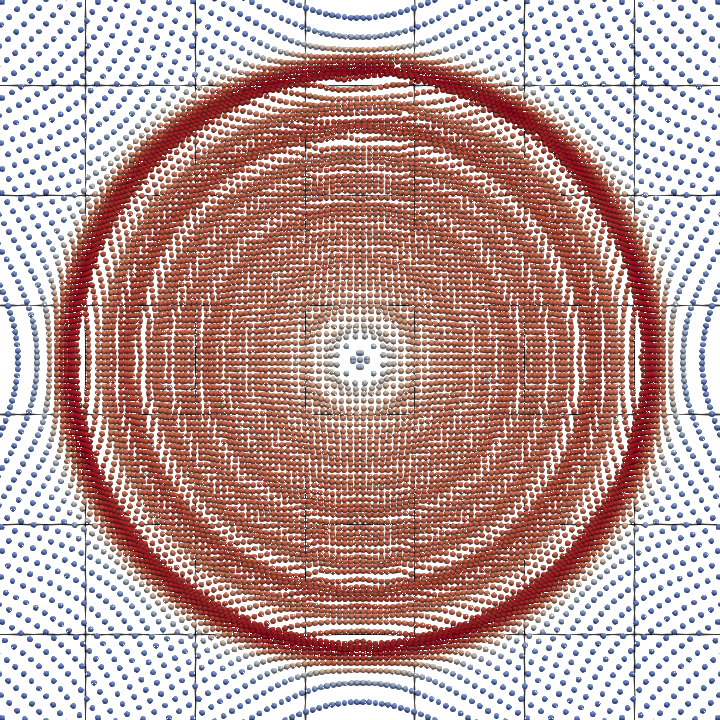}
   \\
   \vspace{1mm}
   \includegraphics[width=0.39\textwidth]{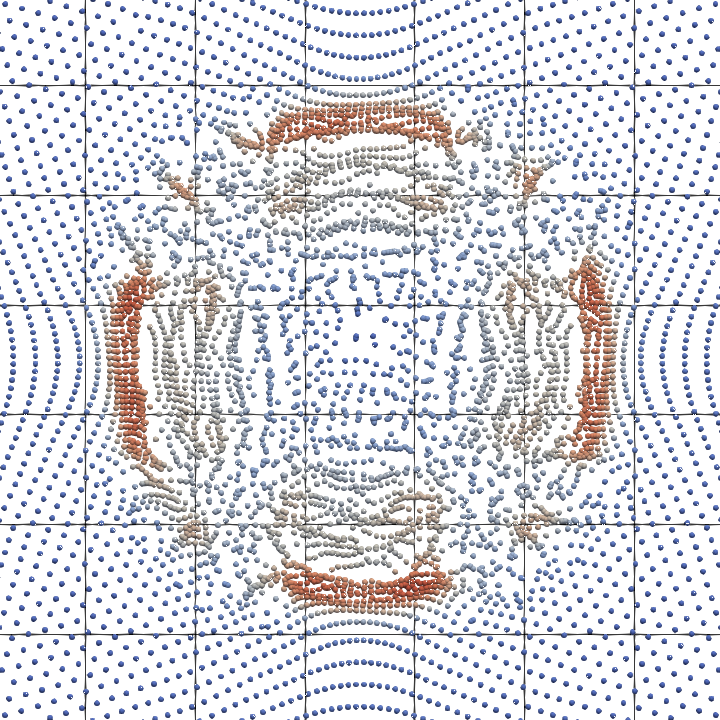}
   \includegraphics[width=0.39\textwidth]{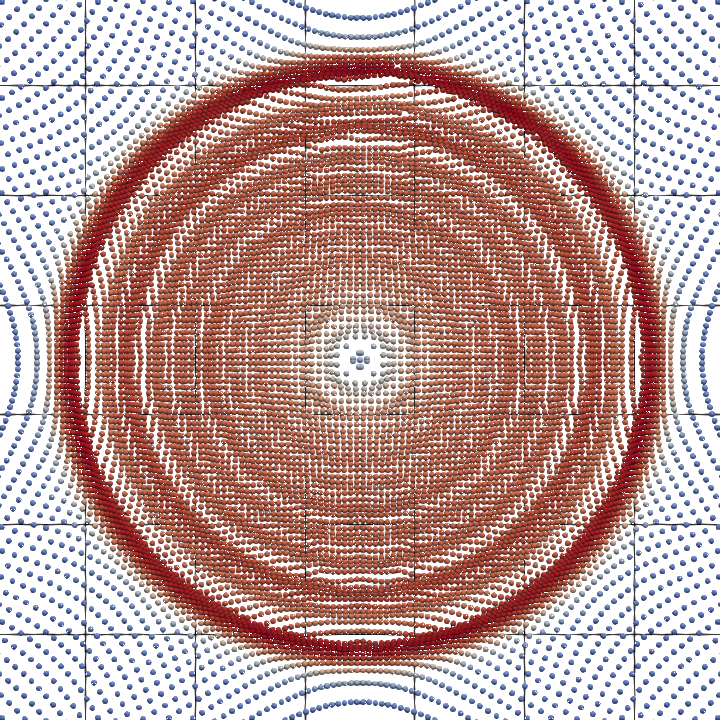}
 \end{center}
 \caption{Central region of the Noh problem at $t=0.1$ for simulations with different
   mantissa size for the particle \replaced[id=R1]{properties}{attributes}. 
   Lexicographically: 
   52 bits (double precision),
   23 bits (single-precision equivalent),
   10 bits (half-precision equivalent), 
   and a mixed precision case, in which 23 bits are used for the
   ``core'' particle \replaced[id=R1]{variables}{attributes} shown in
   Table~\ref{table:demonstrator:particle-core-data-model}, and 10 bits for all
   others.
   The colour map encodes the SPH density field values.
   \label{fig:Noh-accuracy-visualisation}
  }
\end{figure}

\begin{figure}[htb]
  \begin{center}

\includegraphics[width=1.0\textwidth]{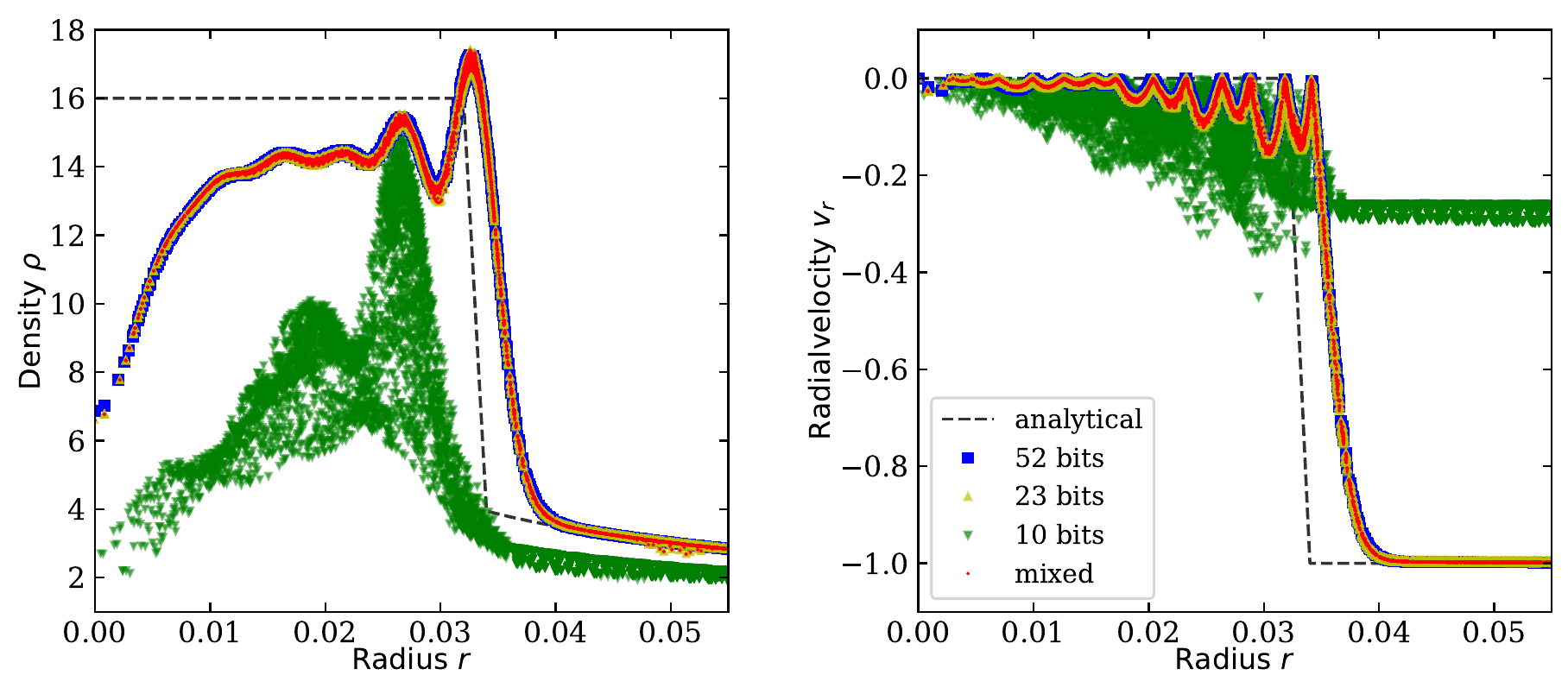}
  \end{center}
  \caption{Radial profiles for the density field (left panel) and the radial
  component of the velocity field (right panel) for the central region of the
  setup in Figure~\ref{fig:Noh-accuracy-visualisation}.
  }
  \label{fig:Noh-accuracy-profiles}
\end{figure}

%
%
A visual comparison of single vs.~double precision suggests that the compression has
no impact at all (Figure~\ref{fig:Noh-accuracy-visualisation}).
However, once we employ only 10 bits per mantissa, the solution is destroyed.
The lack of precision is most noticeable on the diagonals, but we also see
some loss of symmetry within the shock area.
While some particles seem to outrun the shock,
the vanilla SPH version overestimates the shock speed, while very strong
compression yields a shock that propagates too slowly
(Figure~\ref{fig:Noh-accuracy-profiles}).
In a comparison of the radial profiles of the density
and radial velocity fields, the single and double precision case are
indistinguishable.
Besides the late shock arrival time for half precision, 
the solution becomes scattered which is reflected in the loss of symmetry in
the plot, and the velocity outside the shock does not match the initial
condition $v_r =-1$ closely anymore.

%
%
As a final test for the precision trials, we run a ``mixed'' case where we keep the core
particle data from Table~\ref{table:demonstrator:particle-core-data-model} in single
precision and the \replaced[id=R1]{remaining variables}{rest of attributes} in half precision.
$\vec{x}$ remains in native double precision all the time.

%
%
Despite the aggressive compression of non-core particle data, the profiles continue to match
full single precision or double precision calculations quite well
(Figures~\ref{fig:Noh-accuracy-visualisation} and
\ref{fig:Noh-accuracy-profiles}).
The data suggest
the possibility of truncating the mantissa below the 23 bits for at least a
subset of particle \replaced[id=R1]{instance variables}{attributes}, while we retain accurate and stable outcomes. 
At the same time, we recognise that a globally reduced precision is
inappropriate
(Hypothesis~\ref{demonstrator:hypothesis:floating-point-precision}).
Significant work on the numerical and experimental side is
required to understand which \replaced[id=R1]{variables}{attributes} we can compress and by which ratio.
Our annotations can streamline this development work; 
notably as we support user-defined
compression on an \replaced[id=R1]{variable-by-variable}{attribute-by-attribute} basis which can incrementally be
introduced (Rationale~\ref{language:rationale:acceptance}).

\subsection{Tailored MPI datatypes}
\label{sec:results:mpi}

%
%
To highlight the importance of minimalist, tailored MPI datatypes, we strip the
code off any intra-node parallelisation and computation and solely focus
on the data exchange between two MPI ranks.
The ranks are deployed to two nodes and each sends a fixed number $N$ of
particles to their counterpart as we increase the local domains.
This mimicks a ping-pong MPI test.

%
%
In a first run, we exchange whole particles\replaced[id=R1]{i.e.~all variables}{, i.e.~all attributes}.
We study the particle migration or resorting due to position updates or dynamic
load balancing.
In a second run, we exchange solely $f$, $\rho$ and $h$ in line
with (\ref{eq:drho_dh}).
These are the \replaced[id=R1]{values}{attributes} required by the density update iterations.
Third, we exchange all \replaced[id=R1]{variables}{attributes} besides the position and the $f$, $\rho$,
$h$ quantities.
This would be an example of a typical \replaced[id=R1]{data}{attribute} exchange used by the time integrator.
The latter version can also be run with reduced floating-point precision.
As we measure the total communication life span, all data comprise both latency
and bandwidth effects. 
For small $N$, we expect latency to dominate, whereas bandwidth constraints
take over for larger particle counts.

\begin{table}[htb]
  \caption{
    MPI ping-pong test for various particle versions, i.e.~subsets or
    compression factors, with different byte footprint. Time $[t]=s$ per
    particle.
    $N$ is the number of particles exchanged per MPI call, i.e.~per boundary
    exchange.
    The table entries are coloured red or green if the time is bigger or smaller,
    respectively, compared to the column to their left, i.e. compared to the next
    lower level of compression.
    \label{table:experiments:mpi:ping-pong-two-nodes}
  }
  {

\begin{tabular}{|r|rrrr|}
  \hline
  $N$
    & \multicolumn{1}{c|}{288 Bytes}
    & \multicolumn{1}{c|}{272 Bytes}
    & \multicolumn{1}{c|}{168 Bytes}
    & \multicolumn{1}{c|}{144 Bytes}
  \\
  \hline
  1 &
    $1.17\cdot 10^{-3} $ &
        \redcol   $1.20\cdot 10^{-3} $ &
        \redcol   $1.23\cdot 10^{-3} $ &
        \greencol $1.18\cdot 10^{-3} $ \\
  4 &
    $3.61\cdot 10^{-7} $ &
        \redcol   $5.31\cdot 10^{-7} $ &
        \greencol $2.76\cdot 10^{-7} $ &
        \greencol $2.08\cdot 10^{-7} $ \\
  8 &
    $2.89\cdot 10^{-7} $ &
        \greencol $1.44\cdot 10^{-7} $ &
        \redcol $1.83\cdot 10^{-7} $ &
        \redcol $2.07\cdot 10^{-7} $ \\
 32 &
    $3.19\cdot 10^{-7} $ &
        \greencol $2.55\cdot 10^{-7} $ &
        \greencol $8.77\cdot 10^{-8} $ &
        \greencol $8.71\cdot 10^{-8} $ \\
128 &
    $2.04\cdot 10^{-7} $ &
        \redcol   $3.57\cdot 10^{-7} $ &
        \greencol $2.86\cdot 10^{-7} $ &
        \greencol $2.80\cdot 10^{-7} $ \\
512 &
    $1.70\cdot 10^{-7} $ &
        \greencol $1.21\cdot 10^{-7} $ &
        \greencol $8.23\cdot 10^{-8} $ &
        \greencol $7.40\cdot 10^{-8} $ \\
2,048 &
    $1.66\cdot 10^{-7} $ &
        \greencol $1.17\cdot 10^{-7} $ &
        \greencol $7.57\cdot 10^{-8} $ &
        \greencol $6.84\cdot 10^{-8} $ \\
8,192 &
    $1.41\cdot 10^{-7} $ &
        \greencol $1.14\cdot 10^{-7} $ &
        \greencol $6.73\cdot 10^{-8} $ &
        \greencol $6.15\cdot 10^{-8} $ \\
32,768 &
    $1.36\cdot 10^{-7} $ &
        \greencol $1.24\cdot 10^{-7} $ &
        \greencol $6.31\cdot 10^{-8} $ &
        \greencol $5.40\cdot 10^{-8} $ \\
\hline
\end{tabular}

  }
\end{table}
  
%
%
If very few particles are migrated or particles are exchanged
individually---this happens for example when we sort them into cubes incrementally---the size of the particle plays close to
no role (Table~\ref{table:experiments:mpi:ping-pong-two-nodes}).
In some situations, picking a subset of \replaced[id=R1]{variables}{attributes} within the MPI implementation introduces a slight
performance penalty.
The more particles we transfer, the lower the cost per particle.
Further to that, the size of the particle matters, i.e.~exchanging only
\replaced[id=R1]{a subset of variables}{subattributes} or compressed floating-point numbers reduces the runtime.

Picking a subset of \replaced[id=R1]{a struct's instance variables}{attributes} introduces some overhead.
We assume that the MPI implementation internally has to gather and scatter some
data.
If the individual \replaced[id=R1]{variables' memory footprints}{attributes} become smaller due to floating-point compression,
we again profit.
This is likely a memory copy effect.
Once we increase the particle count, the latency penalty is amortised over all
particles, and bandwidth constraints kick in.
Therefore, the particle footprint does matter.
We approach an almost linear regime, where a halving the memory footprint
almost yields a speedup of two.

\subsection{Performance of the \replaced[id=Ours]{algorithmic}{algoritmic} phases}
\label{section:results:scalability}

%
%
We finally assess the performance of the SPH compute phases.
For this, we manipulate two degrees of freedom:
the particle count and the number of threads.
The threads are pinned to cores, and we use \texttt{numactl} with the
\texttt{membind} option to ensure that all data stems from the used cores or
NUMA domains respectively.
That is, the cores use only cache and memory from one socket (Intel) or the
number of NUMA domains employed (AMD).
As each thread is pinned to one core, core and thread are used as synonyms.

Our measurements compare the uncompressed C++ version using the
\texttt{double} data type everywhere with a version where we employ the integer,
enum and bool packing plus annotations of floating point \replaced[id=R1]{values}{attributes}:
Yet, we only restrict the \texttt{double} to \texttt{float}'s mantissa
precision, i.e.~we stick to a regime which is known to be robust and do not study
further gains resulting from below-\texttt{float} storage.
Integer and floating-point compression together bring the particle's memory
footprint down from 256 bytes to 152 bytes.

As we use a leapfrog time integrator, a time step
(Section~\ref{section:use_case_SPH}) decomposes into two kicks (acceleration
update) and one drift (movement, i.e.~position update), interfused by the
density and force calculation.
For the performance studies, we distill a benchmark (miniapp) running through
this sequence.
It allows us to mimick two realisation variants:
In the first variant, we run through the sequence of the time step calculations
one by one, always traversing all particles.
This mirrors classic fork-join parallelism, i.e.~one global parallel for loop
per computational step.
In the second variant, we work on one small chunk of particles at a time, i.e.~run the calculations over this chunk several times before we continue with
the next chunk.
This mirrors a task-based approach
\cite{Cao:2022:GordonBell,Schaller:2023:Swift}, where we traverse the task graph
depth-first:
If a set of particles tied to one vertex has drifted, we immediately kick again,
update the density (with multiple iterations), compute forces, and so forth, all
using minimal data exchange with other tasks handling spatially close particle
sets.
We try to complete as many steps on a small subset of the
data as possible.

All setups are constructed such that the workload resembles the computational load that we obtain when we
hold approximately 64 particles per cell.
We balance these chunks of 64 equally among the involved threads using 
OpenMP's static partitioning.
The benchmark clears all caches prior to the first kernel invocation assessed.

%
%
Two different memory access characteristics arise:
For the sequence of steps, we stream the whole particle set into the cores per
kernel invocation.
The data has to run through the whole memory hierarchy once the total memory
footprint of all particles is big enough, i.e.~once the particles do not fit
into a cache anymore.
Otherwise, they reside within the L3 or L2 cache, respectively.
For a task-like setup, we repeatedly work on the same small chunks of particles.
They likely reside in cache.
\replaced[id=Ours]{Memory-wise}{Memory-wisely}, we work very localised.

Our miniapp breaks the runtime characteristics down
per kernel.
We discuss the kernels with linear internal cost separate from kernels with
quadratic complexity, and use the drift as representative for the former while
the force calculation represents the latter.
Kicks and density iterations exhibit very similar characteristics as those
discussed.
For all kernels, we measure the throughput, i.e.~number of particle updates per
second.
In the case of the density calculations, this corresponds to the cost for one
non-linear iteration.
For the force, it corresponds to the summation over all local neighbours which 
have an impact.
In practice, we do not know how many iterations are required over a set of
particles if we determine the density.
Yet, the characteristics of many iterative updates are covered by the task-based
miniapp execution pattern, i.e.~if some particles trigger many updates, their
memory access characteristics will start to resemble the task-based miniapp,
even though we might globally work with a cascade of for loops over the
individual algorithm phases.

\subsubsection{Kernels with linear computational complexity on Sapphire Rapid}

\begin{figure}[htb]
  \begin{center}
    \includegraphics[width=0.46\textwidth]{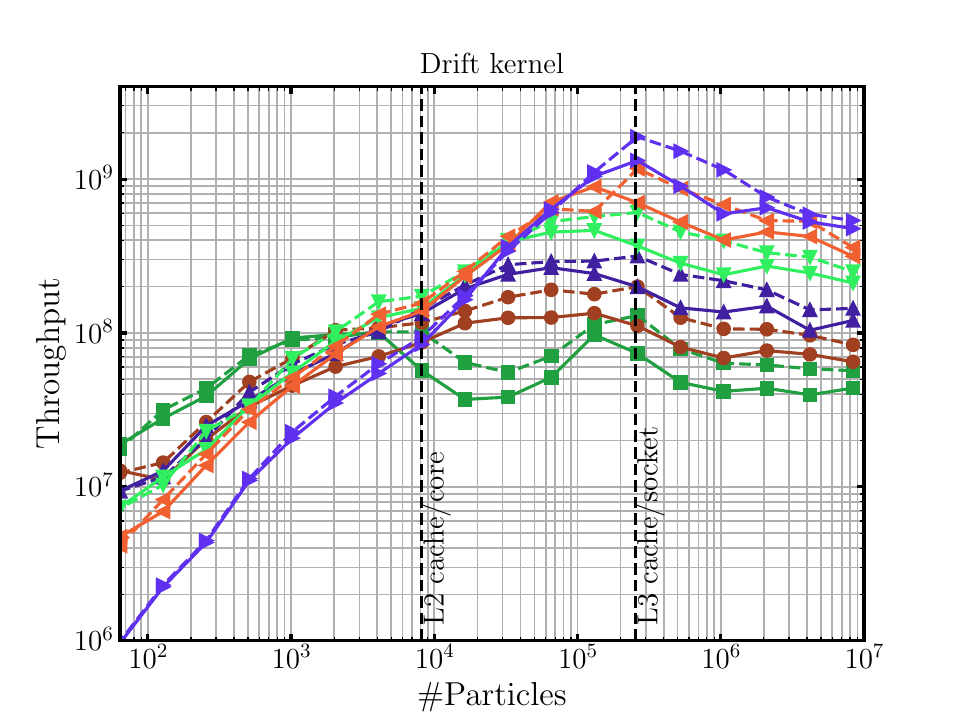}
    \includegraphics[width=0.46\textwidth]{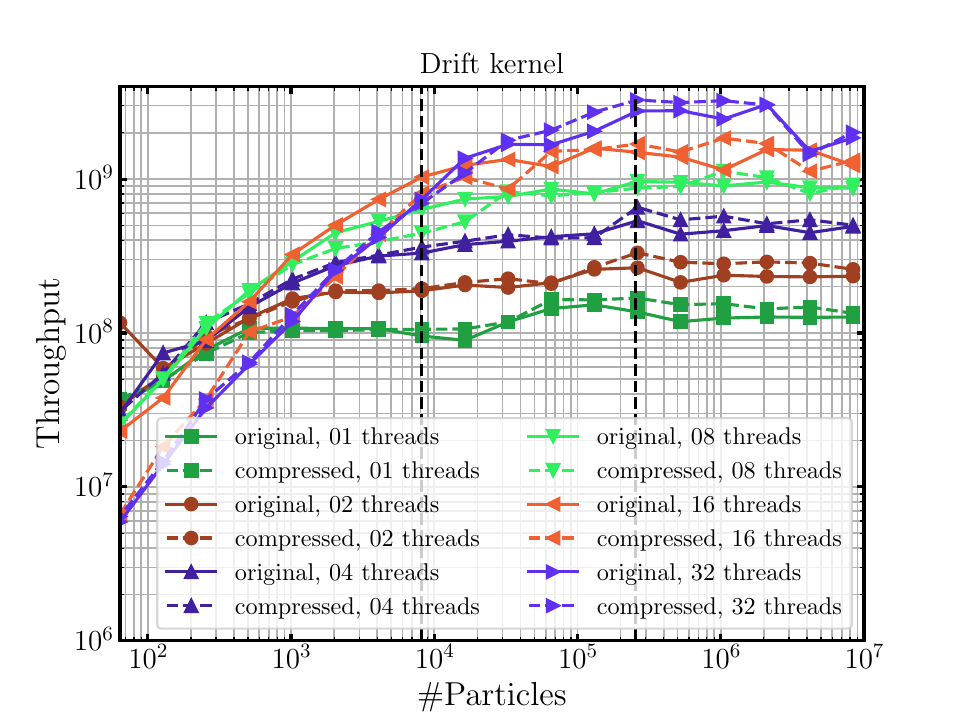}
    \includegraphics[width=0.46\textwidth]{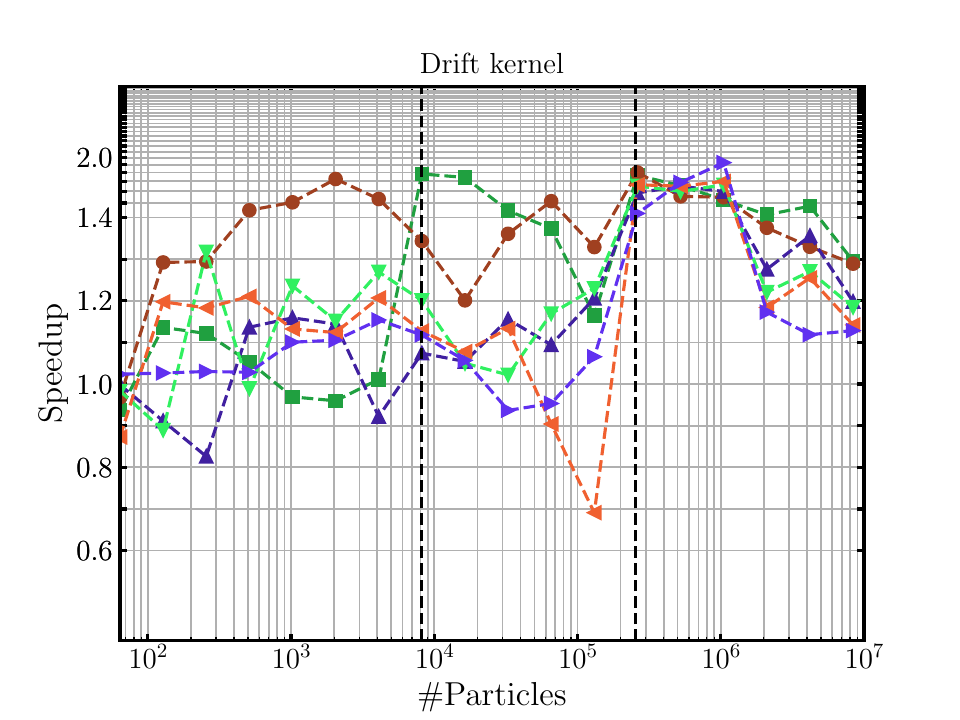}
    \includegraphics[width=0.46\textwidth]{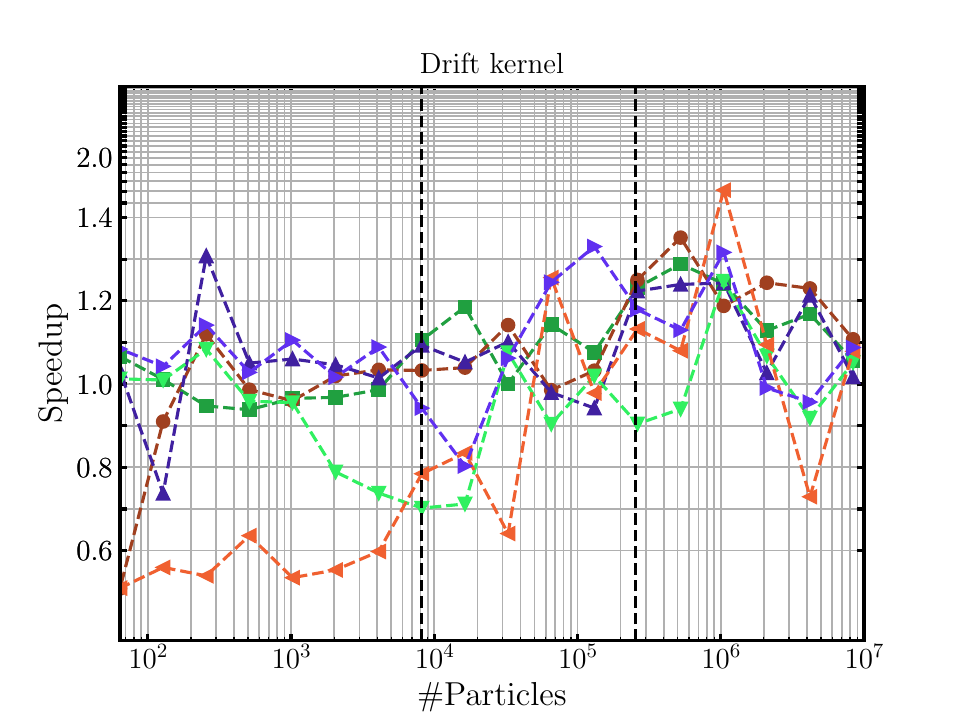}
  \end{center}
  \caption{
    Runtime behaviour for the drift step for various
    particle counts\added[id=Ours]{ measured} on one socket of the Sapphire Rapid. 
    We compare profiles for
    a stream-like access (left) to the profiles resulting from a task-based realisation (right).
    We measure throughput, i.e.~particle updates (top), but translate them into
    speedup of the compressed version over the uncompressed baseline (bottom). 
    \label{figure:throughput:drift:sapphirerapids}
  } 
\end{figure}

%
%
The more threads we use, the lower the throughput for tiny problem sizes
(Figure~\ref{figure:throughput:drift:sapphirerapids}).
As we increase the
particle count, the throughput increases.
Any throughput curve for multiple threads eventually exceeds throughputs
stemming from fewer threads.
The break even point is roughly found around the L2 cache size.
Once the problem size exceeds the L3 cache, the performance of the
stream-like access pattern drops.
If we access data multiple times however, falling out of L3 plays no observable
role unless we put all threads to use.
In this latter case, we pay a minor penalty.
Overall, the throughput resembles a plateau.

%
%
The inverse scaling for very small particle numbers showcases that the OpenMP
parallelisation overhead is not negligible.
This penalty is smaller relative to the runtime if we reuse the
loaded data multiple times due to multiple kernel updates.
As soon as we increase the number of particles sufficiently,
adding more threads becomes advantageous, as
each core contributes its own L2 cache, while the L3 
seems to be well-designed to serve all of the cores at the same time.
If we access data repeatedly before we stream in the next chunk of work, we
obtain a higher throughput.
This is fundamentally a cache blocking effect.
If the main memory serves a stream-like data access pattern, we suffer from its
lower bandwidth relative to the L3 cache.
If the main memory however is only hit occasionally, as we mainly work on
in-cache data, we only pay the price for the latency, which really only
introduces a penalty for very high particle counts.

%
%
The impact of the compression is best studied through the relative speedup
compared to the uncompressed variant.
Even if the problem overall fits into the L2 or L3 cache, we still have to
stream it in from the main memory initially.
We pay for the memory access latency.
The compression increases
this latency logically, as each data access first has to unpack the data and
eventually pack it back.
We add another delay before we can actually compute or store.
Therefore, compression does not pay off for small problems and stream-like data
access where we already suffer from latency constraints.
It only pays off once we stress the memory interconnect due to very large total
problem sizes, and hence become bandwidth-bound.
Consequently, using compression hardly ever is beneficial for the task-based
access characteristics, where we are never bandwidth-bound.

%
%
%
We have to relativise our Hypothesis~\ref{demonstrator:hypothesis:bandwidth}:
Our computationally cheap kernels with linear compute characteristics are not
automatically bandwidth-bound, and we therefore do not uniformly benefit from
the compression.
Instead, compression only pays off robustly for very large problem size, which
is an insight that has to be taken into account by a software performance engineer.

\subsubsection{Kernels with linear computational complexity on Genoa}

\begin{figure}[htb]
  \begin{center}
    \includegraphics[width=0.46\textwidth]{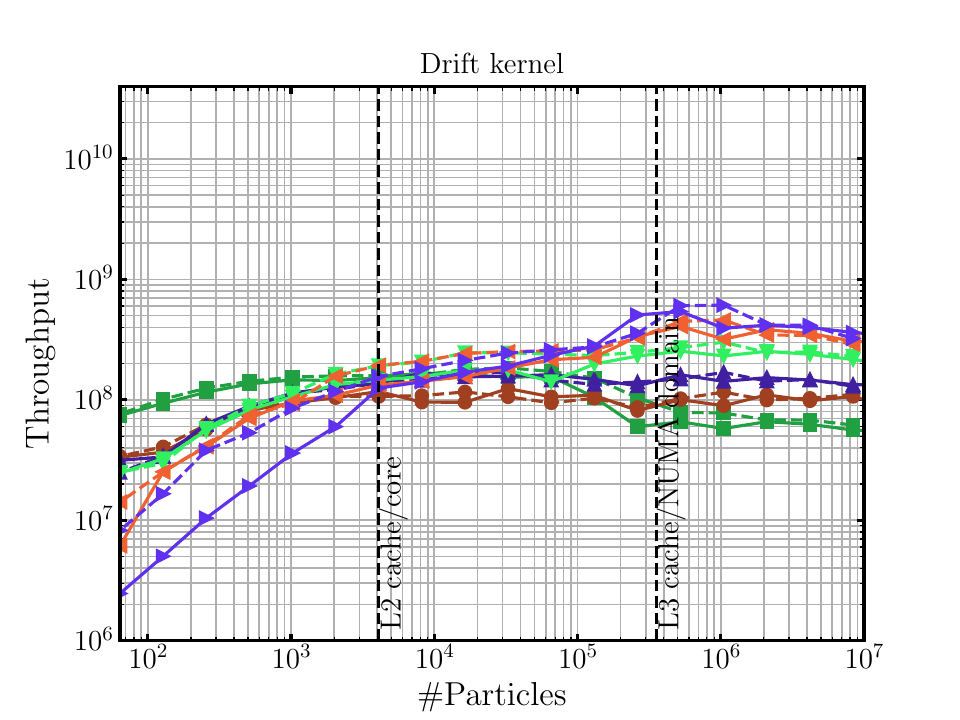}
    \includegraphics[width=0.46\textwidth]{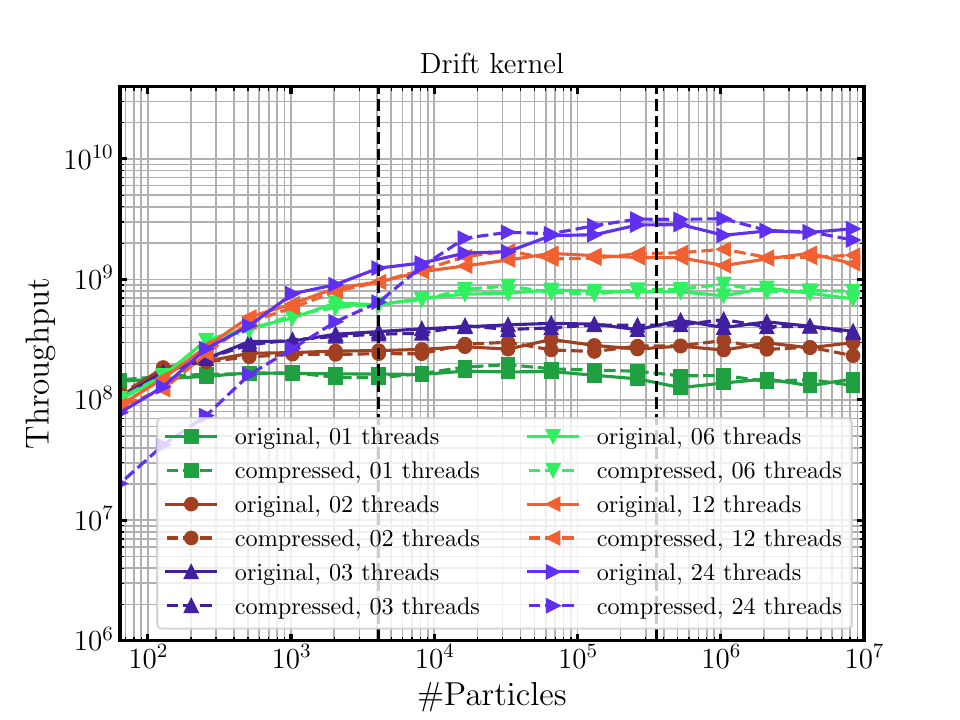}
  \end{center}
  \caption{
    Throughput on Genoa on a single NUMA domain for stream-like access (left) compared to a task-based realisation (right).
    \label{figure:throughput:drift:genoa-1-numa-domain}
  } 
\end{figure}

%
%
The throughput on a single NUMA domain of the Genoa chip is more difficult to
explain (Figure~\ref{figure:throughput:drift:genoa-1-numa-domain}).
For the task-like access patterns, we get qualitatively similar data to the
Sapphire Rapid without any penalty once we fall out of the L3 cache.
The stream access pattern is different.
As long as we stick to very small problem sizes, we again observe that more
threads yield smaller throughput initially, all throughputs increase as we
increase the particle count, and the many threads' measurements catch up with
the single threaded measurements.
Yet, the throughputs all stagnate once we work within the L3 cache.
They only fan out again for bigger setups when we fall out of L3.

%
%
Obviously, the memory controller is well-equipped to serve one NUMA domain.
We never run into a bandwidth issue which would make the 24 thread access
suffer.
This is also a direct consequence of the large L3.
However, the L3 cache in itself seems to become a bottleneck.
It struggles to serve all cores concurrently.
At the same time, once some of the memory accesses hit the main memory, we again
scale with the core count.
The reason for this fan-out behaviour has to be buried within the chip
architecture.
It almost seems as if L3 cache misses allow the L3 to serve more
cache hits while it waits for the main memory.
Lacking in-depth insight into the reasons for this behaviour, we
nevertheless can make statements on the impact
of the compression:

%
%
Due to the L3 bottleneck, the compression is beneficial for all tiny problems
fitting into the local L2 caches.
The L2 cache per core can host more particles in total as we use compression.
It hence reduces pressure on the L3.
We see fewer L3 hits.
This leads to improved throughput.
Compression also hits for problems that can be
hosted completely in L3.
This is a residual effect of the efficiency gains due to the big L2 caches.
In contrast, the compression does not help at all if we stream data all the time
from the main memory, and it has no positive impact on the throughput for the
task-like access pattern.
In many cases, it introduces some overhead. 
Indeed, the speedup strays between 0.8 and 1.2 anarchically.

\begin{figure}[htb]
  \begin{center}
    \includegraphics[width=0.46\textwidth]{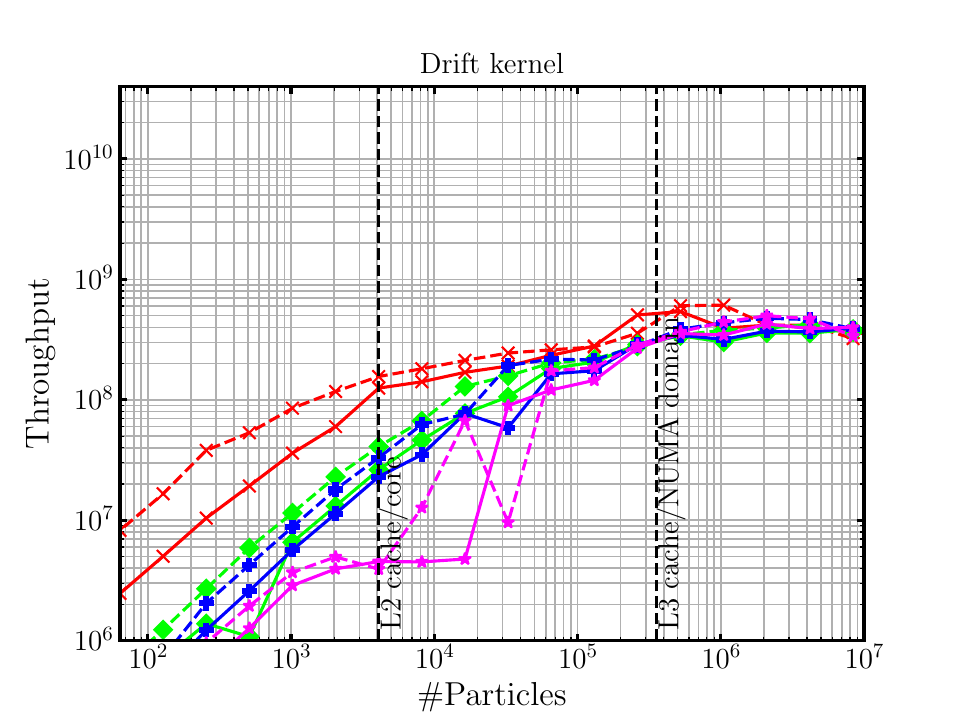}
    \includegraphics[width=0.46\textwidth]{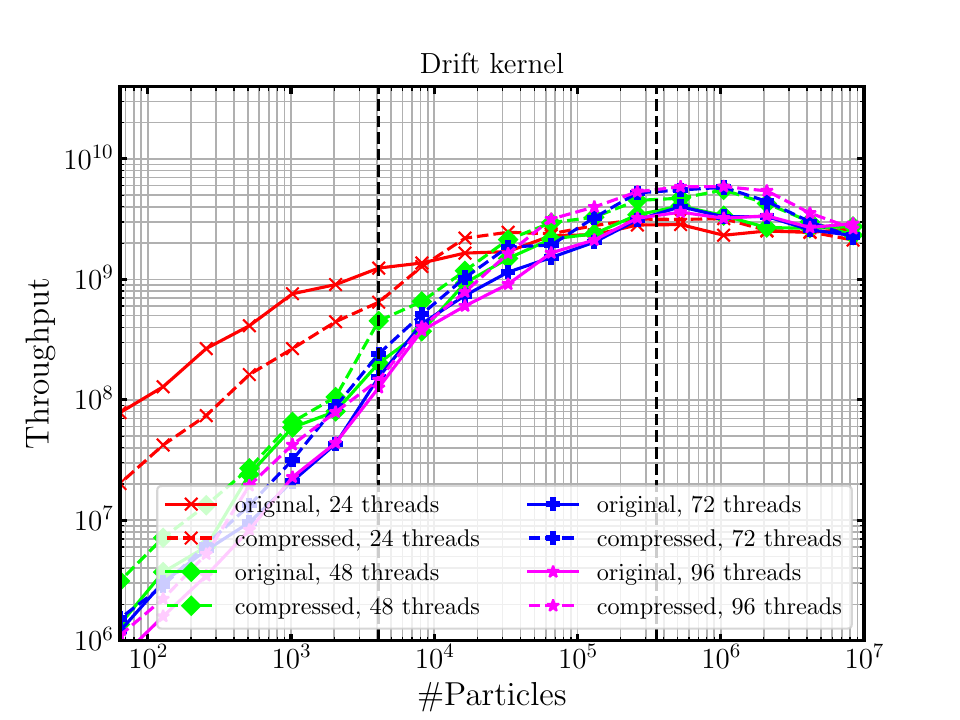}
    \includegraphics[width=0.46\textwidth]{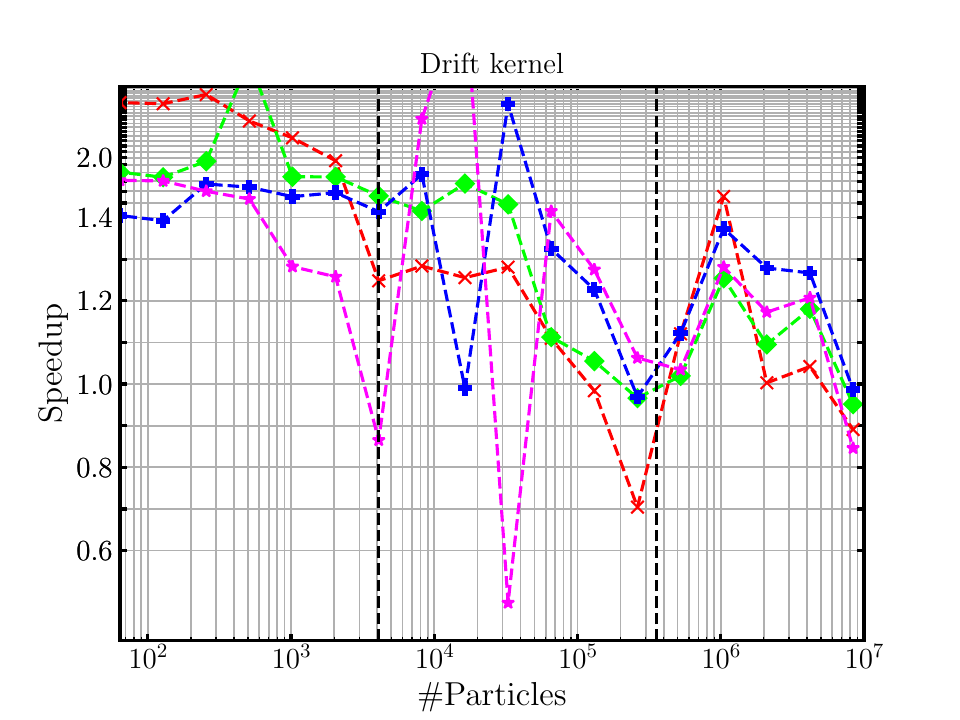}
    \includegraphics[width=0.46\textwidth]{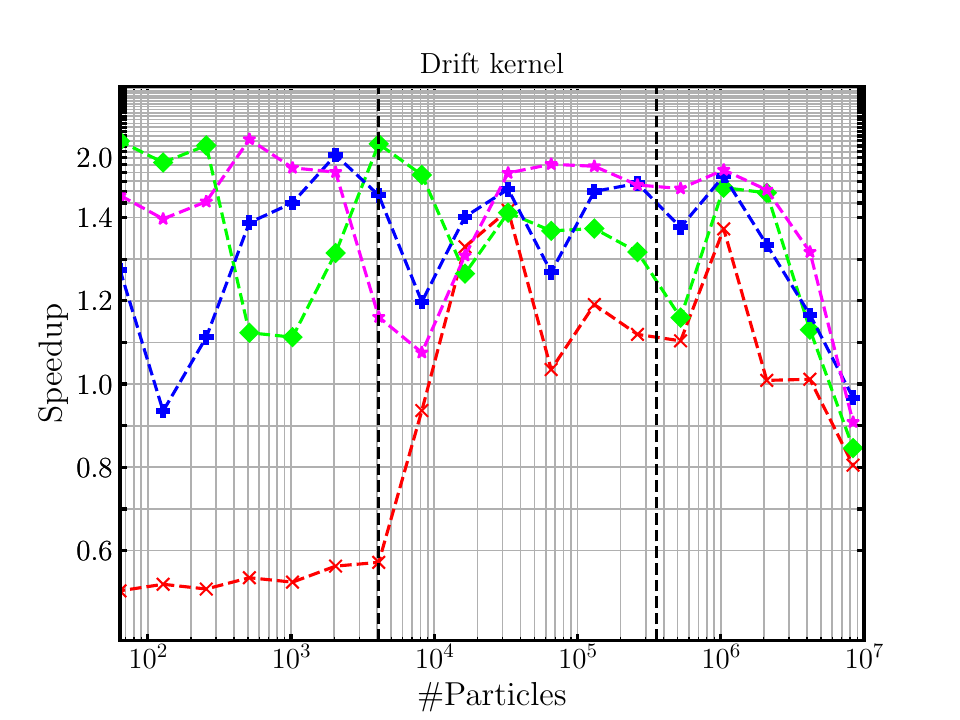}
  \end{center}
  \caption{
    Throughput on Genoa on a multiple NUMA domains for stream-like access
    (left) compared to a task-based realisation (right).
    We present throughputs (top) as well as the speedups obtained through
    compression relative to the uncompressed version (bottom).
    \label{figure:throughput:drift:genoa-many-numa-domains}
  } 
\end{figure}

%
%
Once we run our benchmark over multiple NUMA domains
(Figure~\ref{figure:throughput:drift:genoa-many-numa-domains}), the
throughput and the impact of the compression change character.
We observe that the stream-like access runs into a plateau now as well, while
the task-based access pattern yields a curve which drops once we leave the L3
cache.
The latter localises all data accesses.
Therefore, the drop has to be caused by the comparatively high latency of the
main memory accesses.
We observe an effect similar data to streaming behaviour on Sapphire Rapid, yet
this time for latency rather than bandwidth reasons.

%
%
As we stress the memory hierarchy on all three levels---compare the L2/L3
discussion for a single NUMA domain plus the latency observation
above---compression pays off robustly for all setups, unless we are completely
entering a streaming domain or hit the main memory too often.
Different to the Intel system, where compression pays off for the large particle
counts only, we benefit almost the other way round, i.e.~for the complementary
scenarios.
This is reasonable given the vast L3 cache size of the system, but also the
balancing between memory bandwidth of cores, the complex NUMA architecture and
the total core count.
In this context, it is important to note that all
advantages of compression disappear if we scatter the threads over
NUMA domains, i.e.~use for example 24 threads distributed over two NUMA domains
(not shown).
In such a case, we do not stress the L3 anymore sufficiently.

    

%
%
%
For the Genoa system,
Hypothesis~\ref{demonstrator:hypothesis:subsets-attributes-compute-intense} can
be generalised: 
Compression helps us to release pressure on any bottleneck further down the
memory hierarchy.
Also kernels with low computational load benefit.
Again, our main argument is avoiding latency effects rather than bandwidth
(Hypothesis~\ref{demonstrator:hypothesis:bandwidth}) as we work on a cache
architecture.
The discouraging observation for developers here is that the two systems, though
both x86-based, require completely different strategies regarding the
compression.
This is an important argument to ``outsource'' the actual compression decision
to annotations and a compiler, rather than to realise it manually within source
code.

\subsubsection{Kernels with quadratic computational complexity}

\begin{figure}[htb]
  \begin{center}
    \includegraphics[width=0.46\textwidth]{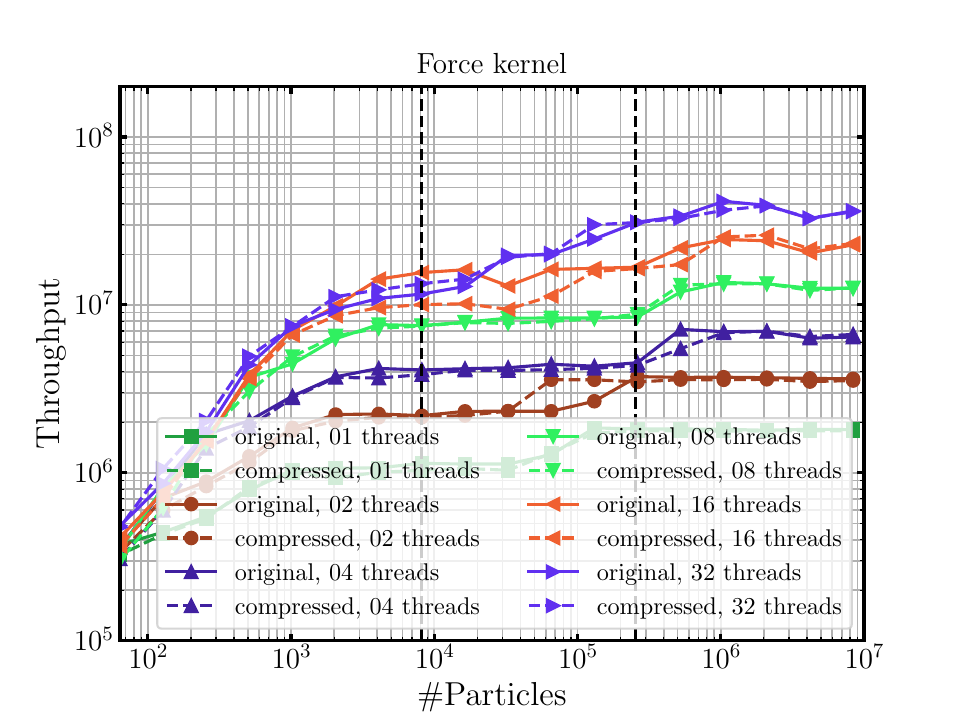}
    \includegraphics[width=0.46\textwidth]{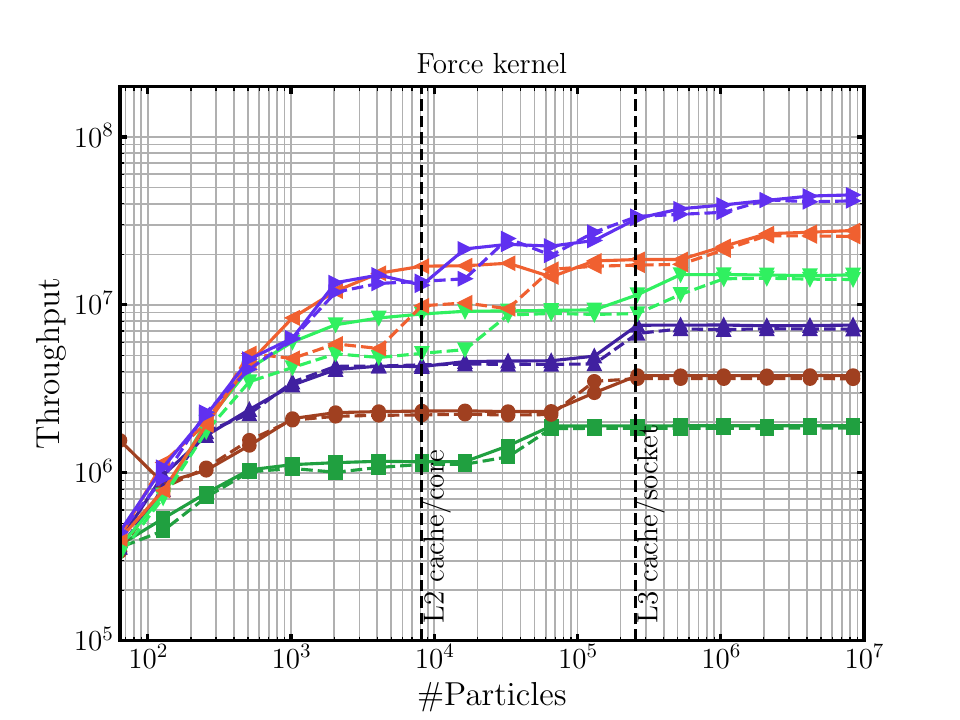}
    \includegraphics[width=0.46\textwidth]{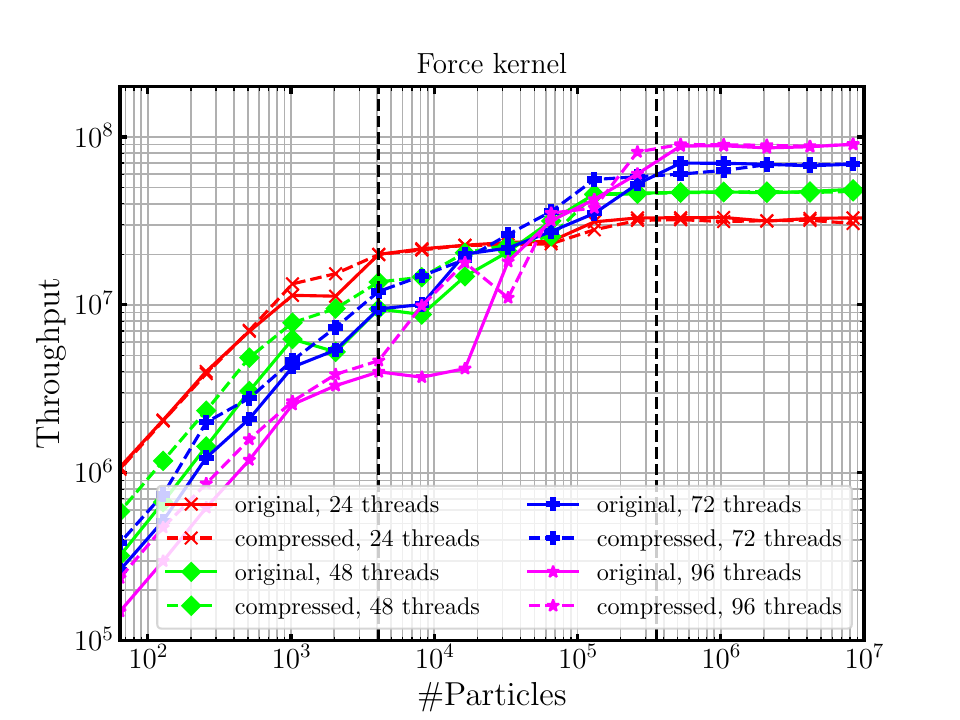}
    \includegraphics[width=0.46\textwidth]{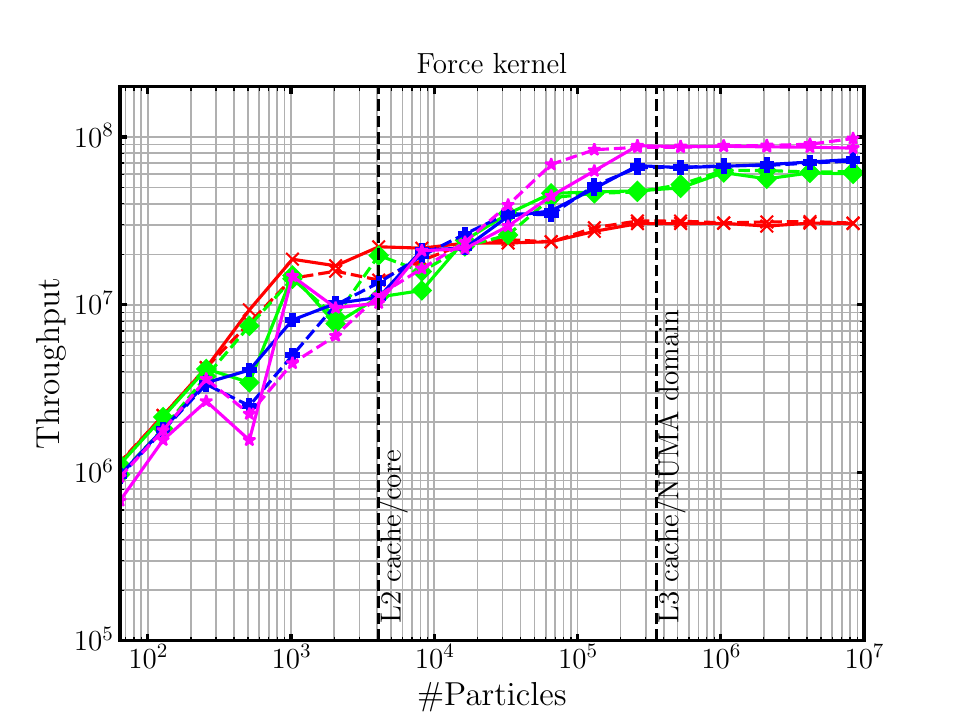}
  \end{center}
  \caption{
    Measurements for the force calculation step for various
    particle counts. 
    Throughput for stream-like access (left) vs.~a task-based
    realisation (right) on Sapphire Rapid (top) and the Genoa testbed (bottom).
    \label{figure:throughput:force}
  } 
\end{figure}

%
%
For the force calculation, both testbed architectures yield qualitatively
similar curves (Figure~\ref{figure:throughput:force}).
More threads pay off, but we hit a plateau once our problem becomes too big to
still fit into the L3 cache.
The data for the Genoa is slightly more ``noisy'', which we can attribute
towards its more complex NUMA architecture.
As we work with a computationally demanding
kernel, the idea of task-based realism plays no significant role for the
throughput:
This statement has to be read with care and only suggests that the force and
density calculations \replaced[id=R4]{per se}{per see} do not require use to break them down into tasks
and to ensure that we work with small data already in caches.
They are already compute-bound.

\begin{figure}[htb]
  \begin{center}
    \includegraphics[width=0.46\textwidth]{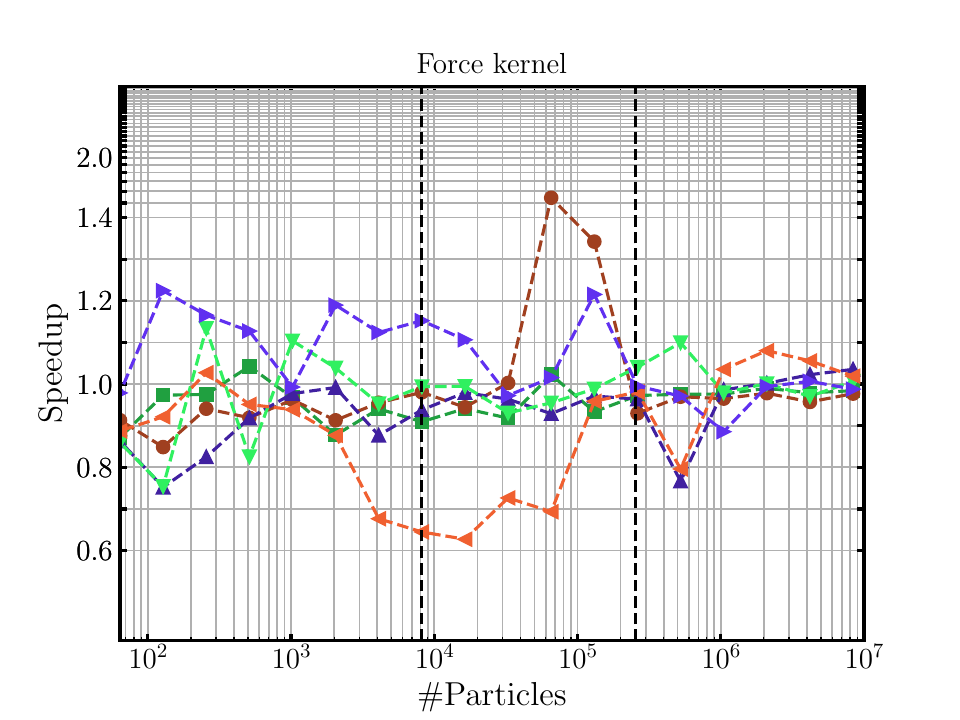}
    \includegraphics[width=0.46\textwidth]{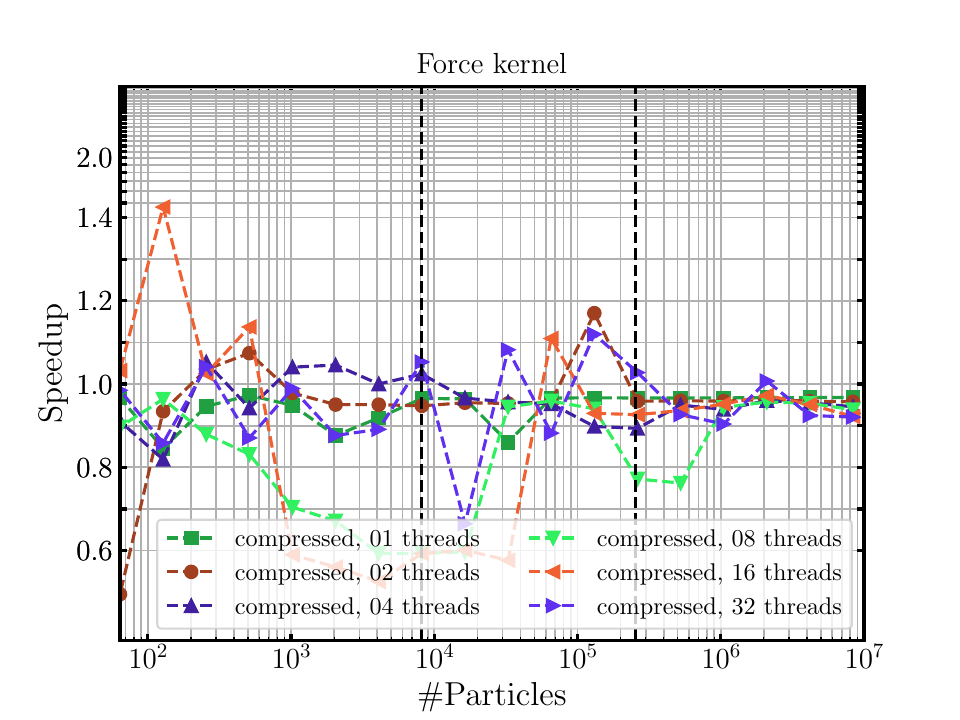}
    \includegraphics[width=0.46\textwidth]{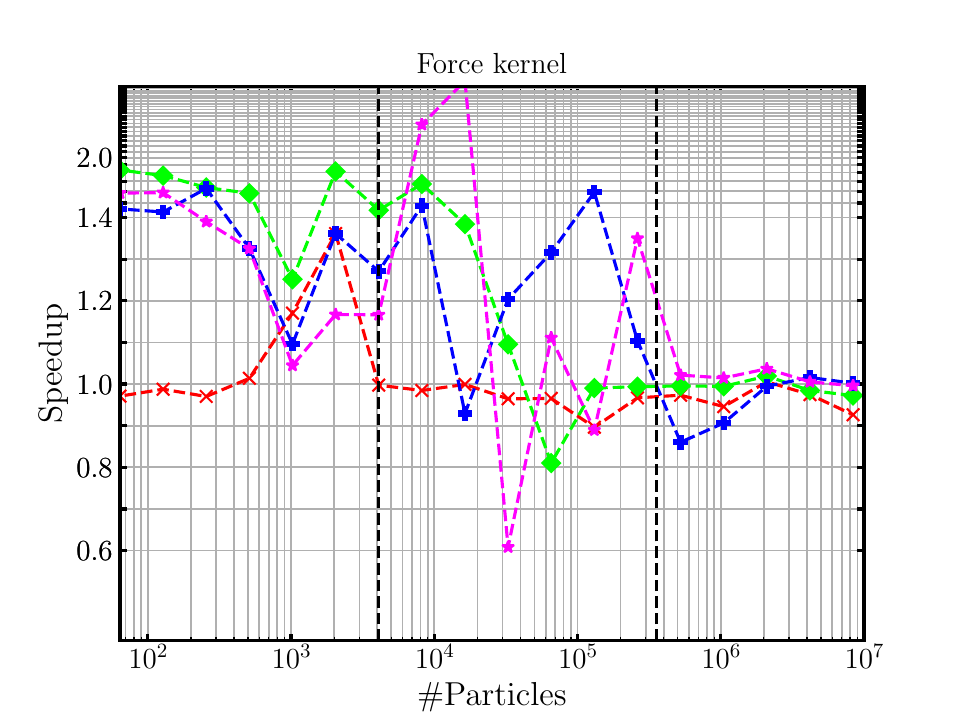}
    \includegraphics[width=0.46\textwidth]{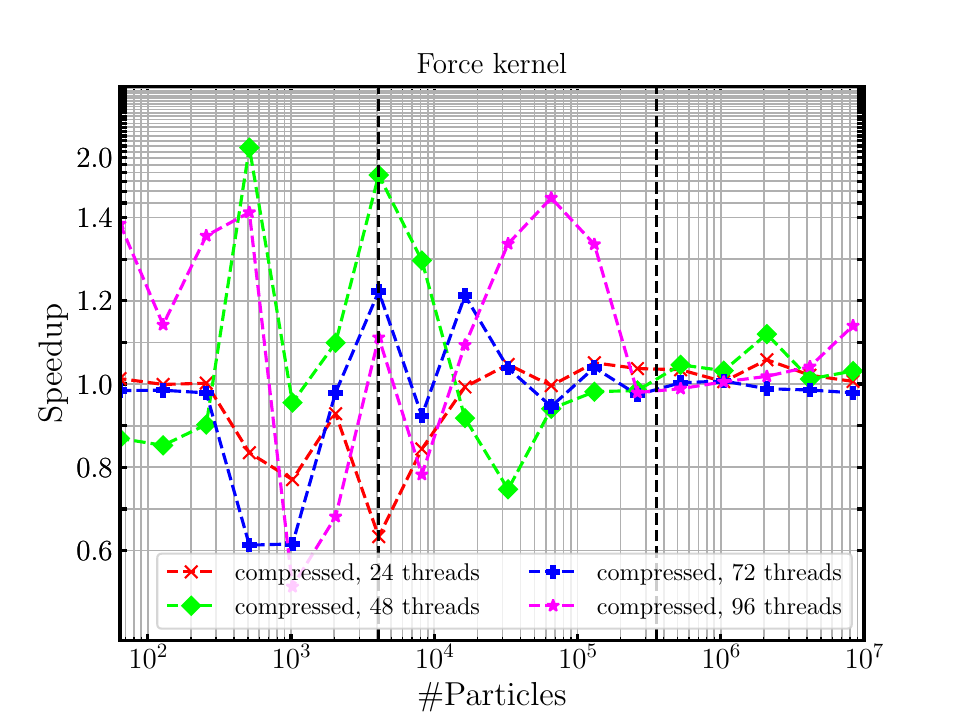}
  \end{center}
  \caption{
    Speedup data for the measurements from Figure~\ref{figure:throughput:force}.
    \label{figure:speedup:force}
  } 
\end{figure}

%
%
The speedup curves for the Sapphire Rapid are rather erratic
(Figure~\ref{figure:speedup:force}) and do not allow us to make robust
statements if or when compression pays off or is detrimental.
It seems that it is reasonably advantageous for the streaming-like kernel as
long as we work within the L2/L3 caches, but if and only if we employ all
threads.
We may assume that this is again a cache effect.

On the Genoa system, the curve peaks are more pronounced, but they also are
``less deterministic'':
A speedup for more than a factor of two can be obtained for some configurations,
while the same thread choice can lead to a performance loss of up to 40\% for a
slightly different particle count.

%
%
We hypothesise in
Hypothesis~\ref{demonstrator:hypothesis:subsets-attributes-compute-intense} that
the ability to hold more data in close caches is beneficial for compute-intense
compute kernels.
However, the data make it clear that the overhead of the floating-point
\replaced[id=R2]{conversions}{conversations} is sometimes too high a price to pay.
For some setups, we benefit from better cache utilisation, for others we pay too
much algorithmic latency (conversion) overhead.
It is a hit or miss.

For this particular type of kernel, a manual conversation from a packed
representation prior to the kernel invocation hence seems to be a natural
modification of the code, making the implementation robust without
giving up on the performance advantages for the cheaper compute kernels.
In this context, developers might consider to convert into SoA, as they have to
copy anyway.

\subsubsection{Resume}

Reduced floating point precision is a widely applied technique in machine
learning and successfully used in large-scale linear algebra setups
\cite{Abdulah:2022:GeostatisticalModeling,Cao:2022:GordonBell,Doucet:2019:MixedPrecision,Ltaief:2023:GordonBell}.
In the context of a complex algorithmic code such as SPH, it however is no
silver bullet. 
Its pros and cons have to be evaluated carefully, and our data suggest that the
major impact of reduced precision results---for compute-intense
applications---from the improved vector efficiency and not the sole bandwidth
savings.
Our compression approach targeting the memory footprint is 
clearly more relevant for compute kernels with low arithmetic intensity.
Otherwise, it has to be used with care.

We reiterate our key statement that our
technique allows for one further optimisation which is not assessed here:
As we bring down the global memory footprint, we can squeeze larger problems
onto a node.
Hence, we can weakly scale to a larger logical problem size, which is typically
advantageous for the parallel efficiency and unfolds it full impact notably for
compute-intense compute kernels.
Our throughputs all stagnate as we leave the L3 cache.
However, at least on the Sapphire Rapids, the reason for this
stagnation is different for the kernels with linear complexity vs.~the kernels
with quadratic cost: 
One suffers from latency effects, one suffers from bandwidth constraints.
If we succeed in running kernels with different characteristics at the same
time---for example through task parallelism---we can construct a code base that
continues to scale in the particle count beyond the L3 threshold.

\subsection{\added[id=R21] { Developer productivity }}

\added[id=R21]{
 We conclude this section by anecdotally quantifying the advantages of the proposed syntax and semantics relative to manual and library-based approaches.
 Our tests compare our proposed C++ language extensions against a purely hand-written implementation, as well as two C++ template-based approaches: 
 Boost.Multiprecision and FloatX~\cite{Flegar2019FloatX}. 
 Throughout the evaluation, we focus exclusively on the compression of scalar floating point data and ignore the packing and reordering over integers, booleans or multiple floating point values,
 as these features are not directly supported by the comparison libraries.
}

\begin{table}[htb]
 \caption{
   \added[id=R21]{
   Quantitative comparison of the proposed attribute-based mechanism with manual and library-based approaches. 
   Boost.MP represents a variant based up on Boost.Multiprecision.	
   }
  \label{table:developer_productivity_comparison}
 }
 \centering
 \begin{tabular}{lcccc}
\hline
\textbf{Metric} &
\textbf{This work} &
\textbf{Manual} &
\textbf{Boost.MP} &
\textbf{FloatX} \\
\hline
Dev effort &
minimal &
very high &
moderate &
moderate \\
Extra LOC &
1 &
$\approx$180 &
1-10 &
1-10 \\
Machine instr./Op &
4 &
4 &
$\approx$ 20 + lib calls &
$\approx$ 100 \\
Branches &
none &
none &
none outside library code &
multiple \\
GPU safe &
yes &
yes &
no &
partial \\
\hline
\end{tabular}

\end{table}

\added[id=R21]{
 Our attribute-based approach requires only one single annotation per compressed field and no auxiliary code. 
 The compiler injects just four additional x86-64 instructions per arithmetic operation to map fields onto their native C++ data type and back, without introducing any branching or library calls. 
 Since the ejected machine code consists solely of arithmetic and bitwise operations, it remains fully compatible with GPU offloading or can be transcribed onto any other ABI (Table~\ref{table:developer_productivity_comparison}).
}

\added[id=R21]{
 A manual implementation of the same functionality requires substantial effort. 
 Even under our simplifications (no fusion of floating-point and packed integer data), handling a single packed field demands around 180 lines of glue code to encode and decode values, maintain const-correctness, handle alignment, and integrate with existing operator syntax. 
 For multiple attributes, functionality can be outsourced into utility routines, but the effort remains high.
 When implemented correctly, the resulting machine instruction count matches that of our compiler-based approach, but correctness is entirely the responsibility of the developer.
 This means that the code remains fragile or high maintenance under refactoring.
}

\added[id=R21]{
 Both library-based approaches incur even higher overheads and lack key capabilities. 
 \linebreak
 Boost.Multiprecision is implemented as a wrapper around MFPR~\cite{Fousse2007MPFRAM}, and as such primarily targets extended---higher than 64-bit---precision. 
 Its bespoke floating-precision data type is always 32 byte in size regardless of the chosen target precision, which indicates possible internal allocations under the hood. 
 Arithmetic operations expand to approximately 20 x86-64 instructions, including 3 MPFR library function calls even under the highest levels of optimisation and LTO.
 Without bespoke Boost support, this approach is unsuitable for GPUs.
}

\added[id=R21]{
 FloatX yields a fairer comparison, as it is a dependency-free template-only alternative to Boost.Multiprecision and supports reduced precision computation and storage.
 It leaves the choice of the native storage format to the user, decoupling it from the exponent and mantissa sizes. 
 Each arithmetic operation yields over 100 x86-64 operations even under the highest levels of compiler optimisation
 and results in heavy branching since the library implements the IEEE-754 corner cases (rounding, sub-normals, NaN handling) in software. 
 Since FloatX does not rely on external library calls, it is, in principle, suited for GPUs.
 Yet, the branching might at least cause SIMT inefficiencies. 
}

\added[id=R21]{
 All approaches suffer from the fact that the resulting reduced precision code becomes ABI incompatible with external libraries.
 However, we note that our compiler extension automatically converts packed data into native formats if they are passed as scalars into library functions.
 Similar functionality can be realised with template solutions through explicit conversion routines. 
}

\added[id=R21]{
 Along the five studied dimensions of interest, our approach is the only one that requires negligible user involvement and delivers the lowest machine instruction overhead.
 The only potential situation where the template-based variants are superior are situations where we combine particular reduced precision variables with each other arithmetically.
 We may assume that this can sometimes be realised directly, without converting into native C++ data types and back.
 However, it is not clear if the present libraries offer such a feature.    
}

\section{Conclusion}
\label{section:conclusion}

Many scientific codes suffer from large memory footprints.
Our annotations of the C++ language allow developers to specify and fine-tune
the information density within a struct by altering the 
accuracy of floating-point numbers and ranges for
integers as well as implicitly removing internal padding and alignment, and
our implementation of these augmentations within LLVM uses the additional
intelligence provided by the developers to reduce the memory
footprint.
Along the same lines, 
we offer a mechanism to develop MPI-based code more
efficiently---at the moment, any change of data layout induces a tedious
alteration of MPI datatypes.
This extension enables developers to exchange only those
\replaced[id=R1]{variables}{attributes} of a struct through MPI which actually change and makes the MPI data
types benefit from our compression technologies, too.

Our experiments with an SPH code show that the extensions 
help to write more memory-modest code.
This allows users to run bigger simulations on machines where memory is limited,
i.e.~to challenge classic strong scaling plateaus or performance degradations.
With the trend to integrate faster yet overall limited High Bandwidth Memory
into chips, this opportunity remains important even though bandwidth penalties might decrease, as we expect the average memory per core to shrink or stagnate.
\added[id=R3]{
 It remains future work to apply our ideas to a wider range of application codes.
}

The correlation of memory modesty with performance \deleted[id=Ours]{however} is a nuanced,
multifaceted one:
Our data suggests that cache-optimised and bandwidth-constrained codes benefit from the
compression most, as we now can squeeze more data into existing caches close to the
chip or transfer more logical data per cache line.
Other codes will have to pay overheads and penalties for the savings
in memory.
Performance engineering hence is not automated or made simpler with our 
approach, but we add an additional level of complexity.

In this context, we
consider the seamless integration into ISO C++ to be pivotal for the realisation
of our intention:
Annotations can be ignored without breaking a code's semantics, 
simple assignments allow the programmer to convert packed and compressed
data into native C++ datatypes which fit directly to machine
instruction sets, and the realisation as additional
compiler pass means that any compiler-internal optimisation further down
the translation pipeline \replaced[id=Ours]{continue to unfold their potential}{remains relevant}.
This way, memory optimisation \added[id=Ours]{also} can be implemented incrementally, and code remains
\deleted[id=Ours]{standard-conform aka} portable \replaced[id=Ours]{across}{for} different machines.
\added[id=R4]{
 Our work focuses on structs as fundamental modelling entity of computing codes. 
 The proposed techniques however might be particularly useful for ``free'' variables and globals as well,
 although it is not clear how intricate such a generalisation is on the compiler side.
}

\added[id=R21]{
 While our MPI extension is key to developer productivity in HPC, and provides opportunities to address bandwidth challenges in scientific codes,
 its discussion highlights a fundamental flaw of the present proposal:
 As we allow the compiler to reorder and compactify memory, code that is written against APIs relying on certain memory layouts runs risk to break.
 We invalidate our own intention to make all annotations optional.
 It is easy to introduce thin compatibility layers ensuring that all external libraries continue to work seamlessly no matter if core compute code is subject to annotation-guided memory rearrangements or not. 
 However, we then miss out on their added value.
 In the context of certain challenges such as I/O and checkpointing, this added value is potentially significant and deserves further attention.
}

\replaced[id=R21]{Beyond that, many}{Many} open questions remain:
Future codes will run on strongly heterogeneous architectures, where
heterogeneous means both heterogeneous memory as well as heterogeneous
compute facilities such as CPU--GPU combinations or special-purpose compute
entities such as large AVX ``subprocessors'' or their matrix extensions (tensor
cores).
While our code transformations reduce the memory footprint and help to write
code with reduced bandwidth needs, they introduce additional computational
work to convert the data representations into each other, and they do not
exploit the reduced precision in any way for the actual computation.
It not clear how work has to be distributed within heterogeneous systems:
Should the conversions be deployed to a GPU if the computations run on the
accelerator, could they be deployed to external smart compute units or
networks once the data is expelled from the local caches,
are the transformations en-bloc operations on all input data that precede the invocation
of an offloaded compute kernel, or can they be triggered lazily on a stream while
a compute unit already starts to process data, can we utilise AVX
co-processors, and so forth?
Further to that, it seems to appealing to use the knowledge about reduced
precision to alter the underlying compute data type of variables:
If the number of significant bits in the annotation is smaller than a
\texttt{float}'s bits, it might seem to be convenient to use 
\texttt{float} as baseline type even though the variable might be modelled as
\texttt{double}.
Such considerations have to be subject of future work.

A second scientific challenge arises from flexible floating-point
storage formats.
Different to lossy compression that is applied only to data prior to
post-processing (\cite{Lindstrom:2014:ZFP}),
our code annotations work in-situ, i.e.~on data potentially used by follow-up
calculations.
They are thus an excellent tool to study mixed-precision algorithms, and to
introduce support for new reduced precision arithmetics in the hardware. 
However, our precision choices are static.
In many applications, the actual information density within floating-point data
changes over time
\cite{Weinzierl:18:AlgebraicGeometric,Eckhardt:15:SPH,Murray:2020:LazyIntegration},
i.e.~the number of significant bits has to be chosen dynamically.
It is an open research question how our extensions can be generalised to support
flexible precision choices.
\added[id=R3]{
 We notably emphasise that our work focuses exclusively on the streamlining of the programming with different storage precisions.
 It is an open question how this storage-centric approach teams up with genuine mixed-precision algorithms, i.e.~codes that mix precisions in their computations.
}

Finally, we assume that our annotations yield a more significant speedup once
they are combined with loop transformations:
If floating point data are stored in reduced, user-defined precisions, our
current compiler realisation wraps each data access into pack and unpack
routines.
This might be convenient for single access loops.
It is likely a poor realisation whenever we work with loops
accessing \replaced[id=R1]{particles' instance variables}{multiple particles' attributes} multiple times.
Here, we may assume that it is advantageous to unpack data once in a preamble to
the loop and to convert it back once the loop has terminated.
\replaced[id=R1]{
 At the moment, such explicit unpacking--packing can be realised manually in source code through native C++.
 However, this is inconvenient and might better be deployed to the compiler, too, in future releases.
 The arising
}{
Such a} prologue-epilogue transformation should be composable with on-the-fly
AoS-to-SoA \replaced[id=R2]{conversions}{conversations} \cite{Radtke:2025:AoStoSoA} and facilitate the usage of
fully vectorised instruction streams including coalesced loads and stores.
A price to pay is an increased temporary memory footprint.

The elephant in the room is the question to
which degree we can automate all of \replaced[id=Ours]{decision making behind the}{these} transformations:
Are there robust, reliable heuristics within a compiler that can guide the
selection of a proper conversion realisation and inform the translation when
in the code to (optimistically) convert data representations?

\section*{\added[id=Ours]{Acknowledgements}}

Cristian's, Mladen's and Tobias' research have been supported by EPSRC's
ExCALIBUR programme through its cross-cutting project EX20-9 \textit{Exposing Parallelism: Task Parallelism}
(Grant ESA 10 CDEL) and the DDWG project \textit{PAX--HPC} (Grant EP/W026775/1).
The latter supplemented our demonstrator code base.
Aspects of the present research arise from Tobias' work for ExCALIBUR's 
\textit{An ExCALIBUR Multigrid Solver Toolbox for ExaHyPE} (EP/X019497/1) made
by EPSRC.
Pawel's PhD studentship is partially supported by Intel's Academic Centre of
Excellence at Durham University.

This work has made use of the Hamilton HPC Service of Durham University as well
as the experimental test nodes installed within  
the DiRAC@Durham facility managed by the Institute for Computational Cosmology
on behalf of the STFC DiRAC HPC Facility
(\href{www.dirac.ac.uk}{www.dirac.ac.uk}). 
The latter equipment was funded by BEIS capital funding via STFC capital grants
ST/K00042X/1, ST/P002293/1, ST/R002371/1 and ST/S002502/1, Durham University and STFC operations grant ST/R000832/1. DiRAC is part of the National e-Infrastructure.

\bibliographystyle{ACM-Reference-Format}
\bibliography{references}

\appendix
\section{Installation of modified LLVM variant}

\replaced[id=R1]{Our compiler extensions are released as LLVM patch, and we tested all compiler modifications with Fedora 42. The patch}{
The release of the compiler extension as LLVM patch is currently under
preparation.
A preview as used for the experiments here} is available from
\href{https://github.com/pradt2/llvm-project.git}{https://github.com/pradt2/llvm-project.git}.

\deleted[id=Ours]{
We have built this compiler version with both Ubuntu 22.04~LTS and Fedora 37,
relying only on relatively mature and old third-party software
(Table~\ref{fig:software-deps}).}

If the use of any of the new attributes leads to a compilation error, a common
starting point for troubleshooting is to inspect the rewritten source code. To
see the rewritten code, add \newline \texttt{-fpostprocessing-output-dump} to the compilation flags. 
The flag causes the post-processed source code be written to the standard
output.


\section{Reproducibility of experimental data}

All experimental data has been produced with the Swift 2 code.
It is shipped as part of the Peano 4 framework \cite{Weinzierl:2019:Peano}
available from \url{https://gitlab.lrz.de/hpcsoftware/Peano.git}.
The repository provides a CMake and autotools build system. The autotools
environment is set up via 

 {\footnotesize
\lstset{language=C}
\begin{lstlisting}
libtoolize; aclocal; autoconf; autoheader
cp src/config.h.in .; automake --add-missing
./configure --with-multithreading=omp --enable-particles --enable-swift \
  --enable-loadbalancing --with-mpi=mpiicpc \  
  CXXFLAGS="-O3 --std=c++20 -fopenmp -g"
make
\end{lstlisting}
 }

\noindent
The \texttt{make} yields all libraries we need for our experiments.
Besides the core libraries, we recommend to create all online documentation
through Doxygen (\texttt{doxygen documentation/Doxyfile}) which also is available from
\url{https://hpcsoftware.pages.gitlab.lrz.de/Peano}.

Each experiment is located within a repository subdirectory and contains a
readme file which automatically is extracted into HTML documentation through
Doxygen.
Every single benchmark is produced through a Python script.
It automatically picks up the compiler settings passed into autotools or CMake,
respectively, and accepts arguments to configure the actual benchmark. 
The benchmark documentation plus the present experimental descriptions provide
information on arguments used.
Eventually, the Python scripts produce a stand-alone executable.

All benchmarks produce human-readable text files as outputs.
Our repository ships Matplotlib scripts to convert them into figures, and
the benchmarks' descriptions provide further information on these postprocessing
scripts. 
The plots in the paper differ from Peano's vanilla benchmarking plots only by
different layout choices and augmented annotations such as cache sizes.

\begin{itemize}
  \item The \emph{grid experiments} (Section~\ref{sec:orbit_experiment}) are
  produces through the benchmarks \linebreak within
  \texttt{benchmarks/swift2/planet-orbit}.
  \item The \emph{mantissa truncation impact}
  (Section~\ref{section:results:solution-accuracy}) is studied through the Noh
  2d benchmark as available through
  \texttt{benchmarks/swift2/hydro/noh-implosion-text}.
  \item The \emph{MPI test case} (Section~\ref{sec:results:mpi}) is a simple
  ping-pong test using the data structures from the Noh 2d benchmark. Its
  integration into the test suite is work in progress.
  \item The \emph{scalability} data (Section~\ref{section:results:scalability})
  are obtained through the benchmarks \linebreak in
  \texttt{benchmarks/swift2/hydro/kernel-throughput}.
\end{itemize}

\section{Comprehensive scalability data}

\begin{figure}[htb]
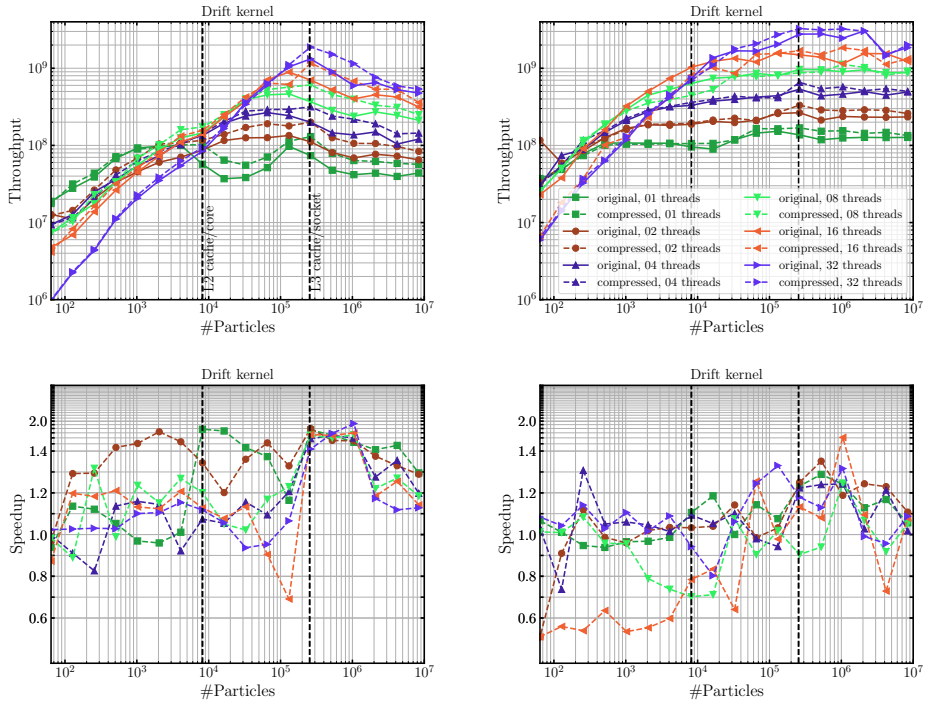

  \begin{center}
    \includegraphics[width=0.46\textwidth]{./experiments/kernel-throughput/sapphirerapids-single-touch/throughput-drift-without-legend.pdf}
    \includegraphics[width=0.46\textwidth]{./experiments/kernel-throughput/sapphirerapids-multitouch/throughput-drift-with-legend.pdf}
    \includegraphics[width=0.46\textwidth]{./experiments/kernel-throughput/sapphirerapids-single-touch/speedup-drift-without-legend.pdf}
    \includegraphics[width=0.46\textwidth]{./experiments/kernel-throughput/sapphirerapids-multitouch/speedup-drift-without-legend.pdf}
  \end{center}
  \caption{
    Measurements for the drift kernel.
    Throughput (top) and speedup relative to uncompressed baseline
    version (bottom) on the Sapphire Rapid testbed for
    stream-like access (left) and task-based access characteristics (right).
    \label{figure:appendix:scalability:drift:sapphirerapids}
  } 
\end{figure}

\begin{figure}[htb]
  \begin{center}
    \includegraphics[width=0.46\textwidth]{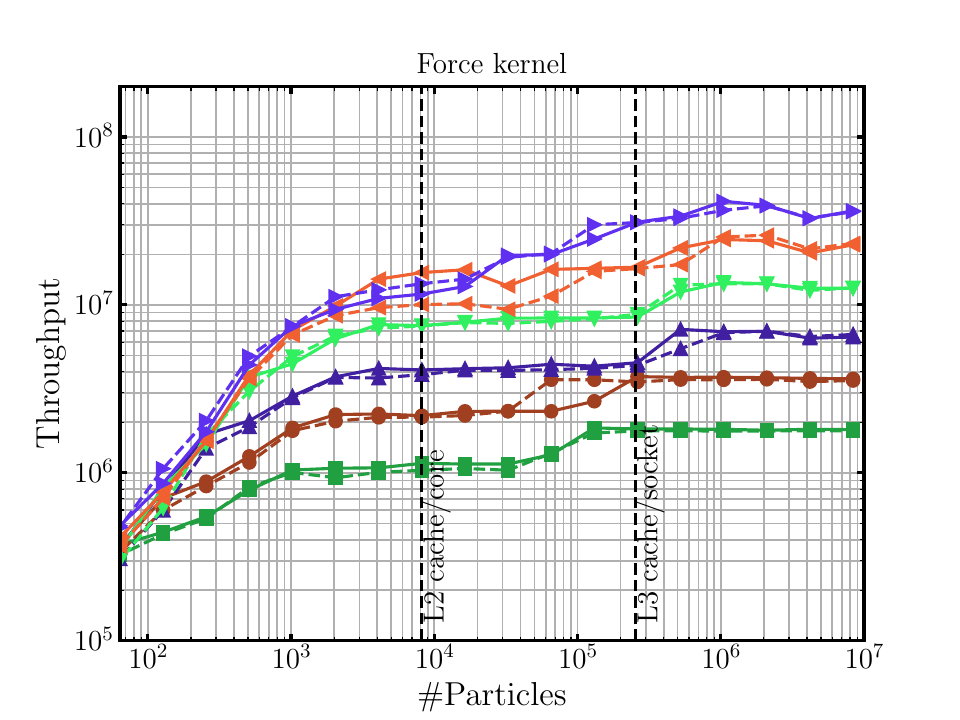}
    \includegraphics[width=0.46\textwidth]{./experiments/kernel-throughput/sapphirerapids-multitouch/throughput-force-without-legend.pdf}
    \includegraphics[width=0.46\textwidth]{./experiments/kernel-throughput/sapphirerapids-single-touch/speedup-force-without-legend.pdf}
    \includegraphics[width=0.46\textwidth]{./experiments/kernel-throughput/sapphirerapids-multitouch/speedup-force-with-legend.pdf}
  \end{center}
  \caption{
    Measurements for the force kernel kernel.
    Throughput (top) and speedup relative to uncompressed baseline
    version (bottom) on the Sapphire Rapids testbed for
    stream-like access (left) and task-based access characteristics (right).
    \label{figure:appendix:scalability:force:sapphirerapids}
  } 
\end{figure}

\begin{figure}[htb]
  \begin{center}
    \includegraphics[width=0.46\textwidth]{./experiments/kernel-throughput/genoa-1numa-single-touch/throughput-drift-without-legend.pdf}
    \includegraphics[width=0.46\textwidth]{./experiments/kernel-throughput/genoa-1numa-multitouch/throughput-drift-with-legend.pdf}
    \includegraphics[width=0.46\textwidth]{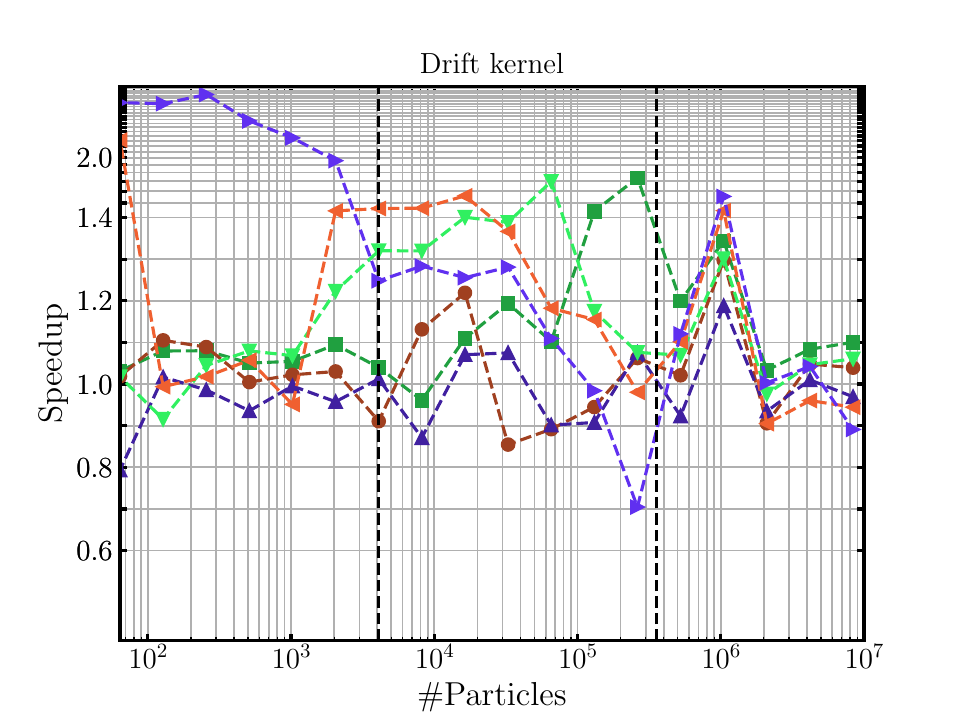}
    \includegraphics[width=0.46\textwidth]{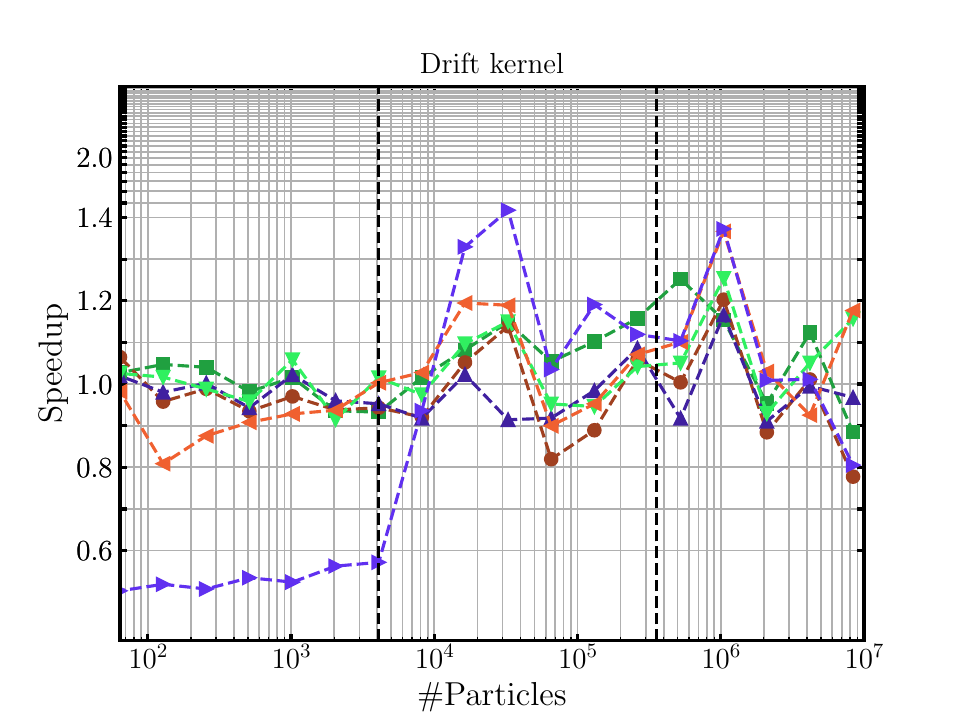}
  \end{center}
  \caption{
    Measurements for the drift kernel.
    Throughput (top) and speedup relative to uncompressed baseline
    version (bottom) on the Genoa testbed for stream-like access (left) and
    task-based access characteristics (right).
    We use one NUMA domain.
    \label{figure:appendix:scalability:drift:genoa-1numa}
  } 
\end{figure}

\begin{figure}[htb]
  \begin{center}
    \includegraphics[width=0.46\textwidth]{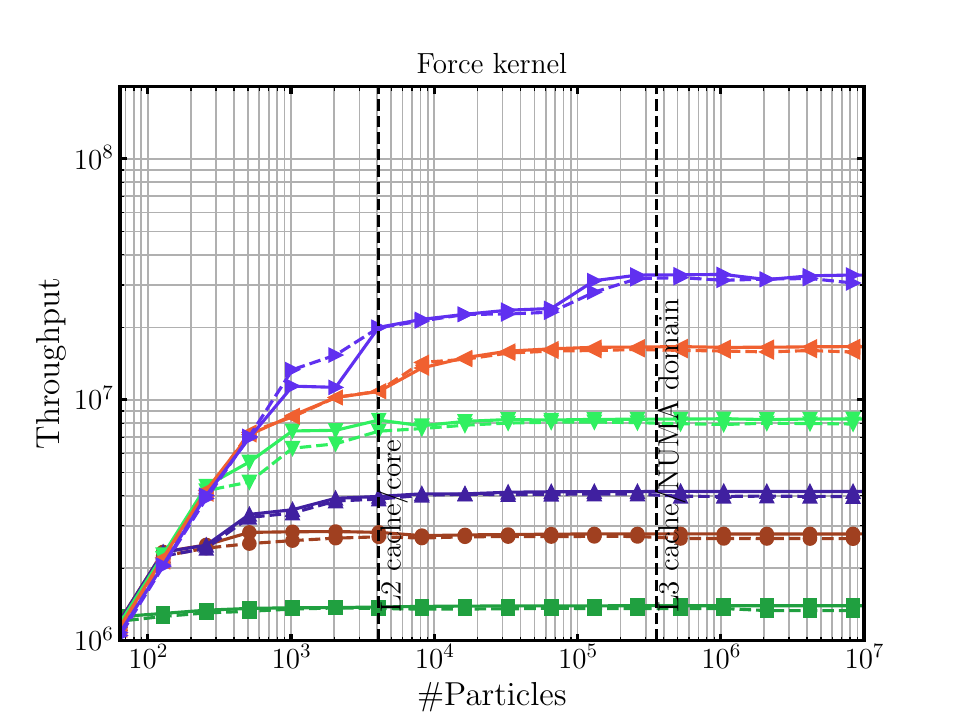}
    \includegraphics[width=0.46\textwidth]{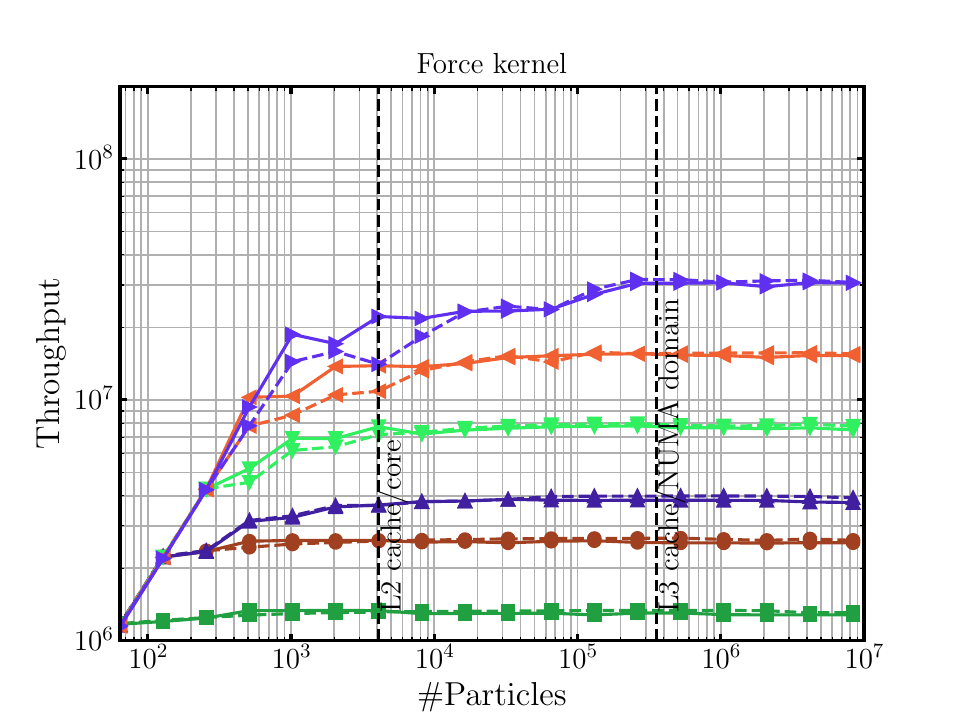}
    \includegraphics[width=0.46\textwidth]{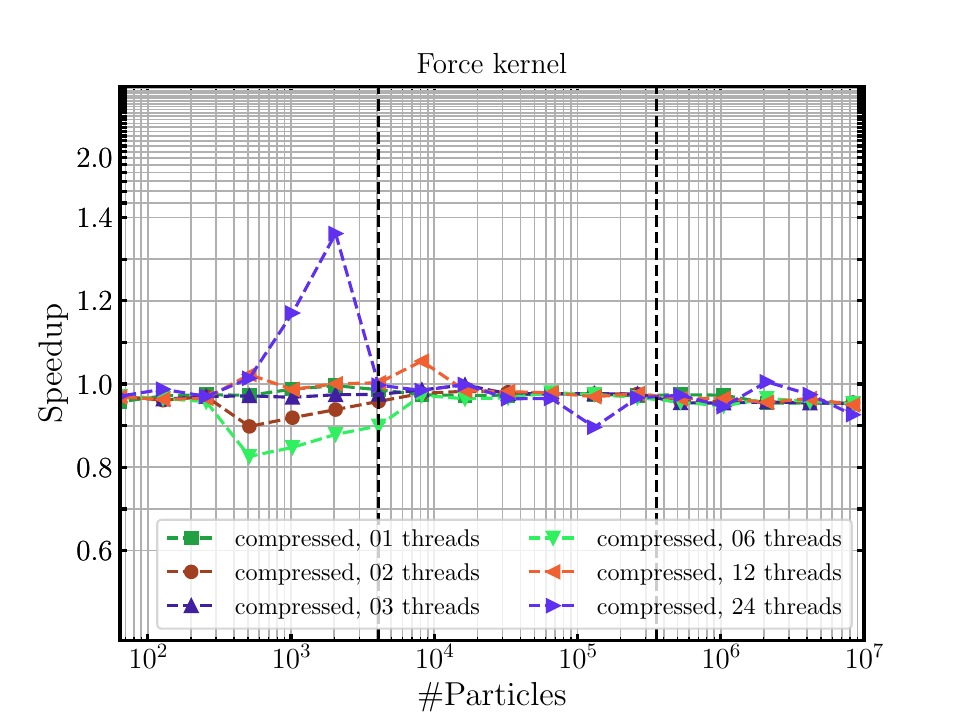}
    \includegraphics[width=0.46\textwidth]{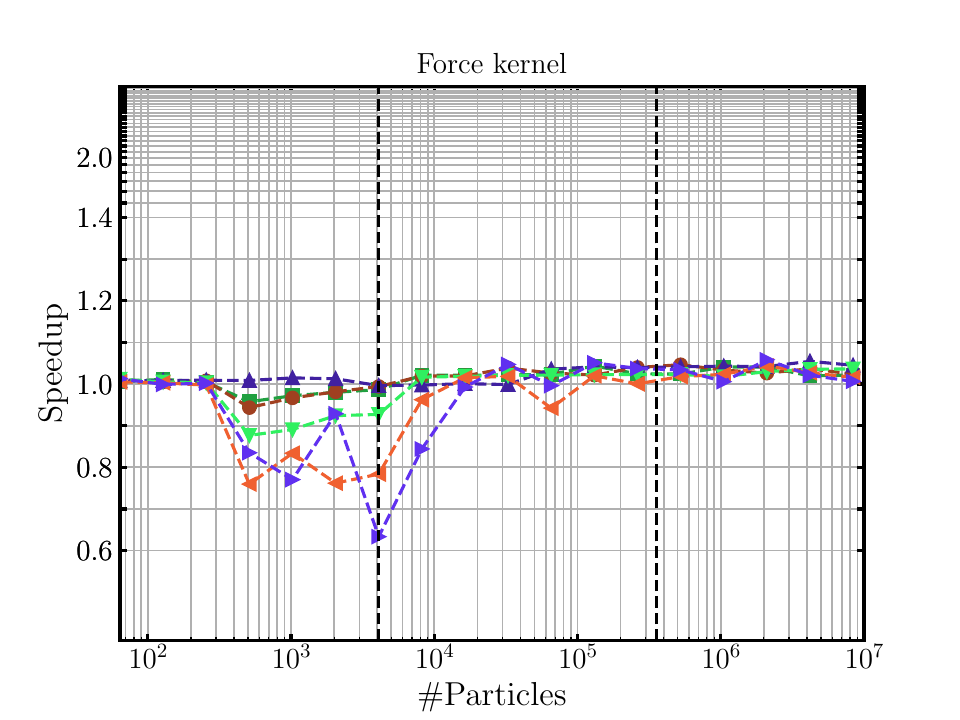}
  \end{center}
  \caption{
    Measurements for the force kernel kernel.
    Throughput (top) and speedup relative to uncompressed baseline
    version (bottom) on the Genoa testbed for stream-like
    access (left) and task-based access characteristics (right).
    We use one NUMA domain.
    \label{figure:appendix:scalability:force:genoa-1numa}
  } 
\end{figure}

\begin{figure}[htb]
  \begin{center}
    \includegraphics[width=0.46\textwidth]{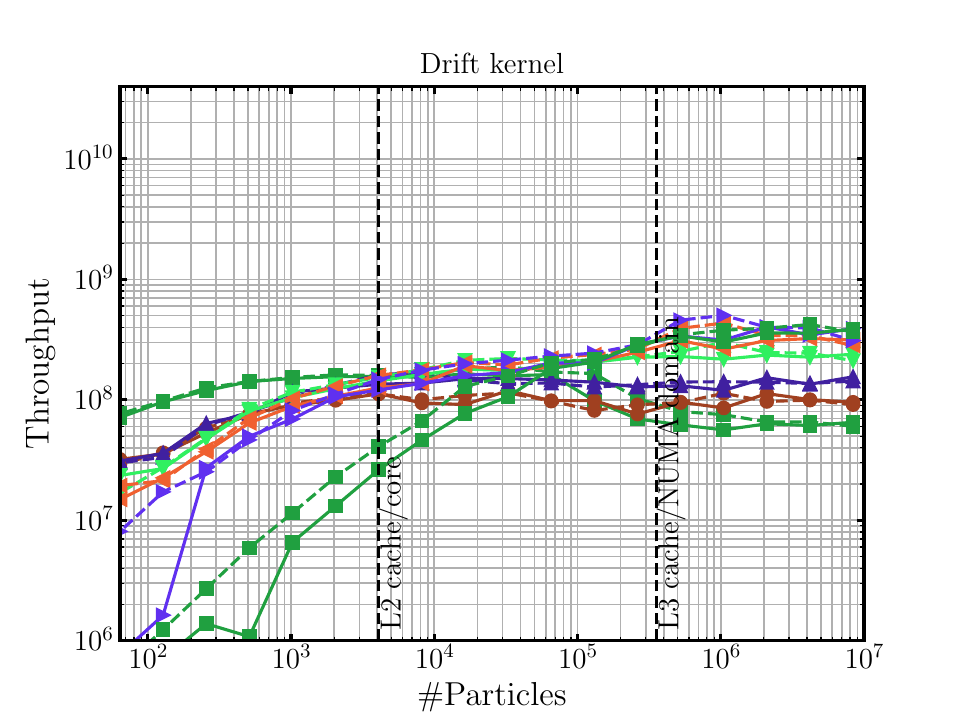}
    \includegraphics[width=0.46\textwidth]{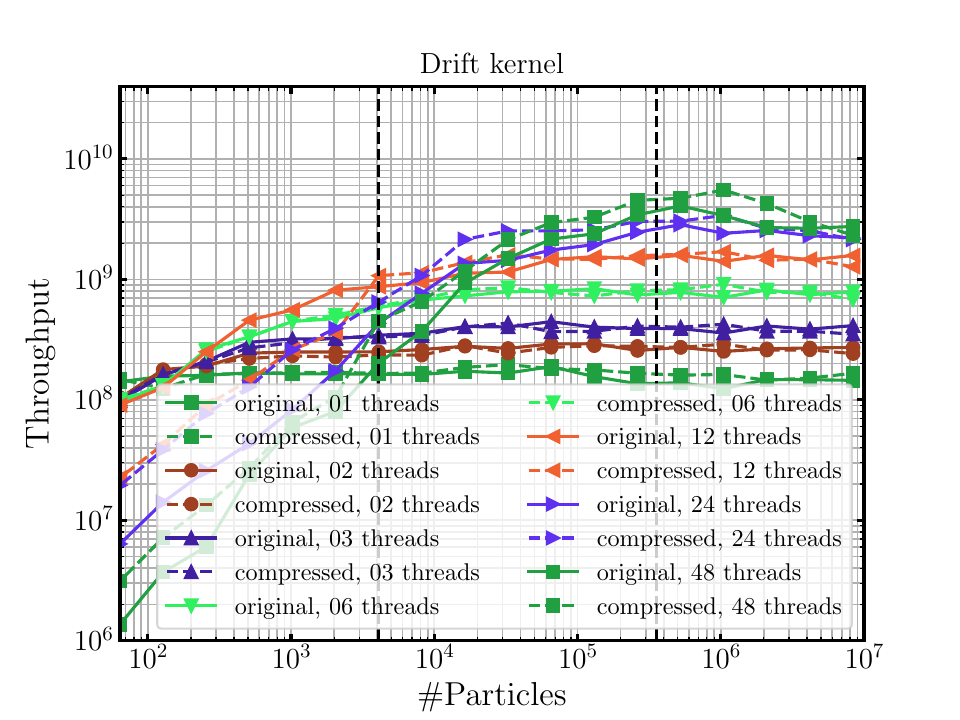}
    \includegraphics[width=0.46\textwidth]{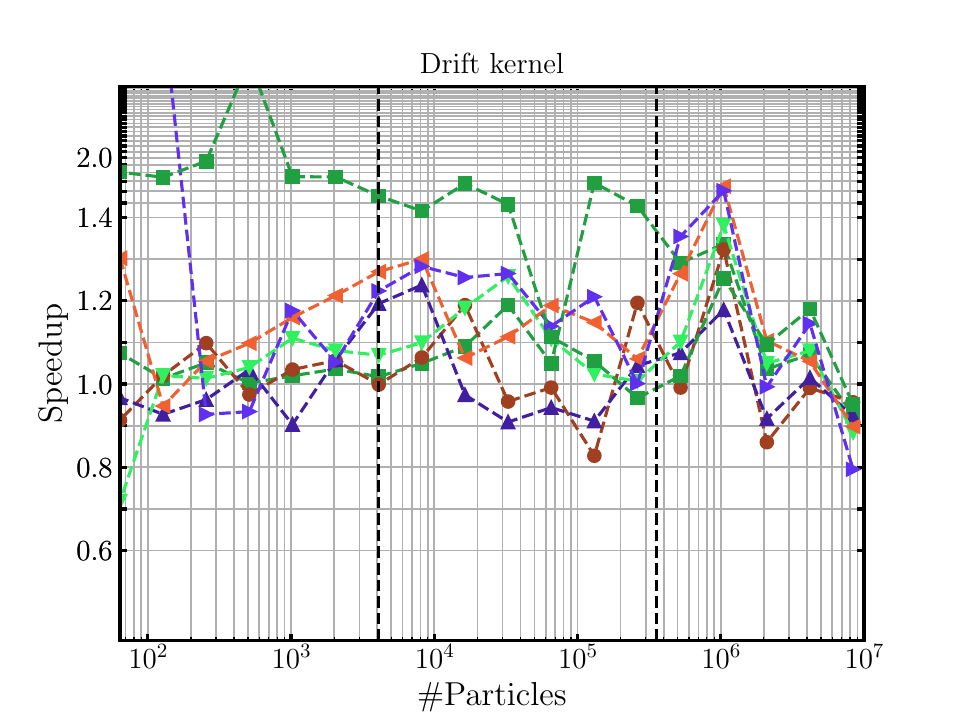}
    \includegraphics[width=0.46\textwidth]{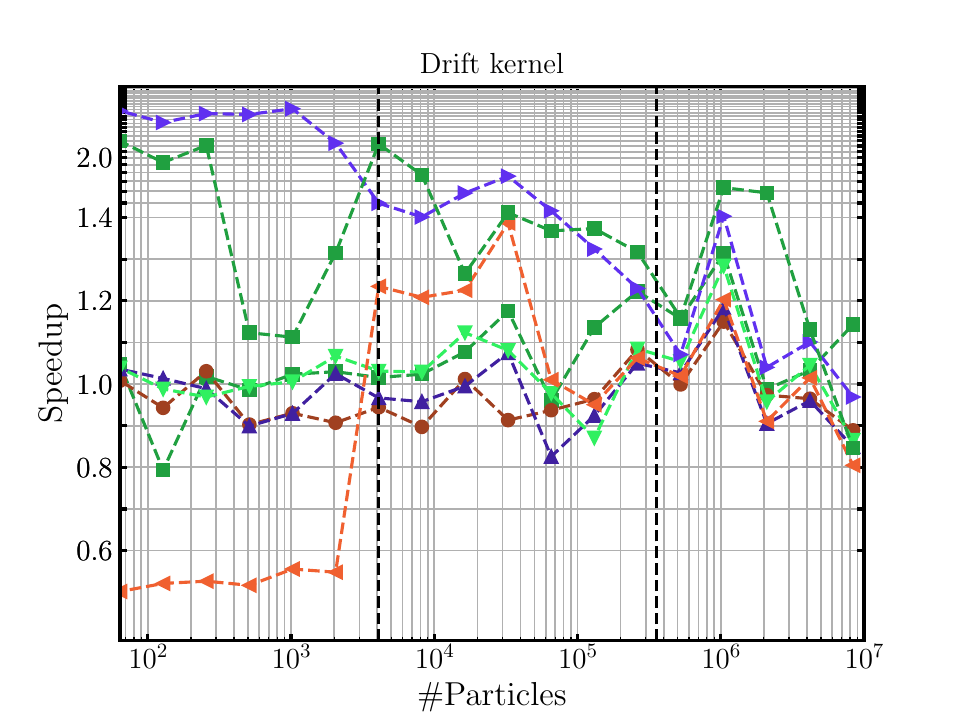}
  \end{center}
  \caption{
    Measurements for the drift kernel.
    Throughput (top) and speedup relative to uncompressed baseline
    version (bottom) on the Genoa testbed for stream-like access (left) and
    task-based access characteristics (right).
    We use two NUMA domains.
    \label{figure:appendix:scalability:drift:genoa-2numa}
  } 
\end{figure}

\begin{figure}[htb]
  \begin{center}
    \includegraphics[width=0.46\textwidth]{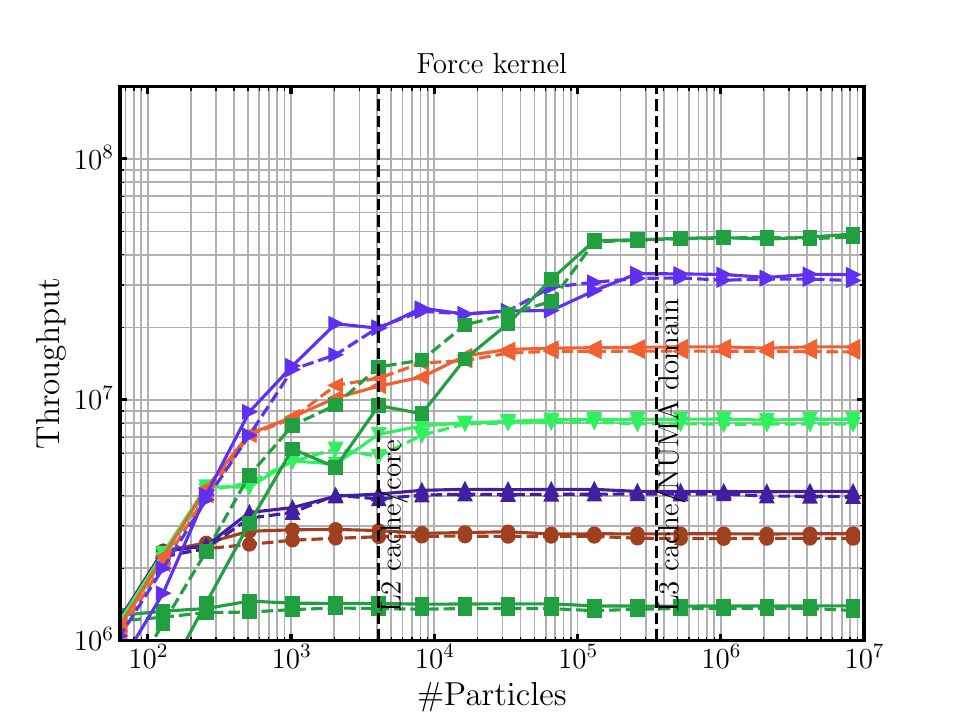}
    \includegraphics[width=0.46\textwidth]{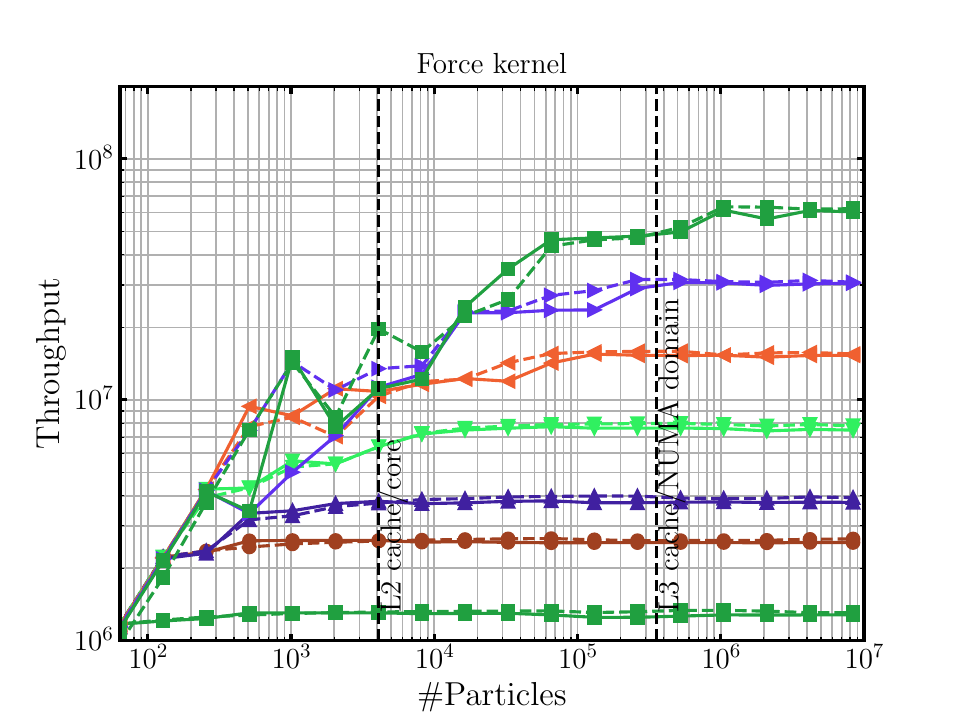}
    \includegraphics[width=0.46\textwidth]{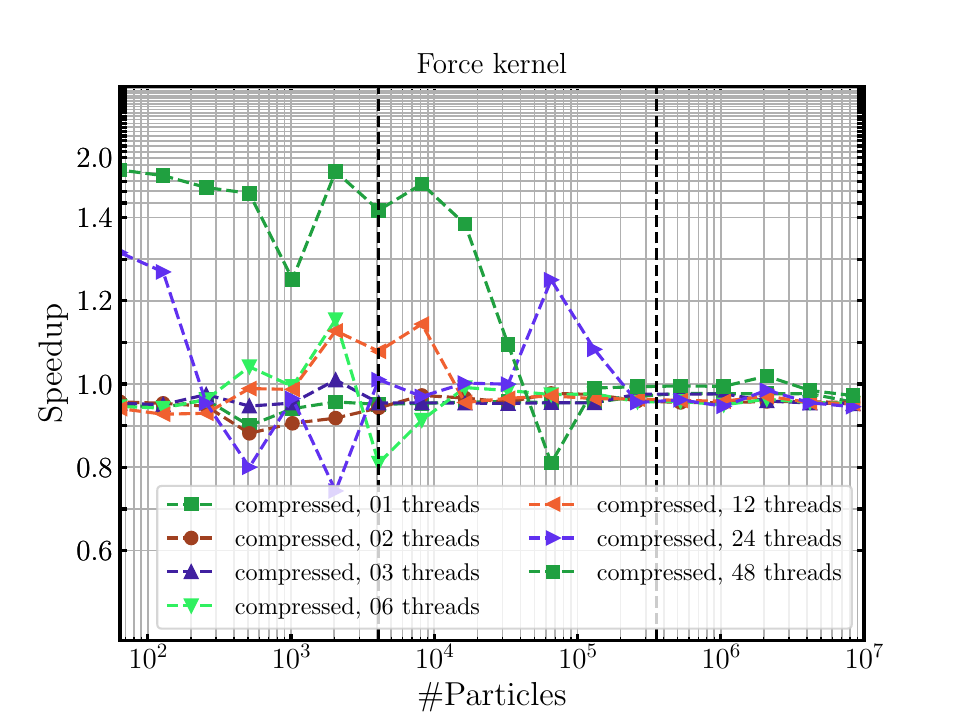}
    \includegraphics[width=0.46\textwidth]{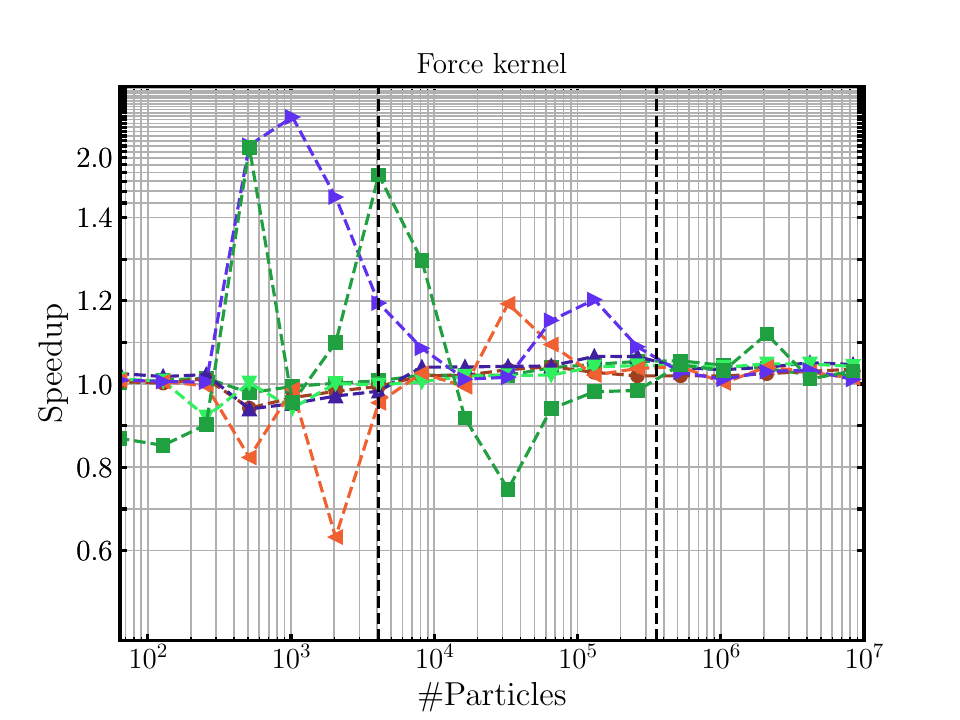}
  \end{center}
  \caption{
    Measurements for the force kernel kernel.
    Throughput (top) and speedup relative to uncompressed baseline
    version (bottom) on the Genoa testbed for stream-like
    access (left) and task-based access characteristics (right).
    We use two NUMA domains.
    \label{figure:appendix:scalability:force:genoa-2numa}
  } 
\end{figure}

\begin{figure}[htb]
  \begin{center}
    \includegraphics[width=0.46\textwidth]{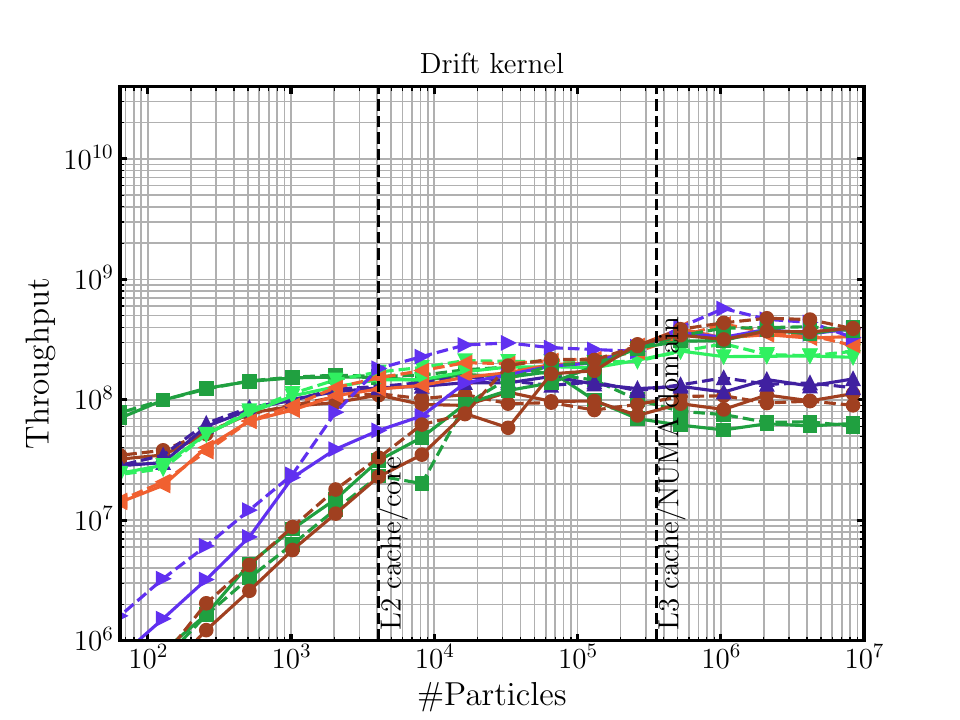}
    \includegraphics[width=0.46\textwidth]{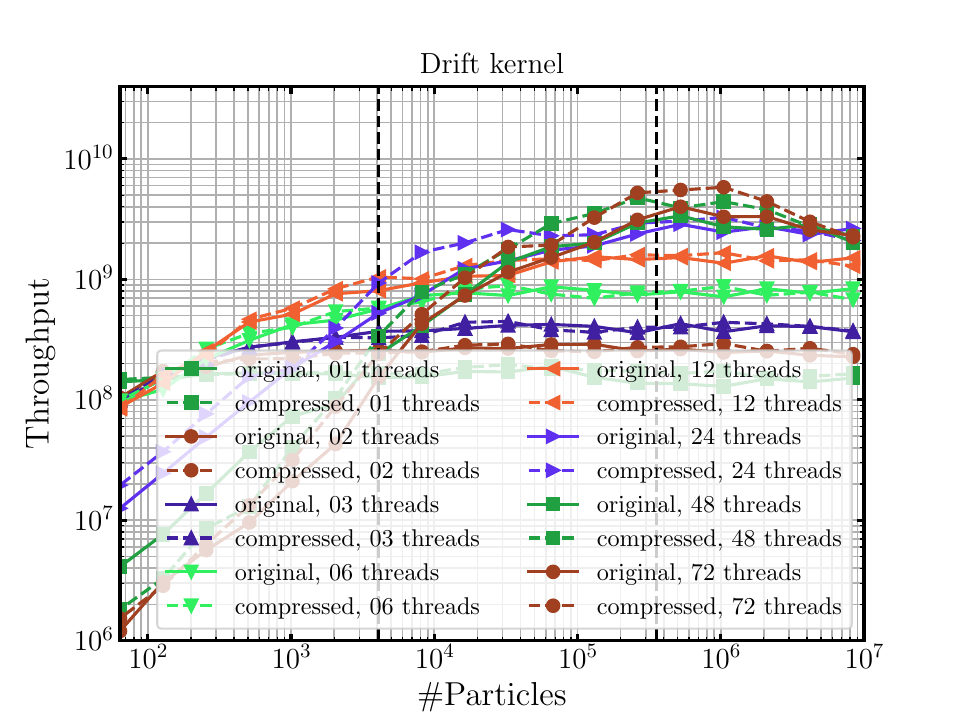}
    \includegraphics[width=0.46\textwidth]{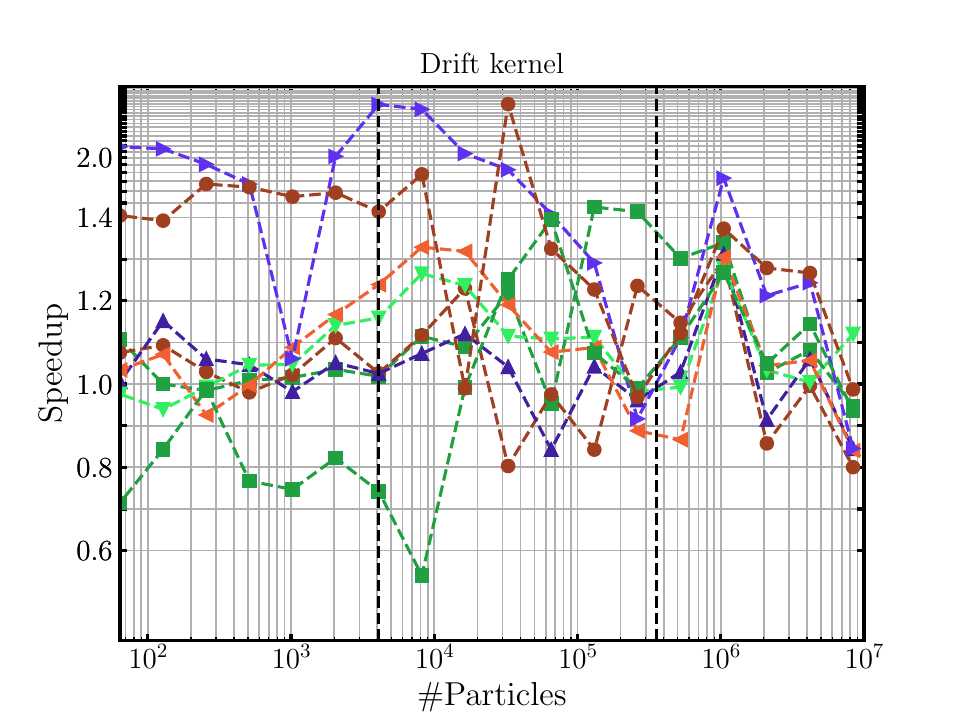}
    \includegraphics[width=0.46\textwidth]{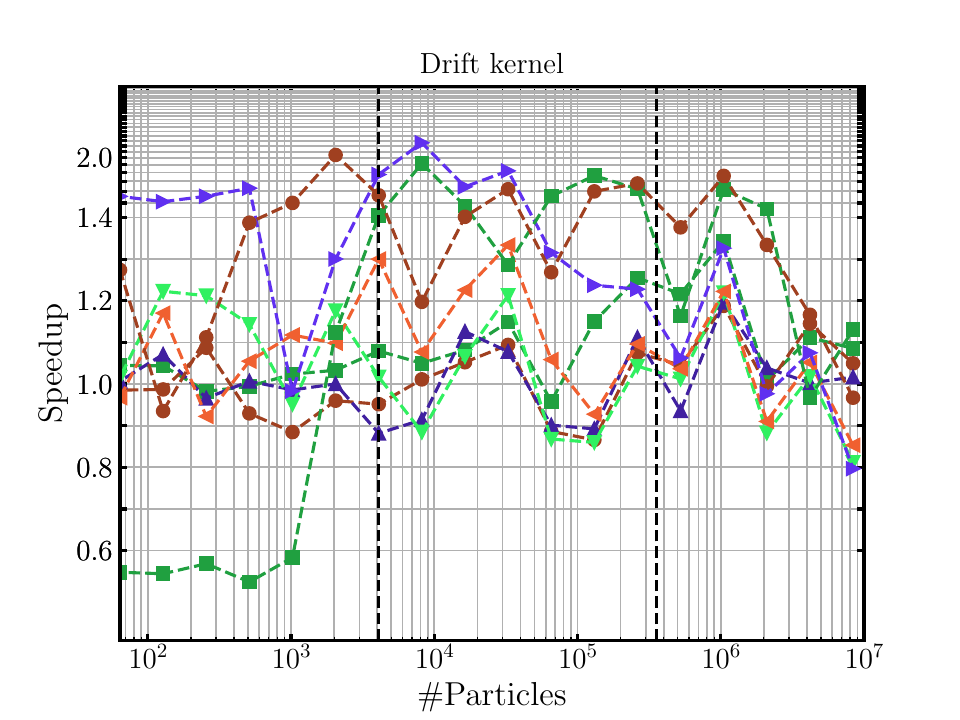}
  \end{center}
  \caption{
    Measurements for the drift kernel.
    Throughput (top) and speedup relative to uncompressed baseline
    version (bottom) on the Genoa testbed for stream-like access (left) and
    task-based access characteristics (right).
    We use three NUMA domains.
    \label{figure:appendix:scalability:drift:genoa-3numa}
  } 
\end{figure}

\begin{figure}[htb]
  \begin{center}
    \includegraphics[width=0.46\textwidth]{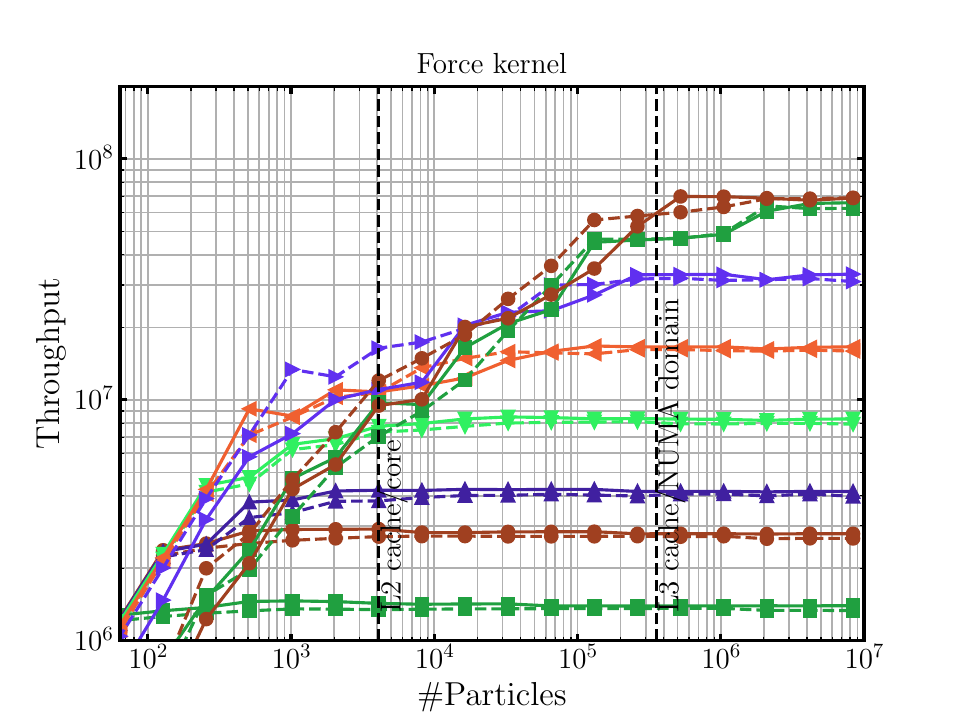}
    \includegraphics[width=0.46\textwidth]{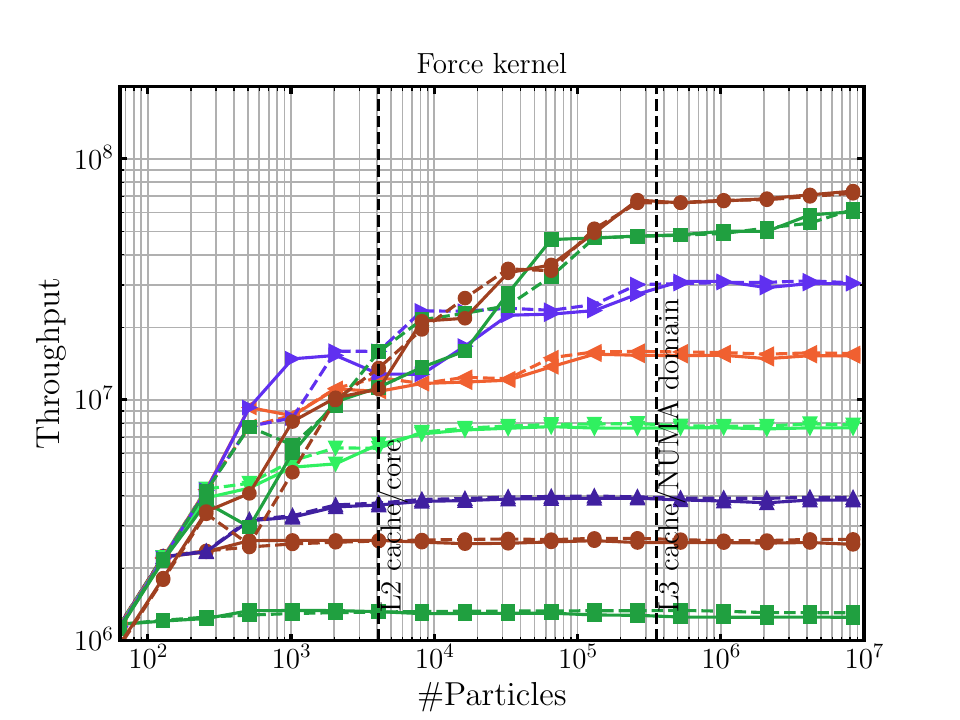}
    \includegraphics[width=0.46\textwidth]{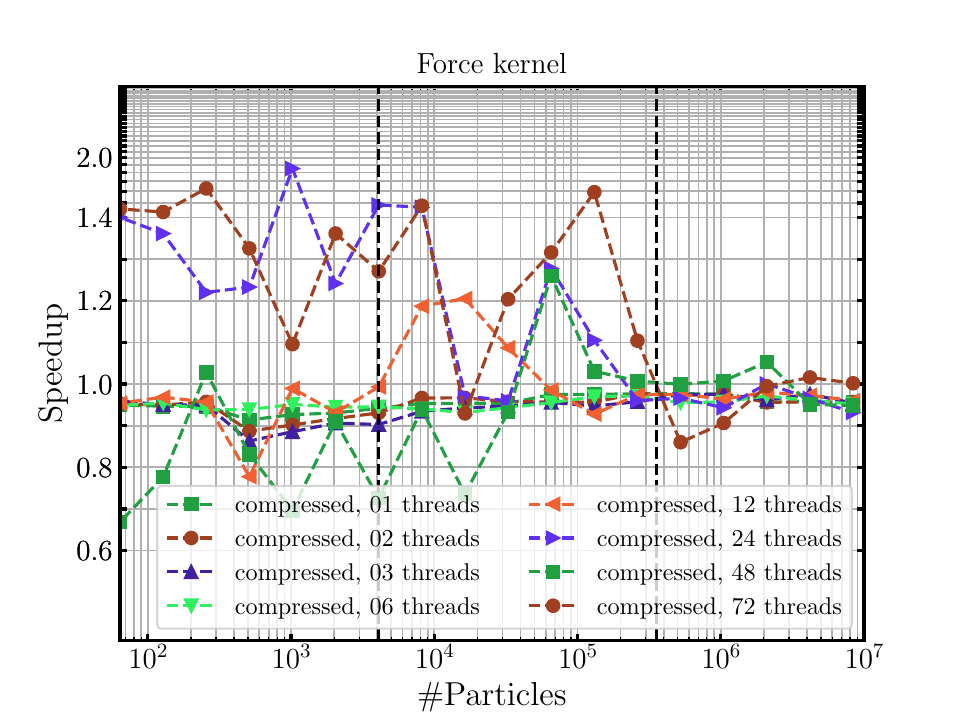}
    \includegraphics[width=0.46\textwidth]{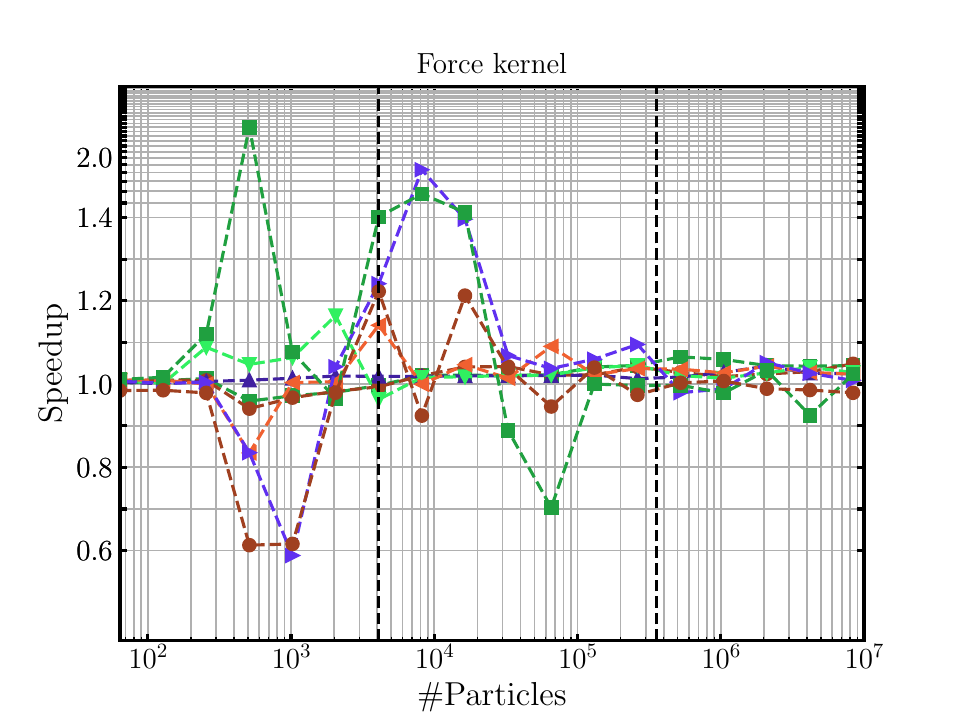}
  \end{center}
  \caption{
    Measurements for the force kernel kernel.
    Throughput (top) and speedup relative to uncompressed baseline
    version (bottom) on the Genoa testbed for stream-like
    access (left) and task-based access characteristics (right).
    We use three NUMA domains.
    \label{figure:appendix:scalability:force:genoa-3numa}
  } 
\end{figure}

\begin{figure}[htb]
  \begin{center}
    \includegraphics[width=0.46\textwidth]{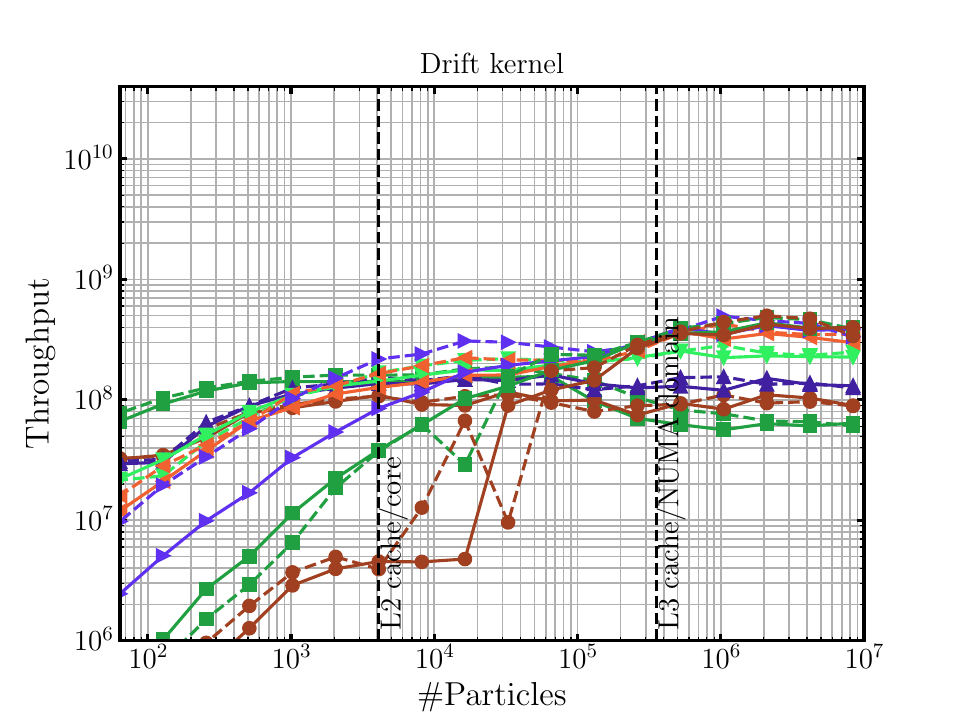}
    \includegraphics[width=0.46\textwidth]{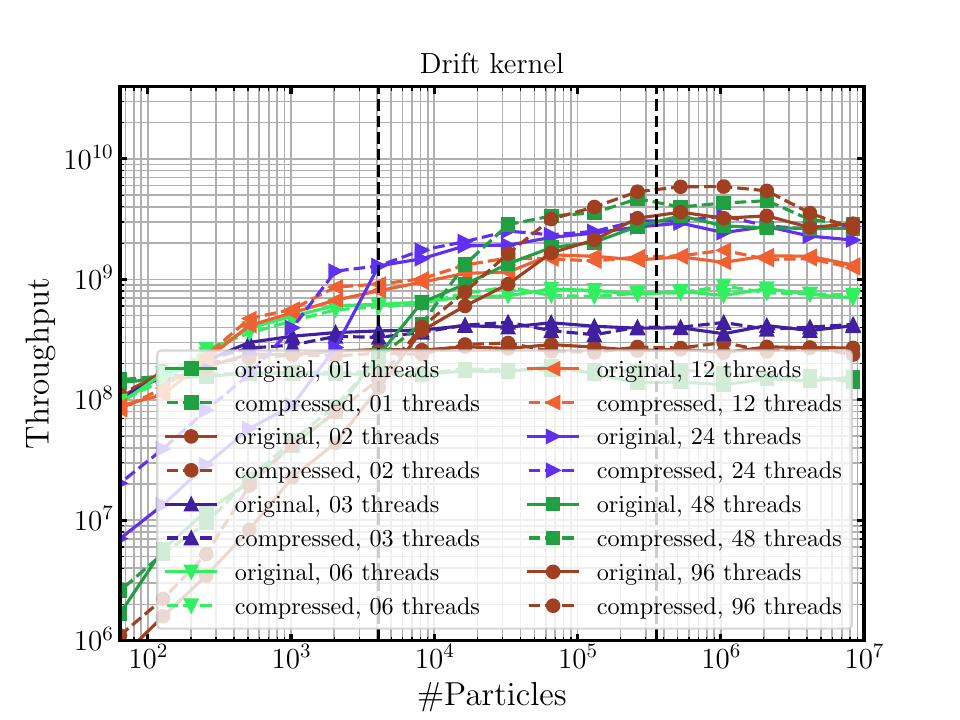}
    \includegraphics[width=0.46\textwidth]{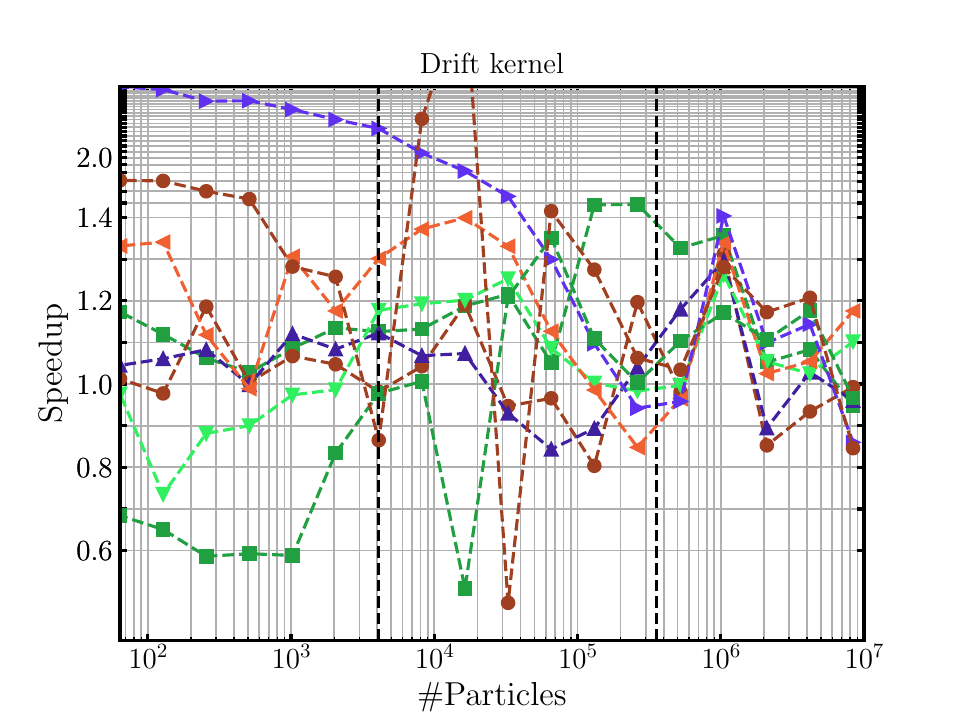}
    \includegraphics[width=0.46\textwidth]{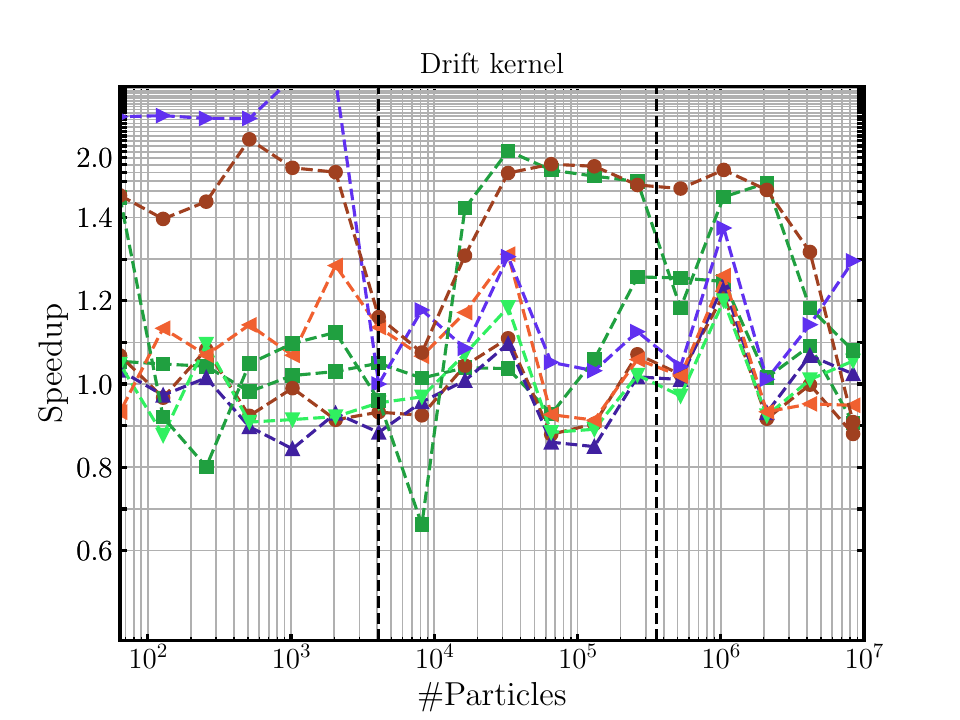}
  \end{center}
  \caption{
    Measurements for the drift kernel.
    Throughput (top) and speedup relative to uncompressed baseline
    version (bottom) on the Genoa testbed for stream-like access (left) and
    task-based access characteristics (right).
    We use four NUMA domains.
    \label{figure:appendix:scalability:drift:genoa-4numa}
  } 
\end{figure}

\begin{figure}[htb]
  \begin{center}
    \includegraphics[width=0.46\textwidth]{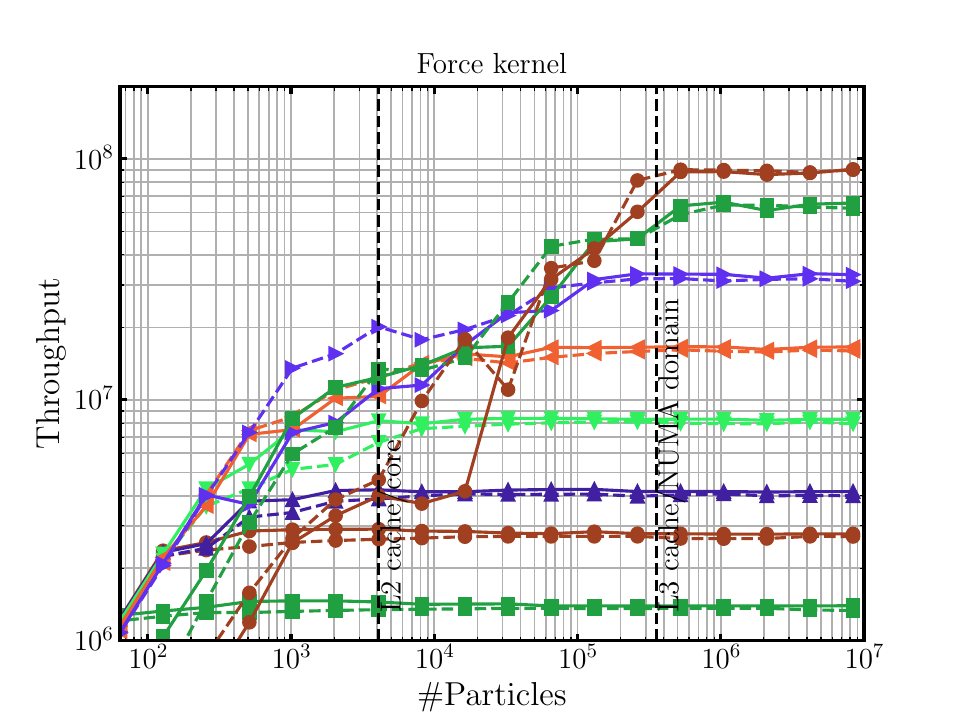}
    \includegraphics[width=0.46\textwidth]{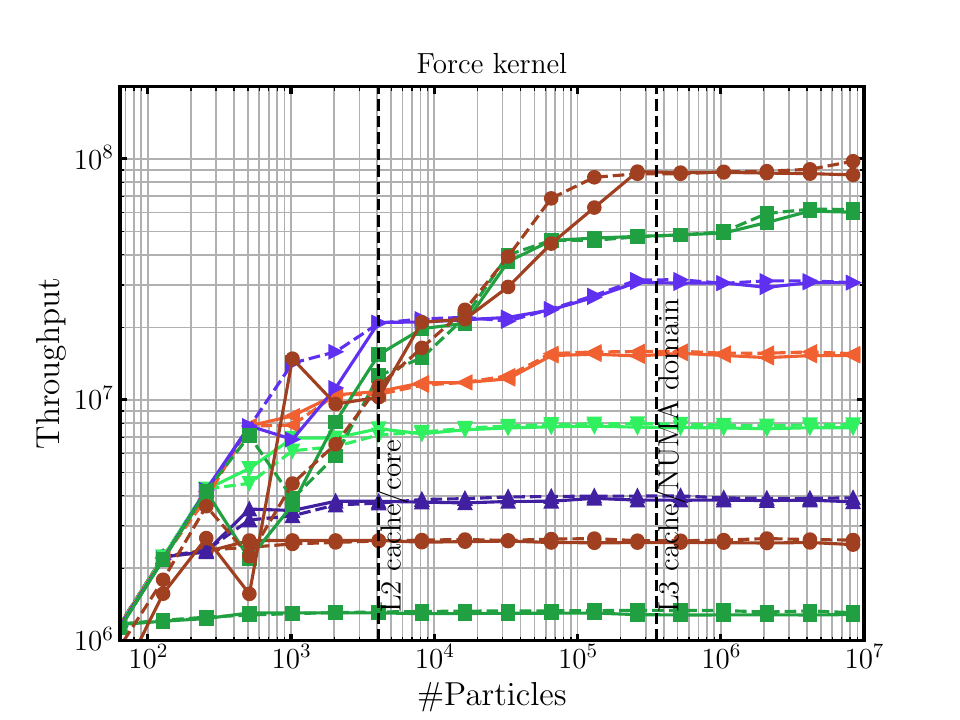}
    \includegraphics[width=0.46\textwidth]{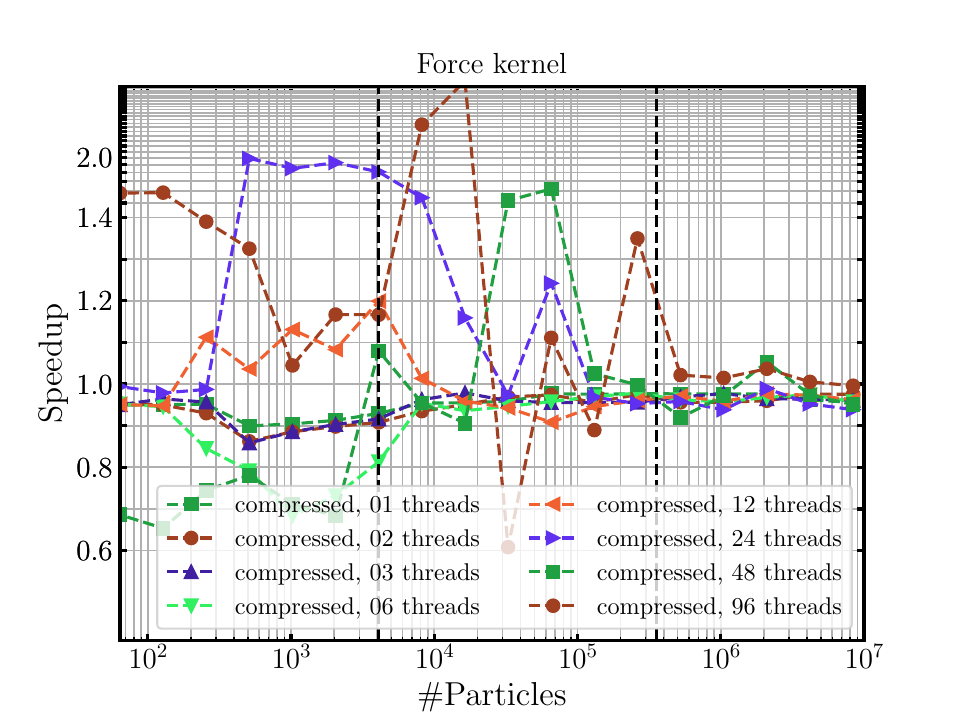}
    \includegraphics[width=0.46\textwidth]{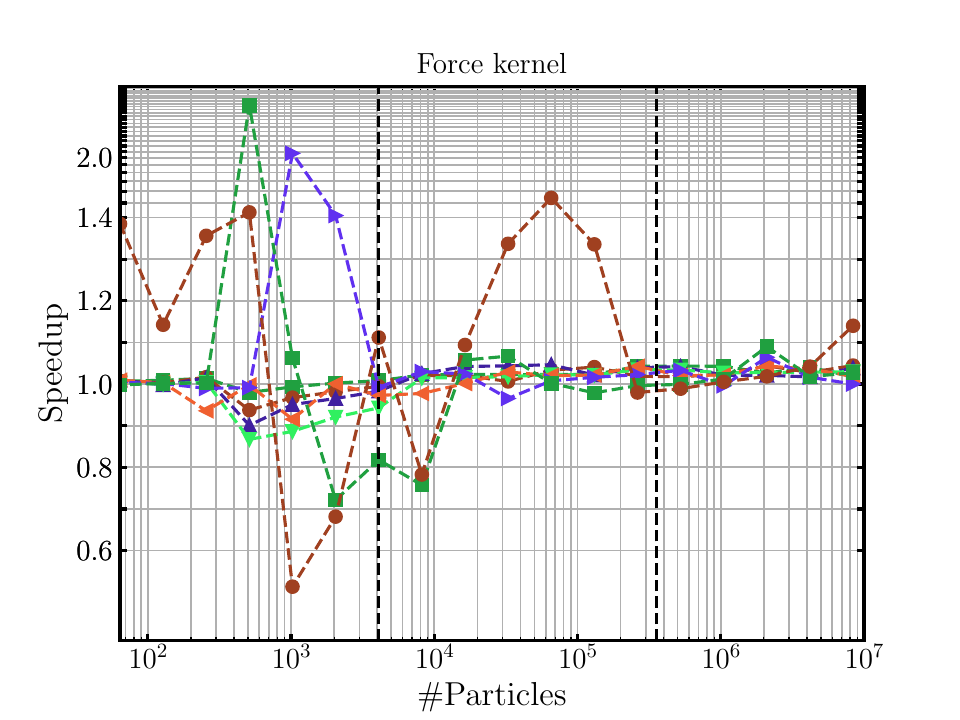}
  \end{center}
  \caption{
    Measurements for the force kernel kernel.
    Throughput (top) and speedup relative to uncompressed baseline
    version (bottom) on the Genoa testbed for stream-like
    access (left) and task-based access characteristics (right).
    We use four NUMA domains.
    \label{figure:appendix:scalability:force:genoa-4numa}
  } 
\end{figure}

Additional benchmark data for the Sapphire Rapid testbed are available from the
Figures
\ref{figure:appendix:scalability:drift:sapphirerapids} and 
\ref{figure:appendix:scalability:force:sapphirerapids}.

For the Genoa testbed, we collect data for one NUMA domain
(Figures~\ref{figure:appendix:scalability:drift:genoa-1numa},
\ref{figure:appendix:scalability:force:genoa-1numa}), two
(Figures~\ref{figure:appendix:scalability:drift:genoa-2numa},
\ref{figure:appendix:scalability:force:genoa-2numa}),
three (Figures~\ref{figure:appendix:scalability:drift:genoa-3numa},
\ref{figure:appendix:scalability:force:genoa-3numa}), and four domains 
(Figures~\ref{figure:appendix:scalability:drift:genoa-4numa},
\ref{figure:appendix:scalability:force:genoa-4numa}).
It is important to note that there are multiple benchmark curves for some thread
choices:
24 threads for example can be distributed over one, two, three or four NUMA
domains.
A spread affinity policy yields significantly worse throughput overall.

Each test is run at least 16 times and data are averaged over these tests.

\section{Smoothed Partichle Hydrodynamics: The Governing Equations}\label{app:SPH}

\subsection{General Remarks}

Smoothed Particle Hydrodynamics (SPH) is a class of meshless methods wherein
the fluid is discretized using particles. Those particles are typically
given some constant mass and are evolved in time using the Lagrangian equations
of fluid dynamics. SPH is based on estimating the local fluid density (and other
quantities) as a weighted sum over neighboring particles, where the
weights are smoothly decreasing functions (kernels) such that the noise in the
density estimate introduced by distant neighboring particles is reduced. More
precisely, let $A(\vec{x})$ be a scalar field of interest of a partial
differential equation. In an SPH description, we write down this quantity as
the convolution

\begin{align}
 A(\vec{x}) =
   \int A(\vec{x}') \delta (\vec{x}- \vec{x}') \d^3 \vec{x}' &\approx
   \int A(\vec{x}') W(\vert \vec{x}- \vec{x}'\vert, H) \d^3 \vec{x}'
   \label{eq:general-convolution-with-dirac}
    \\
   &\approx
    \sum _{i} \frac{m_i}{\rho _i} A(\vec{x}_i) W(\vert \vec{x}- \vec{x}_i\vert, H)\,.
   \label{eq:general-convolution-with-smoothing-kernel}
\end{align}

\noindent
In \eqref{eq:general-convolution-with-dirac}, the Dirac distribution $\delta
(\vec{x})$ is approximated by a smoothing kernel $W(\vec{x},H)$ which is a smooth
differentiable function with compact support $H$. While in principle
\eqref{eq:general-convolution-with-smoothing-kernel} requires a sum over all particles
$i$, the kernel's compact support reduces that problem to a sum over the local neighbourhood
for which $W(|\vec{x}|, H) > 0$.

\noindent
In this work, we use the quartic spline (M5) kernel~\cite{Monaghan:1985:Kernel}. Following
the notation convention of \cite{dehnenImprovingConvergenceSmoothed2012}, the kernel in $\nu$
dimensions is given by

\begin{equation*}
 W(\vec{x}, H) = H^{-\nu} \ w(|\vec{x}| / H)
\end{equation*}
with
\begin{equation}
   w(r) = C_{\rm norm} \times \begin{cases}
            (1 - r)^4 - 5\left(\tfrac{3}{5} - r\right)^4 + 10\left(\tfrac{1}{5} - r\right)^4
            & \text{ if } 0 \leq r \leq \tfrac{1}{5}  \\
            (1 - r)^4 - 5\left(\tfrac{3}{5} - r\right)^4
            & \text{ if } \tfrac{1}{5} \leq r \leq \tfrac{3}{5}  \\
            (1 - r)^4
            & \text{ if } \tfrac{3}{5} < r < 1
          \end{cases}
\end{equation}
where $C_{\rm norm} =\tfrac{5^5}{768},\ \tfrac{5^6 3}{2398\pi}\, \tfrac{5^6}{512 \pi} $ is a
normalisation constant for $\nu = 1, 2, 3$ dimensions, respectively. Table~1 in
\cite{dehnenImprovingConvergenceSmoothed2012} lists other popular SPH kernel choices.

%
%

%

\subsection{Determining the Density and Smoothing Lengths}

The smoothing length, $h$, plays a central role in SPH. Following the definition of
\cite{dehnenImprovingConvergenceSmoothed2012}, it is twice the standard deviation of
the kernel and defines the spatial resolution of the numerical method: Its size determines
the wavelength of acoustic waves that can be resolved.

\noindent
Naturally, the compact support
radius of a kernel and its smoothing length are related. The relation is typically given
as $H = \Gamma_k h$. For the quartic spline kernel that we use,
$\Gamma_k = 1.936492,\ 1.977173,\ 2.018932 $ for $\nu = 1,\ 2,\ 3$, respectively.

Since the smoothing length determines both the spatial resolution and the number
of neighbouring particles incorporated into the weighted sums,
different applications demand different requirements. Hence the smoothing length
can be specified via a free parameter $\eta$:
\begin{equation}
 h_i = \eta \left( \frac{m_i}{\rho_i} \right)^{1/\nu} \,
\end{equation}
%
%
defining $h$ in units of mean inter-particle distance. $\eta$ is typically fixed
in the range 1.2--1.5~\cite{Steinmetz:1993:SPH}. In our work, we use
$\eta = 1.2348$.

\noindent
However, since the particle density $\rho_i$ is determined through the smoothing
\begin{equation}
 \rho_i = \sum\limits_j m_j W_{ij}(h_i) \label{eq:densityCalculation}
\end{equation}
with $W_{ij}(h_i) = W(\vec{x}_i - \vec{x}_j, H(h_i))$, we're left with a circular relation:
$\rho_i$ is required to determine $h_i$, while $h_i$ is needed to estimate $\rho_i$.
As a consequence, the algorithm runs a cascade of Picard Newton-Raphson
iterations per particle to determine $h_i$ and $\rho_i$. Since adaptive and variable smoothing
lengths (in both space and time) are crucial for cosmological applications (given that the
fluid can be compressed over a range spanning several orders of magnitude) this iteration needs
to be performed each time step.

\subsection{Equations of Motion}

With the particles' smoothing lengths and densities determined, we can now turn to the
equations of motion.
For the present SPH demonstrator, we consider an inviscid fluid in the absence
of gravity and external forces or energy sources. Hence, the individual particles tracking
the fluid evolve according to the Euler equation,
\begin{align}
\frac{\d \vec{v}_i}{\d t} &=
    -\sum_j m_j \left[
        f_i\frac{P_i}{\rho_i^2}{\bf \nabla} W_{ij}(h_i) +
        f_j\frac{P_j}{\rho_j^2}{\bf \nabla} W_{ij}(h_j)
    \right] +
    \vec{a}^{\rm AV}_i\,, \label{eq:Euler-equation-particles}
\end{align}
\noindent
while the thermodynamic internal energy per unit mass of the fluid, $u_i$, evolves
according to
\begin{align}
\frac{\d u_i}{\d t} &=
    f_i\frac{P_i}{\rho_i^2}
        \sum_j m_j(\vec{v}_i - \vec{v}_j)
        \cdot{\bf \nabla} W_{ij}(h_i) + \dot{u}^{\rm AV}_i\,.
\label{eq:internal-energy-evolution}
\end{align}

\noindent
$\vec{v}$ is the velocity field, $P$ is the pressure and
${\bf\nabla}\equiv\partial/\partial{\vec{x}}$ is the spatial gradient.
The system is closed by specifying the equation
of state of the fluid, $P=(\gamma - 1)u\rho$, in which $\gamma$ is the adiabatic
index.

The used equations include physical quantities of the fluid plus terms that are intrinsic to the
SPH method. The scalar field

\begin{equation}
f_i = \left(1 + \frac{h_i}{3\rho_i}\frac{\partial\rho_i}{\partial h_i} \right)^{-1}
\quad
\text{with}
\quad
	\frac{\partial\rho_i}{\partial h_i}= 	\sum_j m_j\frac{\partial W_{ij}(h_i)}{\partial h_i}
	\label{eq:drho_dh}
\end{equation}

\noindent
represents the
spatial fluctuations in the smoothing length $h(\vec{x})$ (typically known as
`grad-$h$' terms).
They have to be taken into account whenever $h$ is allowed
to vary over space or time.
Note that the sum in \eqref{eq:drho_dh} can be collected simultaneously with
the density field calculation \eqref{eq:densityCalculation}.

Following~\cite{Monaghan:92:SPH,Balsara:95:SPH}, and \cite{Price:2012:SPH}, we add an artificial viscosity
(AV) to the (physically inviscid) fluid in order to resolve potential discontinuities (e.g.~due to shocks) that
could develop in the fluid. In particular, we adopt the AV model used by
the {\sc gadget}-2 code~\cite{Springel-g2:2005}.
Its contribution to
the acceleration in \eqref{eq:Euler-equation-particles} is given by
\begin{align}
\vec{a}^{\rm AV}_i &=
    -\sum_j m_j \Pi_{ij}\nabla \overline{W}_{ij}\,.
    \label{eq:AV-terms-Euler-equation-particles}
\end{align}
\noindent
We pick $\overline{W}_{ij}\equiv\left[W_{ij}(h_i) + W_{ij}(h_j)\right]/2$, whereas
$\Pi_{ij}$ is the artificial viscosity tensor

\begin{align} \label{eq:viscosity-tensor-minimal-sph}
    \Pi_{ij} = - \frac{\alpha^{\rm AV}}{2}
        \frac{v^{\rm sig}_{ij}\mu_{ij}}{(\rho_i + \rho_j)/2}
        \frac{(B_i + B_j)}{2}\,.
\end{align}

\noindent
In \eqref{eq:viscosity-tensor-minimal-sph}, $\alpha^{\rm AV}=1$ is a (free)
artificial viscosity parameter,
$v^{\rm sig}_{ij} = c_{s,i} + c_{s,j} - \beta^{\rm AV}\mu_{ij}$ is the signal velocity with
$c_{s,i}=\sqrt{\gamma P_i/\rho_i}$ the sound speed of the fluid at position $\vec{x}_i$,
and $\beta^{\rm AV}=3$ is the second viscosity parameter in this model.
$\mu_{ij}$ is given by

\[
  \mu_{ij}=
    \begin{cases}
    \vec{v}_{ij}\cdot\hat{\vec{x}}_{ij}
        & {\rm if\ } {\vec{v}}_{ij}\cdot\hat{\vec{x}}_{ij} < 0\\
    0
        & {\rm otherwise}\\
    \end{cases}
\]

\noindent
where $\hat{\vec{x}}_{ij}$ is the unit position vector separating particles $i$
and $j$ and $\vec{v}_{ij} = \vec{v}_j - \vec{v}_i$.
The term switches the viscous tensor
\eqref{eq:viscosity-tensor-minimal-sph} on whenever two particles
approach each other.
Lastly, the Balsara switch $B_i$~\cite{Balsara:95:SPH} is modelled as

\begin{align}\label{eq:balsara-minimal-sph}
    B_i = \frac{\vert\nabla\cdot{\bf v}_i\vert}
    {\vert\nabla\cdot{\bf v}_i\vert + \vert\nabla\times{\bf v}_i\vert +
    10^{-4}c_{s,i}/h_i}\,.
\end{align}

The divergence and curl of the velocity field are computed using the standard SPH expressions:
\begin{align*}
  \nabla\cdot\vec{v}_i &= \frac{1}{\rho_i}
    \sum_j m_j
    \vec{v}_{ij}\cdot\nabla W(\vec{x}_{ij}, h_i), \\
\nabla\times\vec{v}_i &=
    \frac{1}{\rho_i} \sum_j m_j
    \vec{v}_{ij}\times\nabla W(\vec{x}_{ij}, h_i) .
\end{align*}
\noindent
Likewise, the AV diffusion term for the evolution equation of the internal
energy \eqref{eq:internal-energy-evolution} is
\begin{align} \label{eq:AV-terms-internal-energy-evolution}
u^{\rm AV}_i &=-\frac{1}{2}\sum_j m_j \Pi_{ij}{\bf v}_{ij}\cdot\nabla \overline{W}_{ij}\,.
\end{align}

\subsection{Time Integration}

Finally, the evolution equations \eqref{eq:Euler-equation-particles} and
\eqref{eq:internal-energy-evolution} are supplemented with a well-suited
time integrator.
We use a kick-drift-kick leapfrog in a velocity-Verlet form:
\begin{align*}
    \vec{v}_{i}^{n+1/2} &= \vec{v}_i^n + \vec{a}_i^n \tfrac{\Delta t}{2}  && \text{kick} \\
    \vec{x}_{i}^{n+1} &= \vec{x}_i^n + \vec{v}_{i}^{n+1/2} \Delta t  && \text{drift} \\
    \vec{v}_{i}^{n+1} &= \vec{v}_i^{n+1/2} + \vec{a}_{i}^{n+1} \tfrac{\Delta t}{2}  && \text{kick}
\end{align*}
The evaluation of the updated acceleration $\vec{a}_i^{n+1} = \frac{{\rm d} \vec{v}_i}{{\rm d} t}$
according to \eqref{eq:Euler-equation-particles} as well as the thermodynamic update
\eqref{eq:internal-energy-evolution} are applied in the midpoint in time of the integration step after
the drift operation. The maximally permissible time step size $\Delta t$ is determined using the
CFL condition
\begin{align}
\Delta t_i = 2 C_{\rm cfl} \frac{H_i}{ \max_j v_{ij}^{\rm sig} }
\end{align}
where $0 < C_{\rm cfl} \leq 1$ is a free parameter, typically set to be $0.1$.

\end{document}